\documentclass[a4paper,11pt]{article}
\pdfoutput=1
cm \voffset = -2truecm

\usepackage{graphicx}
\usepackage{amsmath}
\usepackage{amssymb}
\usepackage{latexsym}
\usepackage{color}
\usepackage{cite}
\usepackage{makeidx}
\usepackage{float}
\usepackage{multirow}
\newcommand{\bra}{\begin{array}}
\newcommand{\era}{\end{array}}
\newcommand{\beq}{\begin{equation}}
\newcommand{\eeq}{\end{equation}}
\newcommand{\bqr}{\begin{eqnarray}}
\newcommand{\eqr}{\end{eqnarray}}

\def\BC{\bb C}
\def\_\BC{\bbi C}

\def\( {\left(}
   \def\) {\right)}
\def\[ {\left[}
\def\] {\right]}
\def\no2 {{\textstyle{n\over 2}}}

\def\dag {{\dagger}}
\newcommand{\lga}{\longrightarrow}

\newcommand{\hb}{\hbar}

\newcommand{\lb}{\label}

\begin{document}
\begin{titlepage}
\setcounter{page}{1}
\renewcommand{\thefootnote}{\fnsymbol{footnote}}

\begin{flushright}
%ucd-tpg 30-06-2013\\
%arXiv:yymm.xxxx
\end{flushright}

\vspace{5mm}
\begin{center}

{\Large \bf {%Transmission and Conductance
Band Tunneling through Double Barrier
in Bilayer Graphene }}

\vspace{5mm}

{\bf Hasan A. Alshehab}$^{a,b}$,  {\bf Hocine Bahlouli}$^{a,b}$,
{\bf Abderrahim El Mouhafid}$^{c}$ and {\bf Ahmed
Jellal}$^{b,c}$\footnote{ajellal@ictp.it, a.jellal@ucd.ac.ma}

\vspace{5mm}

{$^a$\em Physics Department,  King Fahd University
of Petroleum $\&$ Minerals,\\
Dhahran 31261, Saudi Arabia}

{$^b$\em Saudi Center for Theoretical Physics, Dhahran, Saudi
Arabia}

{$^{c}$\em Theoretical Physics Group,  %Department of Physics,
Faculty of Sciences, Choua\"ib Doukkali University},\\
{\em PO Box 20, 24000 El Jadida, Morocco}

%{$^d$\em Physics Department, College of Sciences, King Faisal University,\\
%PO Box 9149, Alahssa 31982, Saudi Arabia}

\vspace{3cm}

\begin{abstract}

By taking into account the full four band energy spectrum, we
calculate the transmission probability and conductance of
electrons across symmetric and asymmetric double potential barrier
with a confined interlayer potential difference in
bilayer graphene.  For energies less than the interlayer coupling
$\gamma_{1}$, $E<\gamma_1$, we have one channel for transmission
which exhibits resonances, even for incident particles with
energies less than the strength of barriers, $E< U_j$, depending
on the double barrier geometry. In contrast, for higher energies
$E > \gamma_{1}$, we obtain four possible ways for transmission
resulting from the two propagating modes. We compute the
associated transmission probabilities as well as their
contribution to the conductance, study the effect of the double
barrier geometry.

% \pacs{02.30.Gp, 02,30.Tb, 02.30.Jr}

% \maketitle

%\newpage

\vspace{3cm}

%73.23.-b; 72.80.Rj %11.80.-m

\end{abstract}
\end{center}
\end{titlepage}

%\newpage
%%%%%%%%%%%%%%%%%%%%%%%%%%%%%%%%%%%%%%%%%%%%%%%%%%
\section{Introduction}
%%%%%%%%%%%%%%%%%%%%%%%%%%%%%%%%%%%%%%%%%%%%%%%%%%

In the last few years, graphene \cite{Geim}, a two dimensional one
atom thick sheet of carbon,
 became a hot research topic in the field of  condensed matter physics.
Its exceptional, electronic, optical, thermal, and mechanical
properties have potential future applications. For example, its
thermal conductivity is 15 times larger than that of copper and
its electron mobility is 20 times larger than that of GaAs. In
addition, it is considered as one of the strongest materials with
a Young's modulus of about 1 TPa, and some 200 times stronger than
structural steel\cite{1}.  The most important application of
graphene is to possibly replace silicon in IT-technology. But the
biggest obstacle is to create a gap and control the electron
mobility in graphene taking into account the so called Klein
tunneling, which makes the task more complicated\cite{2,3}.
However, one can create an energy gap in the spectrum in many
different ways, such as by coupling to substrate or doping with
impurities \cite{4,5} or in bilayer graphene by applying an
external electric field \cite{6,7}.

Bilayer graphene is two stacked (Bernal stacking \cite{8})
monolayer graphene sheets, each with honeycomb crystal structure,
with four atoms in the unit cell, two in each layer. In the first
Brillouin zone, the tight binding model for bilayer graphene
\cite{9} predicted four bands, two conduction bands and two
valance bands, each pair is separated by an interlayer coupling
energy of order $\gamma_{1}\approx0.4\ eV$ \cite{10}. At the Dirac
points, one valance band and one conduction band touch at zero
energy, whereas the other bands are split away from the zero
energy by $\gamma_{1}$ \cite{11}. Further details about band
structure and electronic properties of bilayer graphene can be
found in the literature \cite{12, 13, 14, 15, 16, 17, 18, 19, 20}.
Tunneling of quasiparticles in graphene, which mimics relativistic
quantum particles such as Dirac fermions in quantum
electrodynamics (QED), plays a major role in scattering theory. It
allows to develop a theoretical framework, which leads to
investigate different physical phenomena that are not present in
the non relativistic regime, such as the Klein-paradox \cite{2,3}.

In monolayer graphene, there were many studies on the tunneling of
electrons through different potential shapes \cite{22,23, 24, 25}.
While the study of tunneling electrons in bilayer graphene has
been restricted to energies less than the interlayer coupling
parameter $\gamma_1$ so that only one channel dominates
transmission and the two band
model is valid \cite{26, 27, 28, 29}. %$[26-29]$.
Recently, tunneling of electrons in bilayer graphene has been
studied using the four band model for a wide range of energies,
even for energies larger than $\gamma_1$ \cite{30}. New
transmission resonances were found that appear as sharp peaks in
the conductance, which are absent in the two band approximation.

Motivated by different developments and in particular \cite{30},
we investigate the band tunneling through square double barrier in
bilayer graphene. More precisely, the transmission probabilities
and conductance of electrons  will be studied  by tacking into
account the full four band energy spectrum. We analyze two
interesting cases by making comparison between the incident
energies $E$ and interlayer coupling parameter $\gamma_1$. Indeed,
for $E<\gamma_1$ there is only one channel of transmission
exhibiting resonances, even for incident particles with $E$ less
than the strength of barriers $U_j$ ($E< U_j$), depending on the
double barrier geometry. For $E > \gamma_{1}$, we end up with two
propagating modes resulting from four possible ways of
transmission. Subsequently, we use the transfer matrix  and
density of current to determine the transmission probabilities and
then the corresponding conductance. Based on the physical
parameters of our system, we present different numerical results
and make comparison with significant published works on the
subject.

The present paper is organized as follows. In section 2, we
establish a theoretical framework using the four band model
leading to four coupled differential equations. In section 3, by
using the transfer matrix at boundaries together with the
incident, transmitted and reflected currents we end up with eight
transmission and reflection probabilities as well as the
corresponding conductance. We deal with two band tunneling and
analyze their features with and without the interlayer potential
difference, in section 4. We do the same job in section 5 but by
considering four band energy and underline the difference with
respect to other case. In section 6, we show the numerical results
for the conductance and investigate the contribution of each
transmission channel. Finally, we briefly summarize our main
findings in the last section.

%%%%%%%%%%%%%%%%%%%%%%%%%%%%%%%%%%%%%%%%%%%%%%%%%%
\section{Theoretical model}
%%%%%%%%%%%%%%%%%%%%%%%%%%%%%%%%%%%%%%%%%%%%%%%%%%

In monolayer graphene, the unit cell has inequivalent atoms
(usually called A and B). Bilayer graphene on the other hand is a
two stacked monolayer graphene (Bernal stacking) and hence has
four atoms in the unit cell. The relevant Hamiltonian near the $K$
point (the boundary of the Brillouin zone), can be found using the
nearest-neighbor tight binding approximation\cite{16}
\begin{equation}\label{eq1}
H=\left(
\begin{array}{cccc}
  V^{+} & v_{F}\pi^{\dag} & -v_{4}\pi^{\dag} & v_{3}\pi \\
  v_{F}\pi & V^{+} & \gamma_{1} & -v_{4}\pi^{\dag} \\
  -v_{4}\pi & \gamma_{1} & V^{-} & v_{F}\pi^{\dag} \\
  v_{3}\pi^{\dag} & -v_{4}\pi & v_{F}\pi & V^{-} \\
\end{array}%
\right)
\end{equation}
where $ v_{F}=\frac{\gamma_{0}}{\hbar} \frac{3 a}{2}\approx10^{6}\
{m}/{s}$ is the fermi velocity of electrons in each graphene
layer, $a=0.142 \ nm$ is the distance between adjacent carbon
atoms, $v_{3,4}=\frac{v_{F} \gamma_{3,4}}{\gamma_{0}}$ represent
the coupling between the layers, $\pi=p_{x}+ip_{y},
\pi^{\dag}=p_{x}-ip_{y}$ are the in-plan momenta and its conjugate
with $p_{x,y}=-i\hbar
\partial_{x,y}$. %+\frac{e}{c}A$ and $A$ is the vector potential.
$\gamma_{1}\approx 0.4 \ eV$ is the interlayer coupling term and
$V^{+}, V^{-}$ are the potentials on the first and second layer,
respectively. The skew parameters , $\gamma_3\approx0.315\ eV$ and
$\gamma_4\approx0.044\ eV$ have negligible effect on the band
structure at high energy\cite{11,12}. Recently, it was shown that
even at low energy these parameters have also negligible effect on
the transmission\cite{30}, hence we neglect them in our
calculations.

\begin{figure}[h!] \centering
\includegraphics[width=4.2 in]{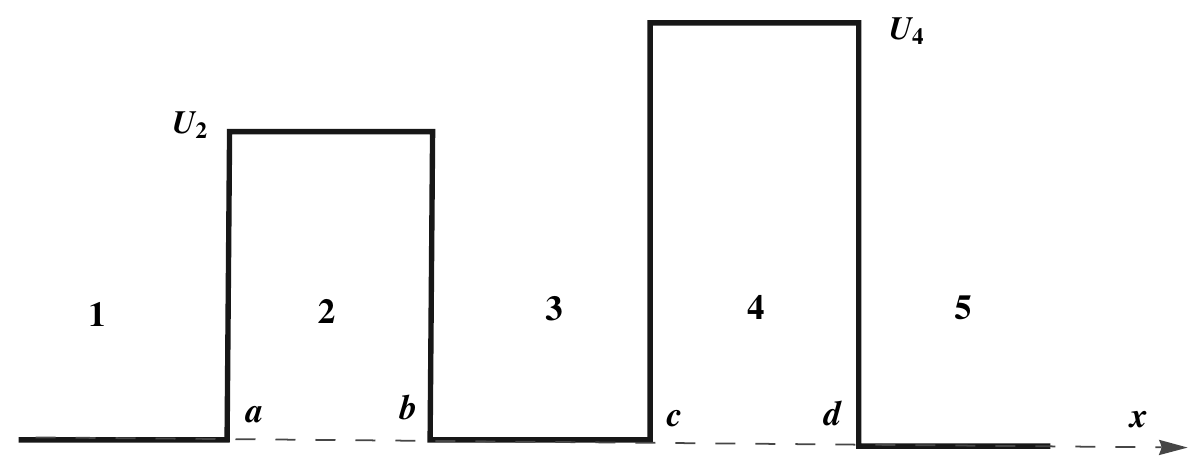} \caption{The parameters
of a rectangular double barrier structure.}\label{fig01}
\end{figure}

Under the above approximation and for double barrier  potential
configuration in Figure \ref{fig01}
our previous Hamiltonian (\ref{eq1}) can be written as follows in
each potential region where we define regions as follows: $j=1$
for $x\leq a$, $j=2$ for $a<x\leq b$, $j=3$ for $b<x\leq c$, $j=4$
for $c<x\leq d$ and $j=5$ for $x>d$ so that in the $j$-th region
we have
\begin{equation}\label{eq2}
H_j=\left(
\begin{array}{cccc}
  V^{+}_j & \nu_{F}\pi^{\dag} & 0 & 0 \\
  \nu_{F}\pi & V^{+}_j & \gamma_{1} & 0\\
  0 & \gamma_{1} & V^{-}_j & \nu_{F}\pi^{\dag} \\
  0 & 0& \nu_{F}\pi & V^{-}_j \\
\end{array}%
\right)
\end{equation}
We define the potential on the first and second layer by
$V^{\pm}_j=U_j \pm \delta_j$, where $U_j$ is the barrier strength
and $\delta_j$ is the electrostatic potential in the $j$-th region
\begin{equation}\label{eq002}
V^{\pm}_j=
\left\{%
\begin{array}{lllll}
    0, & \qquad\hbox{$j=1$} \\
   U_2  \pm \delta_2, & \qquad \hbox{$j=2$} \\
0, & \qquad \hbox{$j=3$}\\
U_4 \pm \delta_4, & \qquad \hbox{$j=4$}\\
0, & \qquad\hbox{$j=5$} \\
\end{array}%
\right.
\end{equation}
($U_2,\delta_2$) and ($U_4,\delta_4$) are the barrier potential
and the electrostatic potential in regions 2 and 4, respectively.
%The label $j=1, \cdots, 5$ denotes different regions forming our
%bilayer graphene as presented in Figure \ref{fig01}:
% system,
 %with $(V_1,V_3)$ for first layer
%and $(V_2,V_4)$ for second one in (first, second barrier).

\begin{figure}[H]
\centering
\includegraphics[width=2 in]{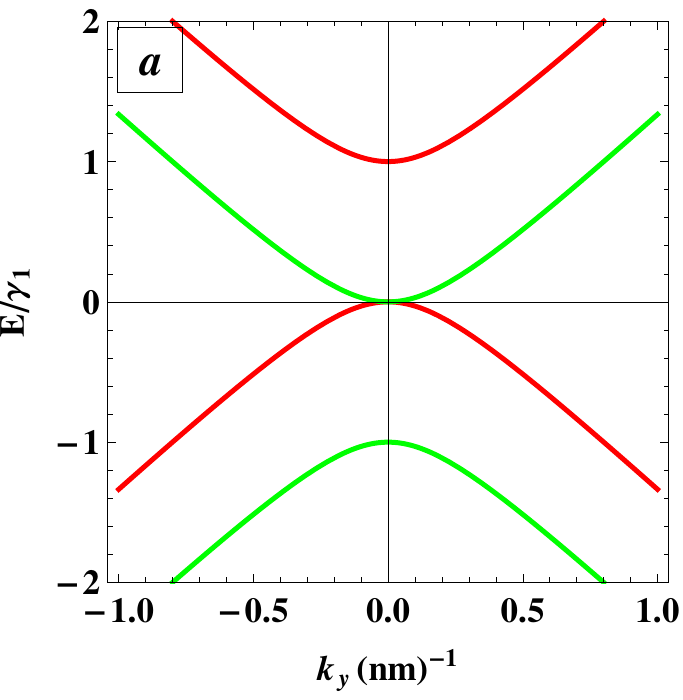}\ \ \
\includegraphics[width=2 in]{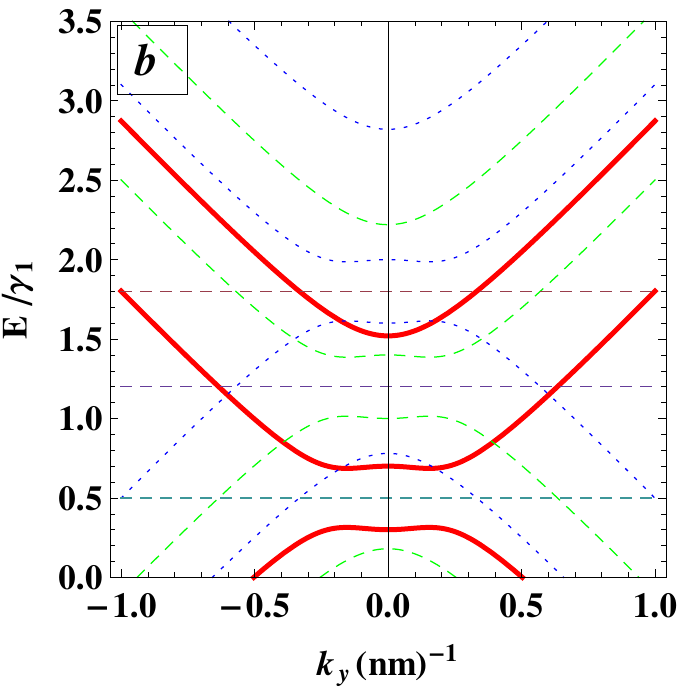}\ \ \
\includegraphics[width=2 in]{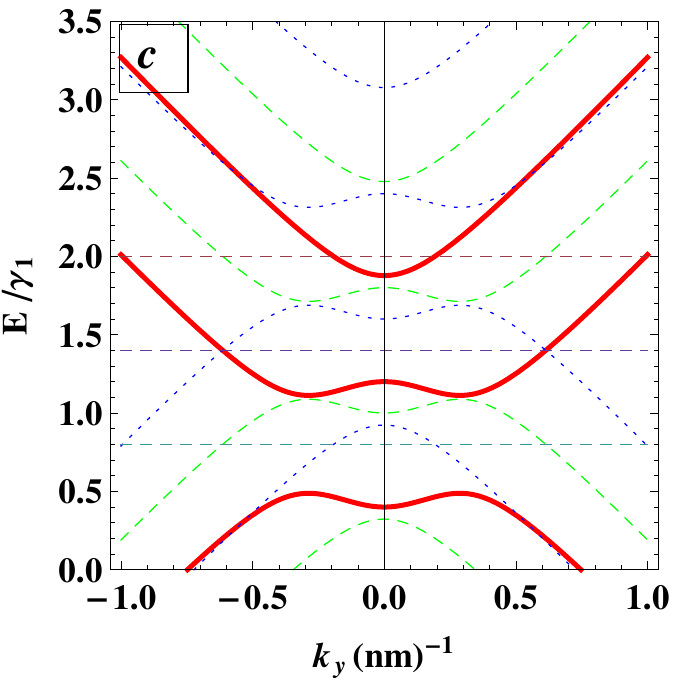}\\
\caption{Energy as function of the momentum $k_y$. (a): $V_1=V_3=V_5=0$.
(b): $\delta_2=0.2 \gamma_1$, $U_2=(0.5, 1.2,
1.8)\gamma_1$ (red, dashed green, dotted blue). (c): $\delta_4=0.4 \gamma_1$, $U_4=(0.8, 1.4, 2)\gamma_1$ (red,
dashed green, dotted blue). The dashed horizontal lines in (b) and
(c) represent the heights of the barriers $U_2$ and $U_4$,
respectively. }\label{fig0103}
\end{figure}

\begin{figure}[H] \centering
\includegraphics[width=4. in]{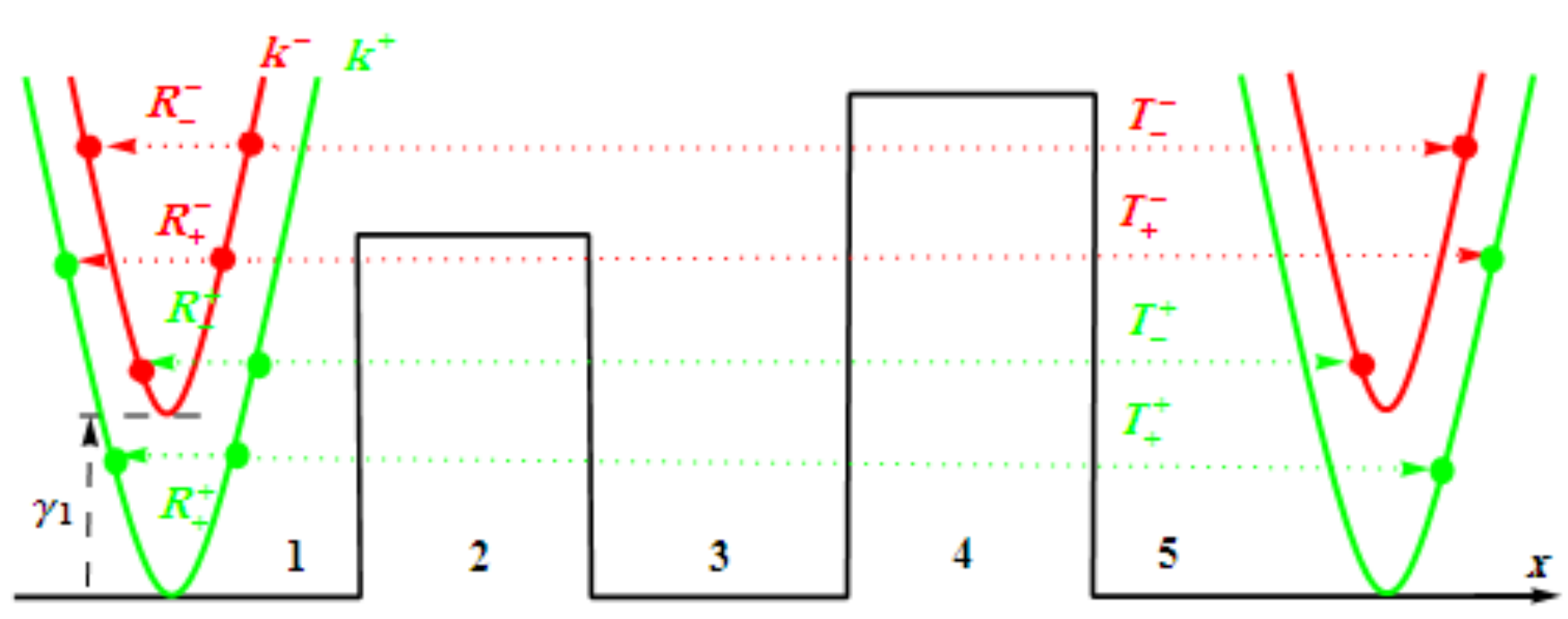} \ \
\caption{Schematic representation of  different modes as well as the
corrresponding transmission and reflection
probabilities.}\label{fig002}
\end{figure}

The eigenstates of (\ref{eq2}) are four-components spinors
$\psi^{j}(x,y) =[{\psi}^j_{A_{1}
},{\psi}^j_{B_{1}},{\psi}^j_{A_{2}},{\psi}^j_{B_{2}}]^{\dag}$,
here $\dag$ denotes the transpose of the row vector. For a double
barrier we need to obtain the solution in each regions as shown in
Figure \ref{fig01}. Since we have basically two different sectors
with zero (1, 3, 5) and nonzero potential (2, 4), a general
solution can be obtained in the second sector and then set the
potential $V^{\pm}_{j}$ to zero to obtain the solution in the
first sector. To simplify the notation, let us  introduce the length
scale $l=\frac{\hbar v_{F}}{\gamma_{1}}\approx 1.76 \ nm$
as well as $E_j \lga \frac{E}{\gamma_1}$ and $V_j
\lga \frac{V_j}{\gamma_1}$.
%allows
%us to define the following dimensionless quantities: $E'=
%\frac{E}{\gamma_1}$ and $V'_j= \frac{V_j}{\gamma_1}$.
%and
%$V_2\rightarrow \frac{V_2}{\gamma_1}$.
Since the momentum along the $y$-direction is a conserved
quantity, i.e $[H,p_y]=0$, and therefore we can write the spinors
as
\begin{equation}\label{eq002}
\psi^j(x,y) =e^{ik_y y}[{\phi}^j_{A_{1}
},{\phi}^j_{B_{1}},{\phi}^j_{A_{2}},{\phi}^j_{B_{2}}]^{T}
\end{equation}
As usual, to derive the eigenvalues and the eingespinors we solve
$H_j\psi_j=E_j\psi_j$. Then, by replacing by (\ref{eq2}) and \eqref{eq002}
we obtain
\begin{equation}\label{eq102}
\left(
\begin{array}{cccc}
  U_j+\delta_j & \frac{l}{\hb}\pi^{\dag} & 0 & 0 \\
  \frac{l}{\hb}\pi & U_j+\delta_j & 1 & 0\\
  0 & 1 & U_j-\delta_j & \frac{l}{\hb}\pi^{\dag} \\
  0 & 0& \frac{l}{\hb}\pi & U_j-\delta_j \\
\end{array}%
\right)\ \left(%
\begin{array}{c}
  \phi^{j}_{A_1} \\
  \phi^{j}_{B_1} \\
  \phi^{j}_{A_2} \\
  \phi^{j}_{B_2} \\
\end{array}%
\right)e^{iky}=E_j\left(%
\begin{array}{c}
  \phi^{j}_{A_1} \\
  \phi^{j}_{B_1} \\
  \phi^{j}_{A_2} \\
  \phi^{j}_{B_2} \\
\end{array}%
\right)e^{iky}
\end{equation}
This gives four coupled differential equations
%acting on its
%eigenstates leads to the following coupled differential equations
\begin{eqnarray}
&&-il\left[\frac{d}{dx}+ky\right]\phi^j_{B1}=(\epsilon_j-\delta_j)\phi^j_{A1}\label{eq0013}\\
&&
-il\left[\frac{d}{dx}-ky\right]\phi^j_{A1}+\phi^j_{A2}=(\epsilon_j-\delta_j)\phi^j_{B1}\label{eq0023}\\
 &&-il\left[\frac{d}{dx}+ky\right]\phi^j_{B2}+\phi^j_{B1}=(\epsilon_j+\delta_j)\phi^j_{A2}\label{eq0033}\\
&&-il\left[\frac{d}{dx}-ky\right]\phi^j_{A2}=(\epsilon_j+\delta_j)\phi^j_{B2}\label{eq0043}
\end{eqnarray}
where $k_{y}$ is the wave vector along the $y$-direction and we
have set $ \epsilon_j=E_j- U_{j}$.
%and $V_{0}=\frac{V_{1}+V_{2}}{2}$ and
%$\delta=\frac{V_{1}-V_{2}}{2}$.
It is easy to decouple the first equations to obtain
\begin{equation}\label{eq4}
\left[\frac{d^{2}}{dx^{2}}+(k^{s}_j)^{2}\right]\phi^j_{B1}=0
\end{equation}
where the wave vector along the $x$-direction is
\beq\lb{ksj}
k^{s}_j=\left[-k^{2}_{y}+\frac{\epsilon^{2}_j+(\delta_j)^{2}}{l^{2}}
+ s\frac{1}{l^{2}} \sqrt{\epsilon^{2}_j(1+
4(\delta_j)^{2})-(\delta_j)^{2}}\right]^{1/2}
\eeq
where $s=\pm$
denotes the propagating modes which will be discussed latter on.
Now for each region one can end up with corresponding wave vector
according to Figure \ref{fig01}. Indeed, for regions 1, 3 and 5 we
have $V^{\pm}_j=0$ and then we can obtain
\begin{equation}\label{eq05}
k^{s}_{0}=\left[-k^{2}_{y}+\frac{\epsilon^2}{l^{2}}+s\frac{\epsilon}{l^{2}}\right]^{1/2}
\end{equation}
with $\epsilon=\epsilon_1=\epsilon_3=\epsilon_5$, as well as the
energy
\begin{equation}\label{eq005}
E^{s}_{\pm}=\pm\frac{1}{2}\left[-s+\sqrt{1+(2lk_0^s)^2+(2lk_y)^2}\right]
\end{equation}
 However generally, for any region we can deduce
energy from previous analysis as
 \beq
\epsilon^{s}_{\pm,j}=\pm\frac{1}{\sqrt{2}}\left[1+2l^2 \left[(k_j^s)^2+k_{y}^2 \right]+2\delta_j^2-
s\sqrt{1+4l^2 \left[(k_j^s)^2+k_{y}^2 \right] \left(1+4\delta_j^2 \right)}\right]^{1/2}.
\eeq The corresponding energy spectrum of the different regions is
shown in Figure \ref{fig0103}.
Associated with each real $k^{s}_0$, the wave vector of the
propagating wave in the first region, there are two right-going
(incident) propagating mode and two left-going (reflected)
propagating mode. For $\gamma_1>E>0$, $k^{+}_0$ is real while
$k^{-}_0$ is imaginary, and therefore the propagation is only
possible using $k^{+}_0$ mode. However when $E>\gamma_1$, both
$k^{\pm}_0$ are real and then the propagation is possible using
two modes $k^{+}_0$ and $k^{-}_0$. In Figure \ref{fig002} we show
these different modes and the associated transmission
probabilities through double barrier structure.\\

\begin{figure}[h!]
\centering
\includegraphics[width=3. in]{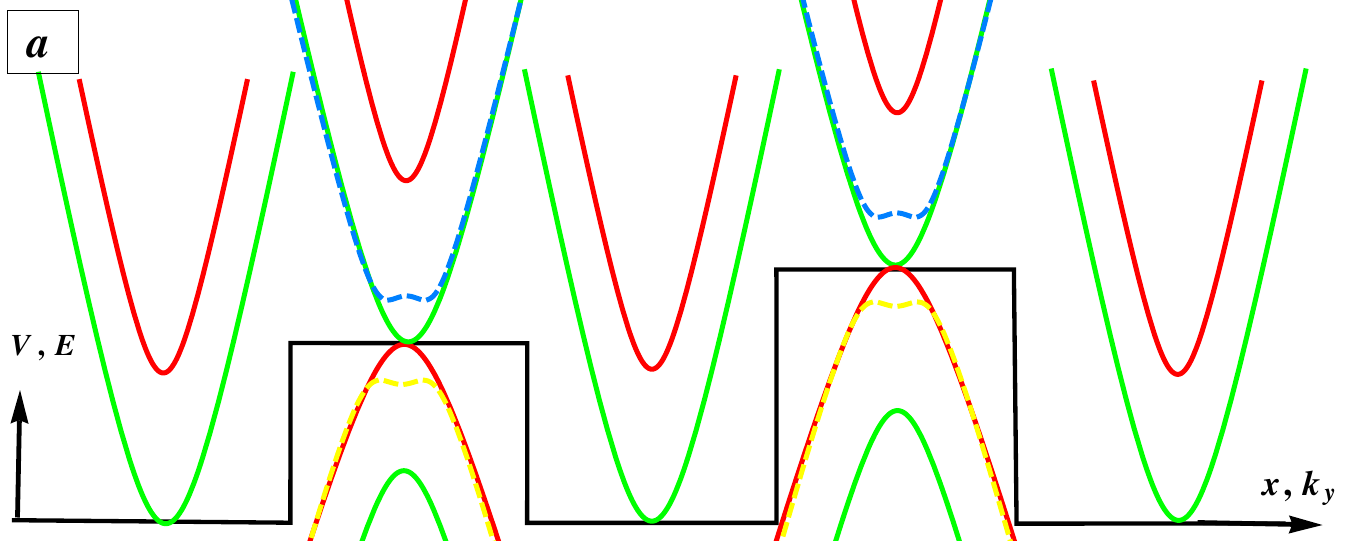}\ \ \ \
\includegraphics[width=3. in]{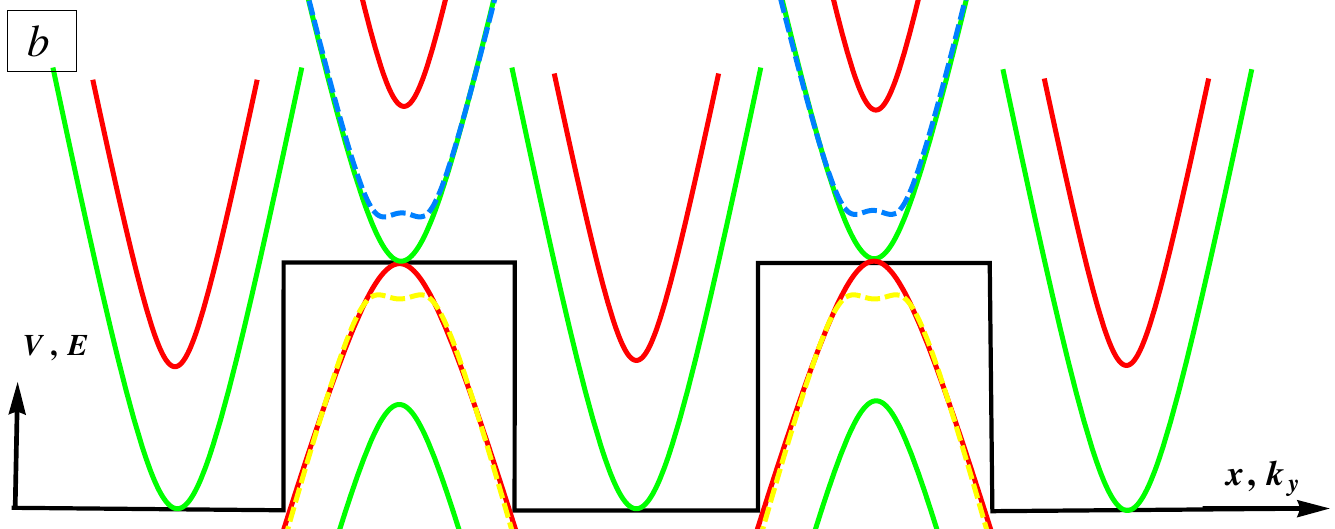}
\caption{Scheme represents the bands inside and outside the
barriers for the same interlayer potential difference. (a):
asymmetric for $U_2<U_4$. (b): symmetric for
$U_2=U_4$.}\label{figvs4}
\end{figure}

%In Figure \ref{fig06} we address two different cases: (a) the
%height of the two barriers is the same ($U_2=U_4$) and (b) with
%different height of the two barriers ($U_2<U_4$) for similar and
%different interlayer potential difference on both barriers in both
%cases. Indeed, the case (a) in Figure \ref{fig06} when $U_2=U_4$
%and $\delta_2=\delta_4$ should be reduced into the previous work
%\cite{30} in the limit of zero width between the two barriers
%($\Delta=0$).
Figure \ref{figvs4} presents two different cases. (a): asymmetric
double barrier structure for $U_2<U_4$, $\delta_2=\delta_4$ and
(b): another symmetric one for  $U_2=U_4$, $\delta_2=\delta_4$. It is
interesting to note that the Ben results \cite{30} can be
recovered from our results by considering the case (b) and requiring
$b=c$ in our double barrier structure.
We notice that different channels of transmission and reflection in Figure
\ref{fig002} can be mapped into all cases in Figure \ref{figvs4}
since they are related to the band structure on the both sides of
the barriers. However, the effect of the different structure of
the two barriers should appear in the transmission and reflection
probabilities.\\

The solution of (\ref{eq4}) can be written as a linear combination
of plane waves
\begin{equation}\label{eq5}
\phi_{B1}^j=a_1e^{ik^{+}_jx}+a_2e^{-ik^{+}_jx}+a_3e^{ik^{-}_jx}+a_4e^{-ik^{-}_jx}
\end{equation}
where $a_m$  $(m=1,2,3,4)$ are coefficients of normalization.
Substituting (\ref{eq5}) into (\ref{eq0013}\ -\ref{eq0043}) we
obtain the rest of the spinor components:
\begin{eqnarray}\label{eq6}
\phi_{A1}^j&=&a_1 A^{+}_{-}e^{ik^{+}_jx}-a_2 A^{+}_{+}e^{-ik^{+}_jx}+a_3A^{-}_{-}e^{ik^{-}_jx}-a_4A^{-}_{+}e^{-ik^{-}_jx}\\
\phi_{A2}^j&=&a_1 \rho^{+}e^{ik^{+}_jx}+a_2 \rho^{+}e^{-jk^{+}_jx}+a_3\rho^{-}e^{ik^{-}_jx}+a_4\rho^{-}e^{-ik^{-}_jx}\\
\phi_{B2}^j&=&a_1 \zeta^{+}_{+}e^{ik^{+}_jx}-a_2
\zeta^{+}_{-}e^{-ik^{+}_jx}+a_3\zeta^{-}_{+}e^{ik^{-}_jx}-a_4\zeta^{-}_{-}e^{-ik^{-}_jx}
\end{eqnarray}
where we have set
\beq
A^{s}_{\pm}=\frac{l(k^{s}_j\pm
ik_{y})}{\epsilon_j-\delta_j},\qquad
\rho^{s}=(\epsilon_j-\delta_j)\left[1-\frac{l^{2} \left[(k^{s}_j)^{2}+k^{2}_{y}\right]}{(\varepsilon_j-\delta_j)^{2}}\right],
\qquad
\zeta^{s}_{\pm}=\frac{\epsilon_j-\delta_j}{\epsilon_j+\delta_j}\rho^{s}A^{s}_{\pm}.
\eeq
Now, we can write the general solution
\begin{equation}\label{eq9}
\psi^j(x,y)=G_j M_j(x)C_j e^{ik_{y}y}
\end{equation}
in terms of the matrices
\begin{equation}\label{eq10}
G_j=\left(%
\begin{array}{cccc}
  A^{+}_{-} & -A^{+}_{+} & A^{-}_{-} & -A^{-}_{+} \\
  1 & 1 & 1 & 1 \\
  \rho^{+} & \rho^{+} & \rho^{-} & \rho^{-} \\
  \zeta^{+}_{+} & -\zeta^{+}_{-} & \zeta^{-}_{+} & -\zeta^{-}_{-} \\
\end{array}%
\right),M_j=\left(%
\begin{array}{cccc}
  e^{ik^{+}_jx} & 0 & 0 & 0 \\
  0 & e^{-ik^{+}_jx} & 0 & 0 \\
  0 & 0 & e^{ik^{-}_jx} & 0 \\
  0 & 0 & 0 & e^{-ik^{-}_jx} \\
\end{array}%
\right),C_j=\left(%
\begin{array}{c}
  a_1 \\
  a_2 \\
  a_3 \\
  a_4 \\
\end{array}%
\right)
\end{equation}
Since we are using the transfer matrix, we are interested in the
normalization coefficients, the components of $C$, on the both
sides of the double barrier. In other words, we need to specify
our spinor in region 1
\begin{eqnarray}\label{eq016}
\phi^1_{A1}&=&\delta_{s,1} A^{+}_{-}e^{ik^{+}_0x}-r^{s}_+ A^{+}_{+}e^{-ik^{+}_0x}+\delta_{s,-1}A^{-}_{-}e^{ik^{-}_0x}-r^{s}_-A^{-}_{+}e^{-ik^{-}_0x}\\
\phi^1_{B1}&=&\delta_{s,1}e^{ik^{+}_0x}+r^{s}_+e^{-ik^{+}_0x}+\delta_{s,-1}e^{ik^{-}_0x}+r^{s}_-e^{-ik^{-}_0x}\\
\phi^1_{A2}&=&\delta_{s,1} \rho^{+}e^{ik^{+}_1x}+r^{s}_+ \rho^{+}e^{-ik^{+}_0x}+\delta_{s,-1}\rho^{-}e^{ik^{-}_0x}+r^{s}_-\rho^{-}e^{-ik^{-}_0x}\\
\phi^1_{B2}&=&\delta_{s,1} \zeta^{+}_{+}e^{ik^{+}_0x}-r^{s}_+
\zeta^{+}_{-}e^{-ik^{+}_0x}+\delta_{s,-1}\zeta^{-}_{+}e^{ik^{-}_0x}-r^{s}_-\zeta^{-}_{-}e^{-ik^{-}_0x}
\end{eqnarray}
%and in region 3 are
%\begin{eqnarray}\label{eq0106}
%\phi^3_{A1}&=&a_1 A^{+}_{-}e^{ik^{+}_0x}-a_2 A^{+}_{+}e^{-ik^{+}_0x}+a_3A^{-}_{-}e^{ik^{-}_0x}-a_4A^{-}_{+}e^{-ik^{-}_0x}\\
%\phi^3_{B1}&=&a_1e^{ik^{+}_0x}+a_2e^{-ik^{+}_0x}+a_3e^{ik^{-}_0x}+a_4e^{-ik^{-}_0x}\\
%\phi^3_{A2}&=&a_1 \rho^{+}e^{ik^{+}_0x}+a_2 \rho^{+}e^{-ik^{+}_0x}+a_3\rho^{-}e^{ik^{-}_0x}+a_4\rho^{-}e^{-ik^{-}_0x}\\
%\phi^3_{B2}&=&a_1 \zeta^{+}_{+}e^{ik^{+}_0x}-a_2
%\zeta^{+}_{-}e^{-ik^{+}_0x}+a_3\zeta^{-}_{+}e^{ik^{-}_0x}-a_4\zeta^{-}_{-}e^{-ik^{-}_0x}
%\end{eqnarray}
as well as region 5
\begin{eqnarray}\label{eq026}
\phi^5_{A1}&=&t^{s}_+ A^{+}_{-}e^{ik^{+}_0x}+t^{s}_-A^{-}_{-}e^{ik^{-}_0x}\\
\phi^5_{B1}&=&t^{s}_+e^{ik^{+}_0x}+t^{s}_-e^{ik^{-}_0x}\\
\phi^5_{A2}&=&t^{s}_+ \rho^{+}e^{ik^{+}_0x}+t^{s}_-\rho^{-}e^{ik^{-}_0x}\\
\phi^5_{B2}&=&t^{s}_+
\zeta^{+}_{+}e^{ik^{+}_0x}+t^{s}_-\zeta^{-}_{+}e^{ik^{-}_0x}
\end{eqnarray}
Since the potential is zero in regions 1, 3 and 5, we have the
relation \beq\lb{144} G_{1} M_{1}(x)=G_{3} M_{3}(x)=G_{5}
M_{5}(x). \eeq We will see how the above results will be used to
determine different physical quantities. Specifically we focus on
the reflection and transmission probabilities as well as related
matters.

%%%%%%%%%%%%%%%%%%%%%%%%%%%%%%%%%%%%%%%%%%%%%%%%%%%%%%%%%%%
\section{Transmission  probabilities and conductance}
%%%%%%%%%%%%%%%%%%%%%%%%%%%%%%%%%%%%%%%%%%%%%%%%%%%%%%%%%%%

Implementing the appropriate boundary condition in the context of
the transfer matrix approach, one can obtain the transmission and
reflection probabilities. Continuity of the spinors at the
boundaries gives the components of the vector $C$ which are  given
by
\begin{equation}\label{eq17}
C^{s}_{1}=\left(%
\begin{array}{cccc}
 \delta_{s,1} \\
  r^{s}_{+} \\
  \delta_{s,-1} \\
  r^{s}_{-} \\
\end{array}%
\right), \qquad
C^{s}_{5}=\left(%
\begin{array}{c}
  t^{s}_{+} \\
  0 \\
  t^{s}_{-} \\
  0 \\
\end{array}%
\right)
\end{equation}
where $\delta_{s,\pm1}$ is the Kronecker delta symbol. The
coefficients in the incident and reflected regions can be linked
through the transfer matrix $M$
\begin{equation}\label{eq18}
C^{s}_{1}=M C^{s}_{5}
\end{equation}
which can be obtained explicitly by applying the continuity at the
four boundaries of the double barrier structure (Figure \ref{fig01}). These are given by
\begin{eqnarray}\label{eq19}
    G_{1} M_{1}(a)C_{1}&=&G_{2} M_{2}(a)C_{2}\\
    G_{2} M_{2}(b)C_{2}&=&G_{3} M_{3}(b)C_{3}\\
    G_{3} M_{3}(c)C_{3}&=&G_{4} M_{4}(c)C_{4}\\
    G_{4} M_{4}(d)C_{4}&=&G_{5} M_{5}(d)C_{5}
\end{eqnarray}
Now solving the above system of equations and taking into account of the
relation \eqref{144}, one can find
the form of $M$.
%the fact that $G_{I}
%M_{I}(x)=G_{III} M_{III}(x)=G_{V} M_{V}(x)$
%we can obtain the transfer matrix $M$.
%\begin{equation}\label{eq20}
%N=M^{-1}_{I}(a)G^{-1}_{I}G_{II}M_{II}(a)M^{-1}_{II}(b)G^{-1}_{II}G_{I}M_{I}(b)
%M^{-1}_{I}(c)G^{-1}_{I}G_{IV}M_{IV}(c)M^{-1}_{IV}(d)G^{-1}_{IV}G_{I}M_{I}(d)
%\end{equation}

Then we can specify the complex coefficients of the transmission
$t^{s}_{\pm}$ and reflection $r^{s}_{\pm}$ using the transfer
matrix $M$. Since we need the transmission $T$ and reflection $R$
probabilities and because the velocity of the waves scattered
through the two different modes is not the same, it is convenient
to use the current density $\mathbf{J}$ to obtain the transmission
and reflection probabilities. \beq \mathbf{J}=\nu_{F}
{\mathbf{\Psi}}^{\dagger}\vec{\alpha}{\mathbf\Psi} \eeq to end up
with
\begin{equation}\label{eq21}
T=\frac{|\mathbf{J}_{\sf tra}|}{|\mathbf{J}_{\sf inc}|}, \qquad
R=\frac{|\mathbf{J}_{\sf ref}|}{|\mathbf{J}_{\sf inc}|}
\end{equation}
where $\vec{\alpha}$ is a $4\times4$ diagonal matrix, on the
diagonal  2 Pauli matrices $\sigma_{x}$. From \eqref{eq17} and
\eqref{eq21}, we show that
%Then,
the eight transmission and reflection probabilities are given by
\cite{33}
\begin{equation}\label{eq23}
T^{s}_{\pm}=\frac{k^{\pm}_0}{k^{s}_0}|t^{s}_{\pm}|^{2},\qquad
R^{s}_{\pm}=\frac{k^{\pm}_0}{k^{s}_0}|r^{s}_{\pm}|^{2}
\end{equation}
These expressions can be explained as follows.
Since we have four band, the electrons can be scattered between them
and
then we need to take into account the change in their velocities.  With
that, we find four channels in transmission and reflection such that
$k_0^{\pm}$ is also given by
\eqref{eq05}. More precisely,
at low energies $(E < \gamma_1)$, we have just
one mode of propagation $k_0^{+}$ leading to one transmission $T$
and reflection $R$ channel through the two conduction bands
touching at zero energy on the both sides of the double barrier.
Whereas at higher energy $(E > \gamma_1)$, we have two modes of
propagation $k_0^{+}$ and $k_0^{-}$ leading to four transmission
$T^{\pm}_{\pm}$ and reflection $R^{\pm}_{\pm}$ channels, through
the four conduction
bands.

Since we have found transmission probabilities, let see how these will effect
the conductance of our system.  This actually can be obtained
through the Landauer-B\"{u}ttiker formula \cite{31} by summing
on all channels to end up with
\begin{equation}\label{eq24}
\mathbf{G}(E)=G_{0}\frac{L_y}{2
\pi}\int_{-\infty}^{+\infty}dk_{y}\sum_{s,n=\pm}T^{s}_{n}(E,k_y)
\end{equation}
where $L_y$ is the width of the sample in the $y$-direction and
$G_0=4\ \frac{e^2}{h}$, the factor $4$ is due to the valley and
spin degeneracy in graphene.

The obtained results will be numerically analyzed to discuss the
basic features of
our system and also make link with other published results. Because of the nature of our system,
we do our task by  distinguishing two different cases
in terms of the band tunneling.

%%%%%%%%%%%%%%%%%%%%%%%%%%%%%%%%%%%%%%%%%%%%%%%%%%
%\section{Band tunneling analysis }
%%%%%%%%%%%%%%%%%%%%%%%%%%%%%%%%%%%%%%%%%%%%%%%%%%

%%%%%%%%%%%%%%%%%%%%%%%%%%%%%%%%%%%%%%%%%%%%%%%%%%%
\section{Two band tunneling} % ($E < \gamma_{1}$)}
%%%%%%%%%%%%%%%%%%%%%%%%%%%%%%%%%%%%%%%%%%%%%%%%%%%%

%%%%%%%%%%%%%%%%%%%%%%%%%%%%%%%%%%%%%%%%%%%%%%%%%%%
%\subsection{Two band tunneling ($E < \gamma_{1}$)}
%%%%%%%%%%%%%%%%%%%%%%%%%%%%%%%%%%%%%%%%%%%%%%%%%%%

%Let us analyze  two interesting cases related to our energy
%spectrum. Indeed,  Therefore we consider each case %the results will be discussed
%separately and underline their relevant properties.

Barbier \cite{27} investigated the transmission and conductance
for single and multiple electrostatic barriers with and without
interlayer potential difference and for $E<\ \gamma_1$, however
the geometry dependance of the transmission was not done. In this
section, we briefly investigate the resonances resulting from the
available states in the well between the two barriers and how they
influence by the geometry of the system.
\begin{figure}
\centering
\includegraphics[width=3.14 in]{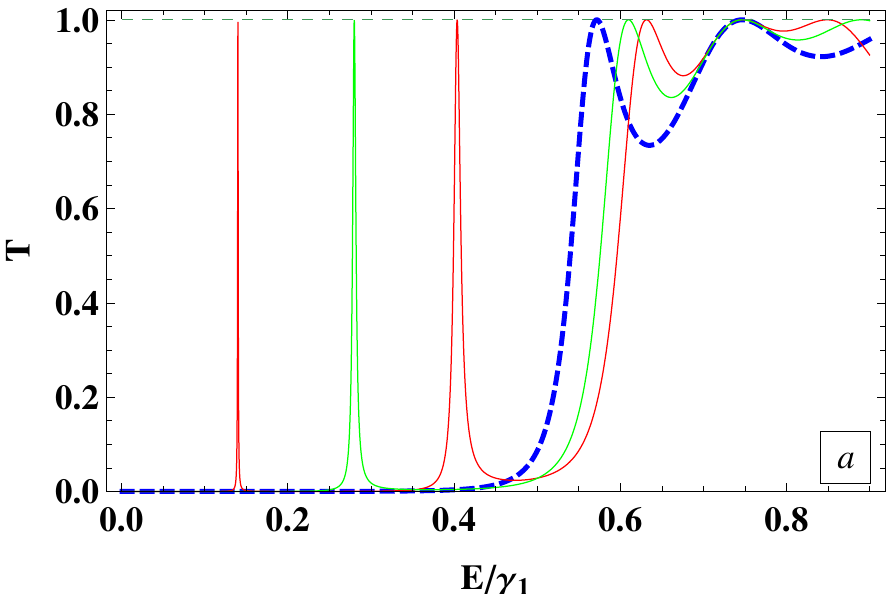}\ \
\includegraphics[width=3.14 in]{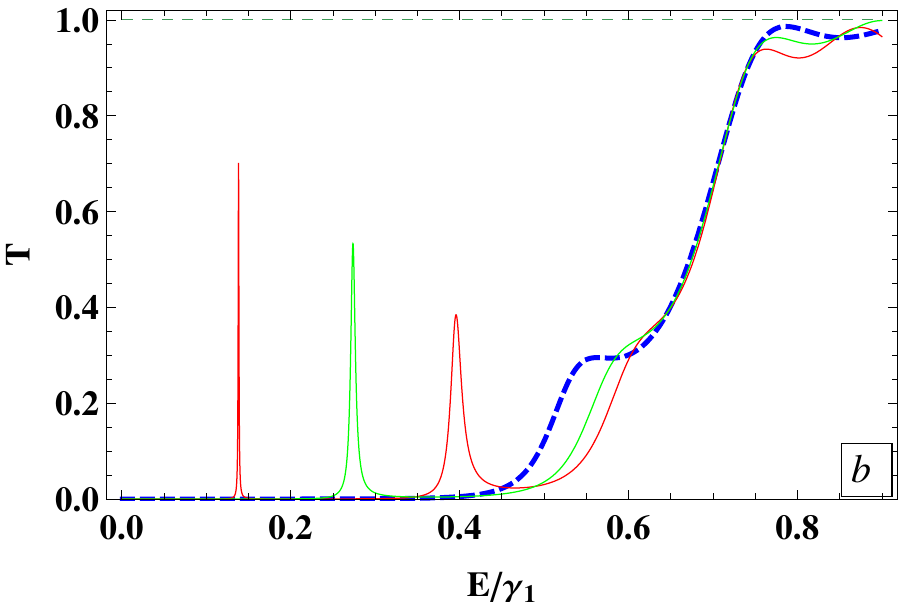}\\
\includegraphics[width=3.14 in]{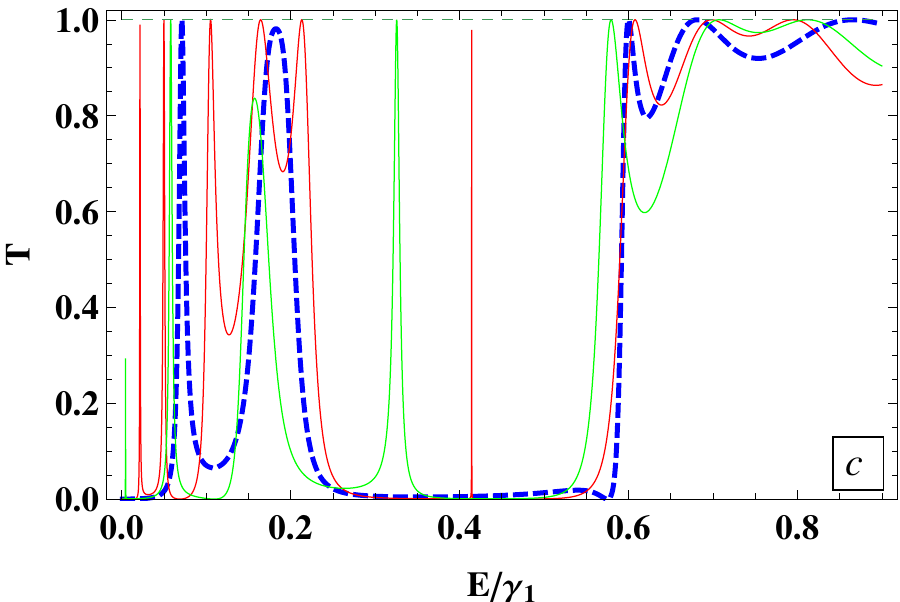}\ \
\includegraphics[width=3.14 in]{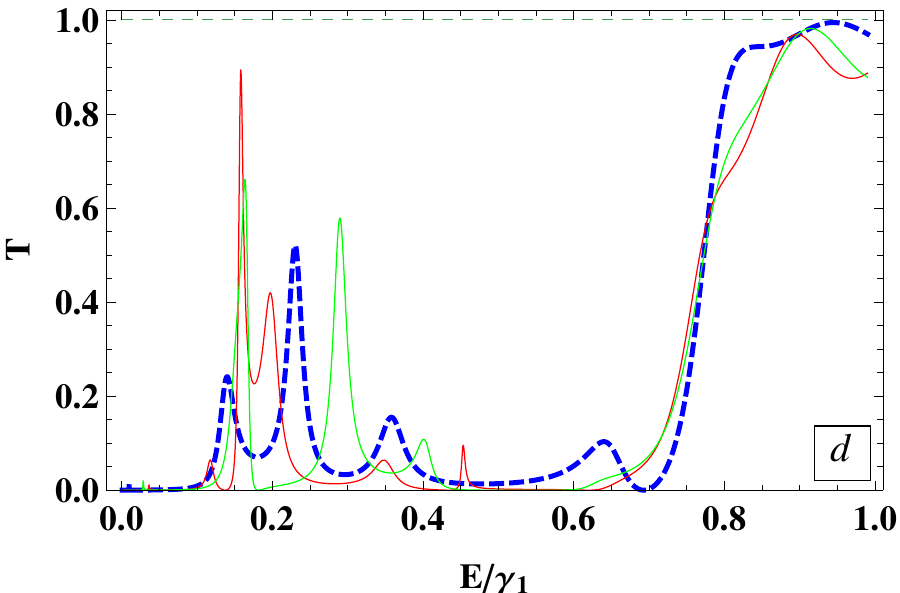}
\caption{ Transmission for normal incidence with $b_1=b_2=10\ nm$,
and $\Delta=0$ (blue dashed), $\Delta=5\ nm$ (green), $\Delta=10\
nm$ (red). (a):   $U_{2}=U_{4}=0.4\ \gamma_{1}$. (b):
$U_{2}=0.4\ \gamma_{1}$ and $U_{4}=0.6\ \gamma_{1}$. (c,d):
the same parameters as in (a,b), respectively, but with
$\delta_2=\delta_4=0.2\ \gamma_1$.}\label{fig2}
\end{figure}

For a normal incidence
and for $\delta_2=\delta_4=0$ the transmission amplitude is shown
in Figure \ref{fig2}a for different values of the distance
$\Delta$ between the barriers. The dashed blue curve is for a
single barrier with $(\Delta=0)$ and with width $(b_1+b_2=20\
nm)$, we note that the transmission is zero and there are no
resonances in this regime of energy $(E < U_{2} = U_{4})$. Unlike
the case of the single barrier, the double barrier structure has
resonances in the above mentioned range of energy. These full
transmission peaks can be attributed to the bound electron states
in the well region between the barriers. In agreement with
\cite{133}, the number of these resonances depends on the distance
between the barriers. Indeed, for $\Delta=5\ nm$ we have one peak
in the transmission amplitude, increasing the distance allows more
bound states to emerge in the well, and for $\Delta=10\ nm$ there
are two peaks (green and red curves in Figure \ref{fig2},
respectively). Figure \ref{fig2}b shows the same results in
\ref{fig2}a but with different height of the two barriers such
that $U_2=0.4\ \gamma_1$ and $U_4=0.6\ \gamma_1$. We see that the
asymmetric structure of the double barrier reduces those
resonances resulting from the bound electrons in the well between
the two barriers. For $\delta_2=\delta_4=0.2\ \gamma_1$, we show
the transmission probability by choosing $U_2=U_4=0.4\ \gamma_1$
in Figure \ref{fig2}c and for $U_2=0.4\ \gamma_1$, $U_4=0.6\
\gamma_1$ in Figure \ref{fig2}d. For single barrier, there are no
resonant peaks inside the induced gap which is not the case for
the double barrier as clarified in Figure \ref{fig2}c.

%%%%%%%%%%%%%%
\begin{figure}
\centering
\includegraphics[width=2 in]{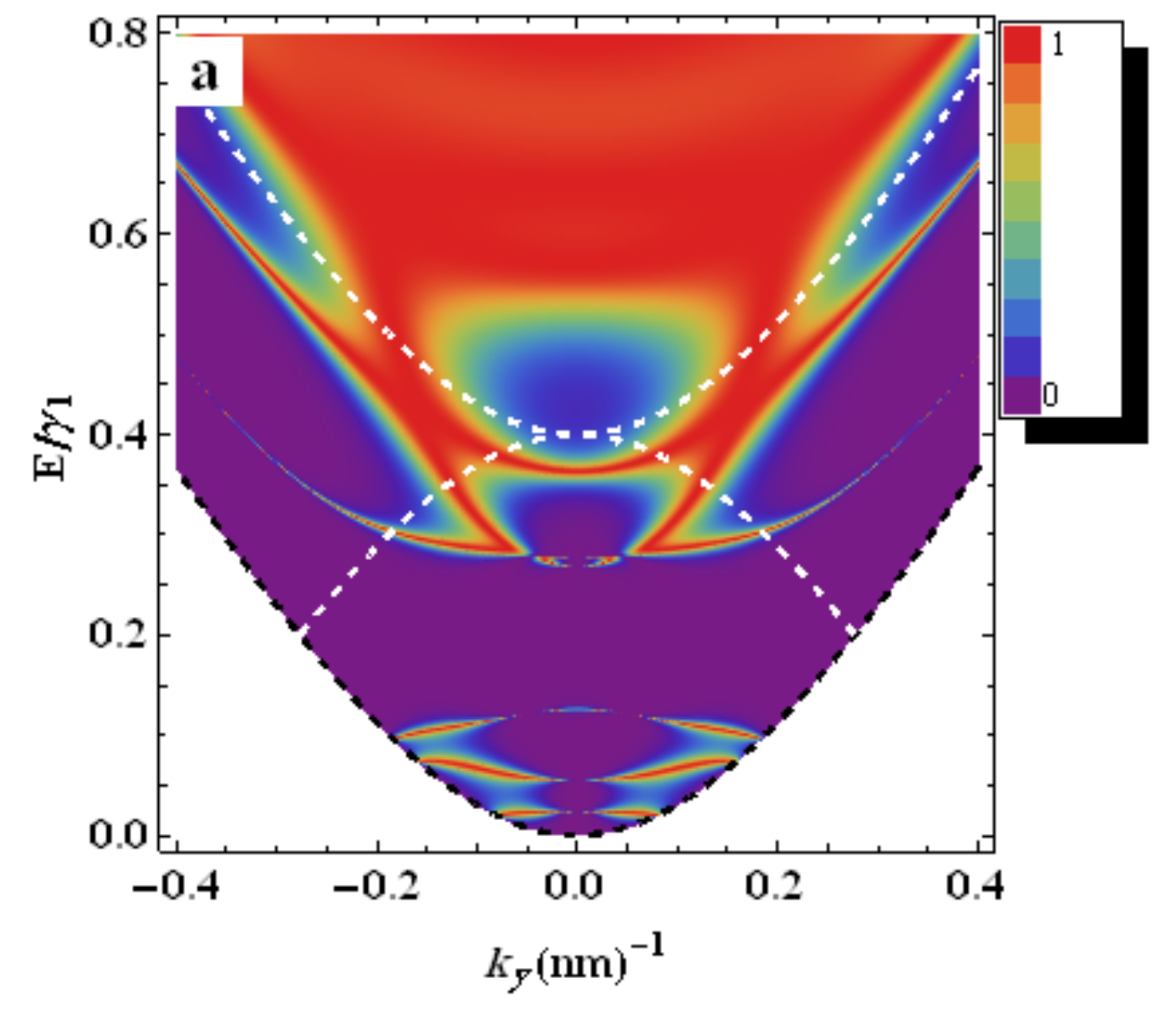}\ \ \
\includegraphics[width=2 in]{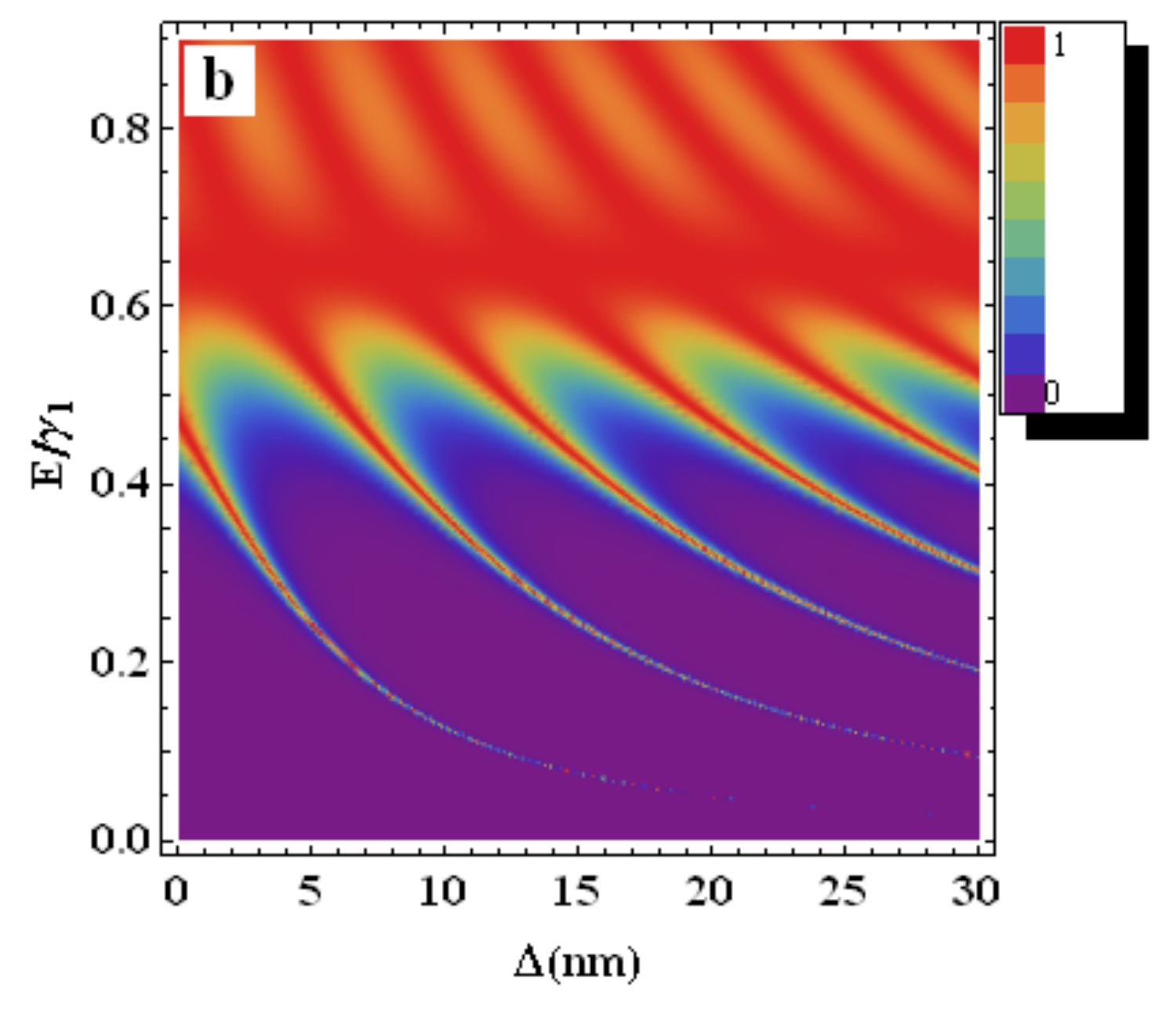}\\
\includegraphics[width=2 in]{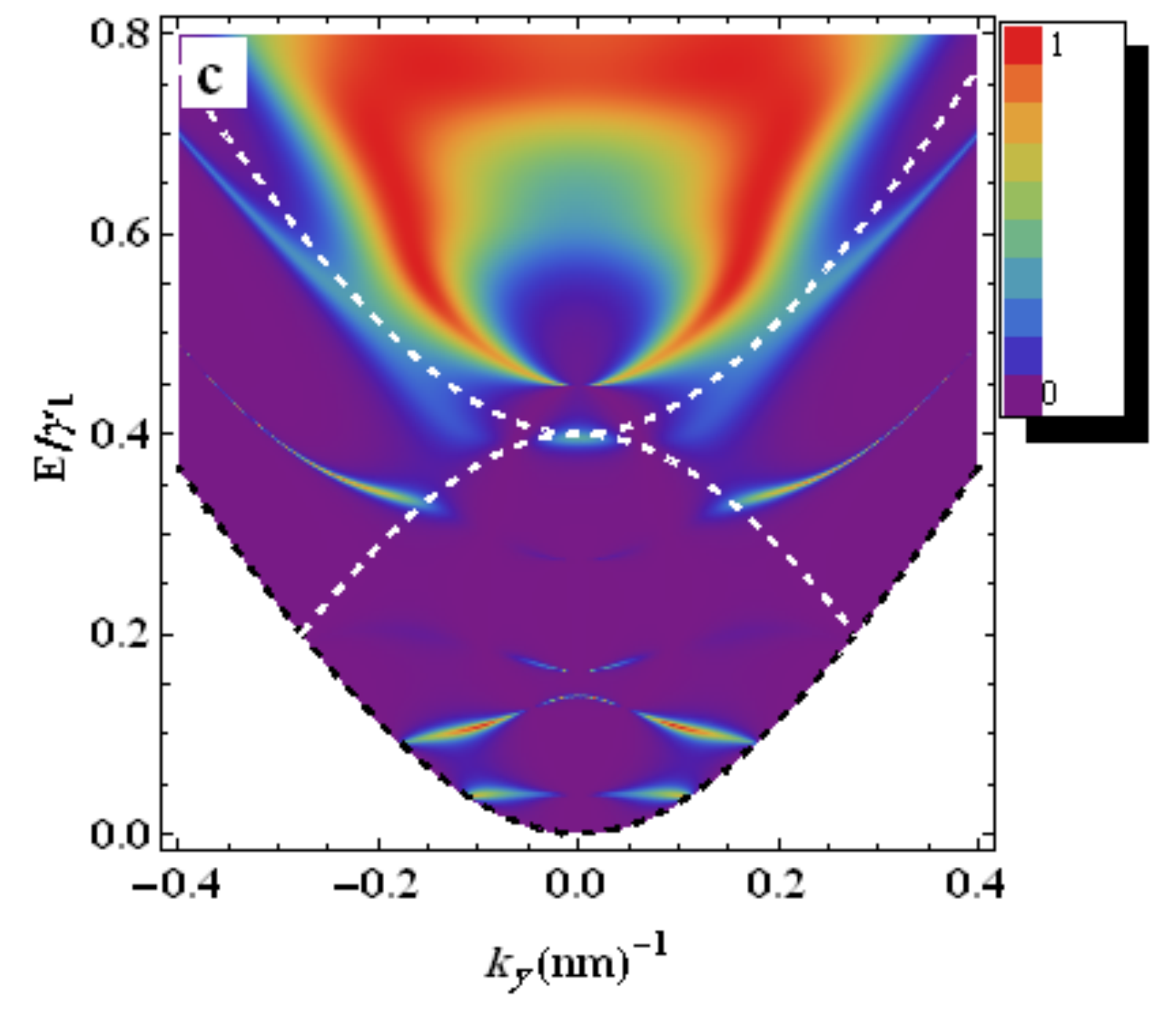}\ \ \
\includegraphics[width=2 in]{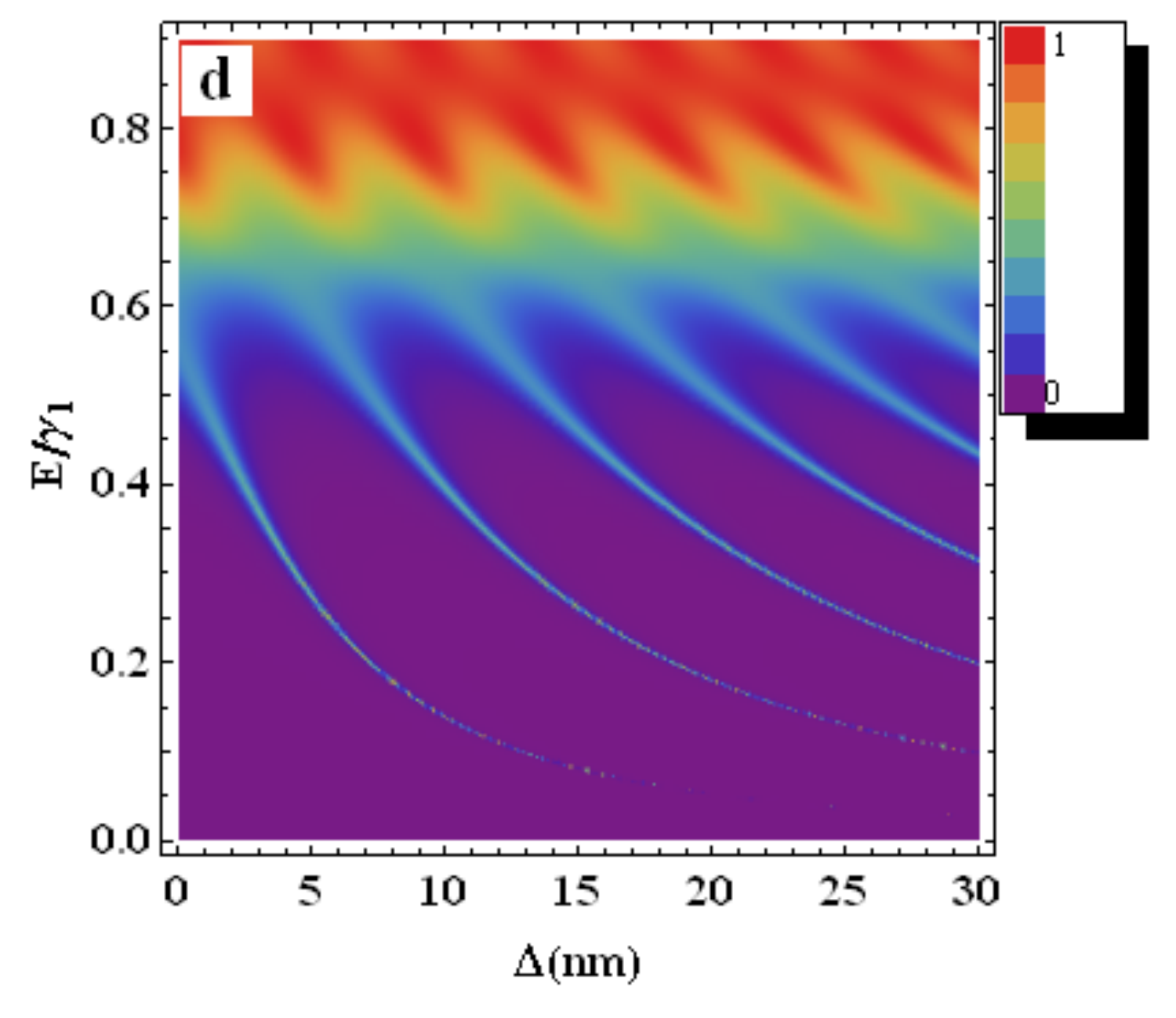}
\caption{ Density plot of transmission probability, for
$\delta_2=\delta_4=0$, versus (a): $E$ and $ky$ for
$U_{2}=U_{4}=0.4\ \gamma_{1}$, and $b_1=b_2=\Delta=10\ nm$, (b):
$E$ and $\Delta$ with $ky=0$ and $b_1=b_2=10\ nm$, (c): $E$ and
$ky$ with $U_{2}=0.4\ \gamma_{1}$, $U_{4}=0.6\ \gamma_{1}$ and
$b_1=b_2=\Delta=10\ nm$, (d): $E$ and $\Delta$ with $U_{2}=0.4\
\gamma_{1}$, $U_{4}=0.6\ \gamma_{1}$, $ky=0$ and $b_1=b_2=10\ nm$.
White and black dashed lines represent the band inside and outside
the first barrier, respectively. }\label{fig3}
\end{figure}
Figures \ref{fig3}a,\ref{fig3}c present a comparison of the
density plot of the transmission probability as a function of the
transverse wave vector $k_y$ of the incident wave and its energy
$E$ between different structure of the double barrier with
$U_2=U_4=0.4\ \gamma_1$ and $U_2=0.4\ \gamma_1<U_4=0.6\ \gamma_1$,
respectively, and for $\delta_2=\delta_4=0$ in both. For
non-normal incidence in Figure \ref{fig3}a ($k_y\neq 0$) we still
have a full transmission, even for energies less than the height
of the barriers, which are symmetric in $k_y$. Those resonances
are reduced and even disappeared in Figure \ref{fig3}c due to the
asymmetric structurer of the double barrier.
In Figures \ref{fig3}b,\ref{fig3}d we show the density plot of
transmission probability, for normal incidence, as a function of
$\Delta$ and $E$ for the same parameters as in Figure \ref{fig3}a
and \ref{fig3}c, respectively. We note that the number of
resonances in Figure \ref{fig3}b, due to the bounded electrons in
the well between the barriers, increases as long as the distance
is increasing. They are very sharp for the low energies and become
wider at higher energies. In contrast to Figure \ref{fig3}d and as
a result of the asymmetric structure of the double barrier these
resonances do not exist anymore for
$E<U_4=0.6\ \gamma_1$.
\begin{figure}
\centering
\includegraphics[width=2 in]{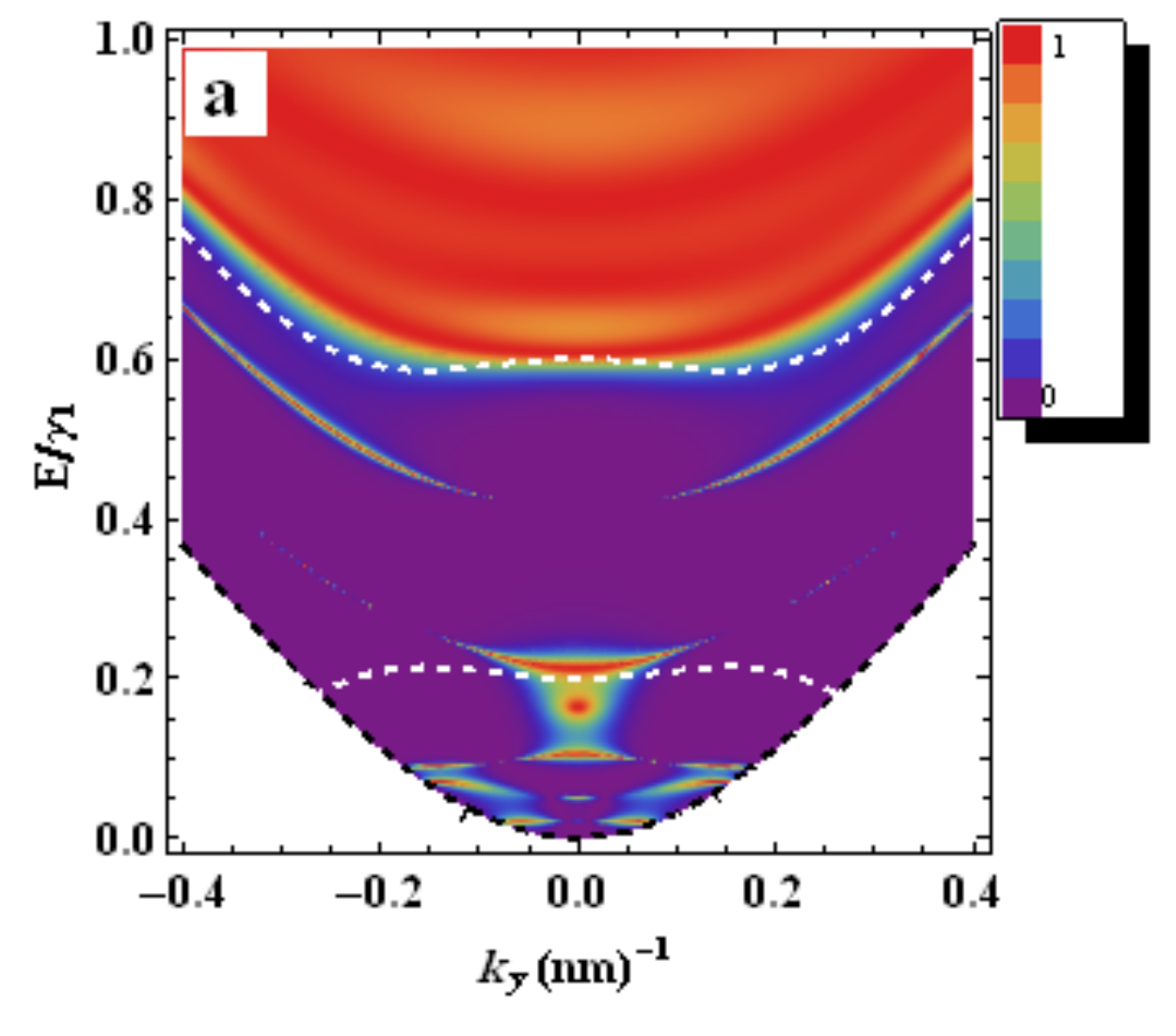}\ \
\includegraphics[width=2 in]{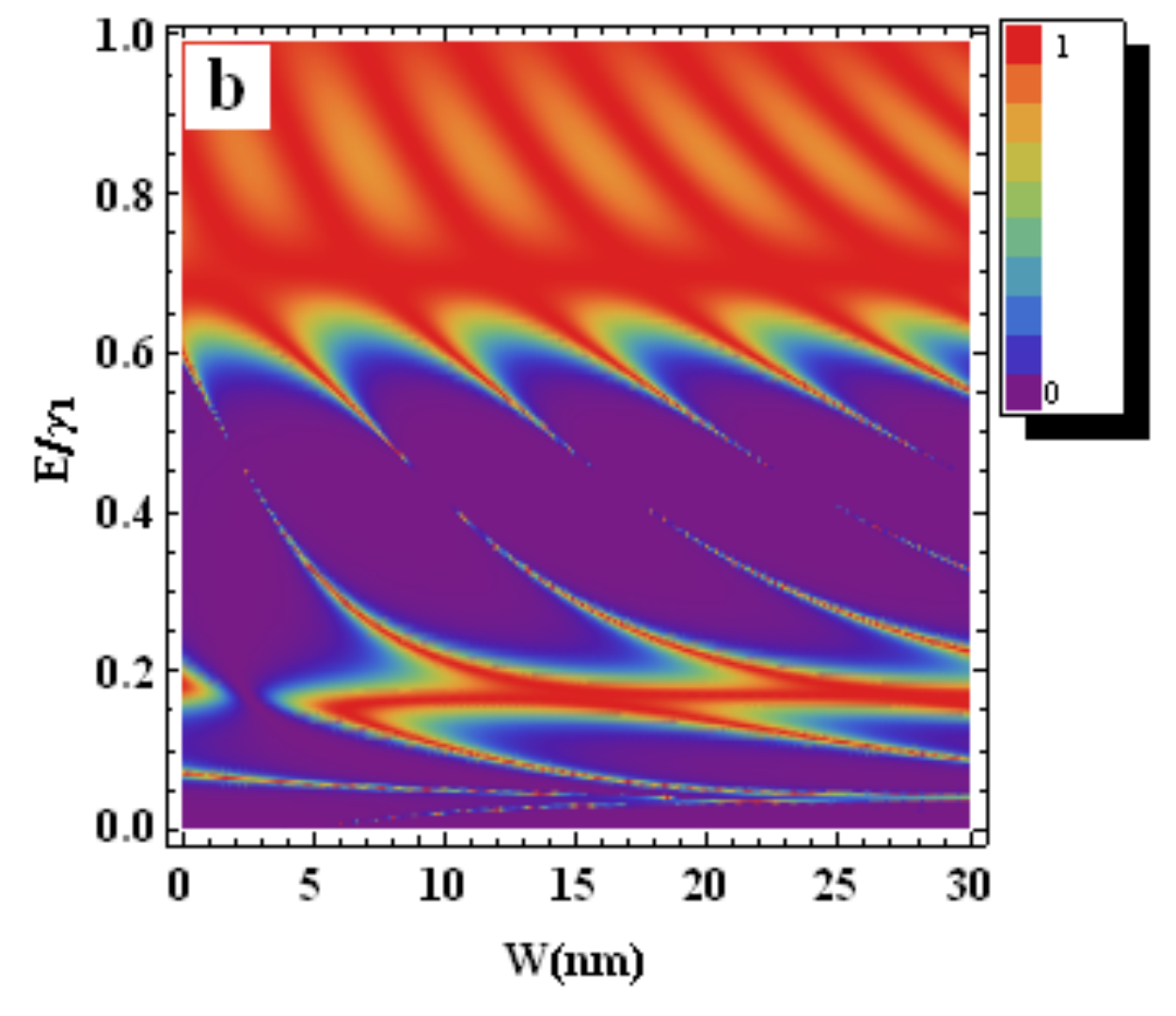}\\
\includegraphics[width=2 in]{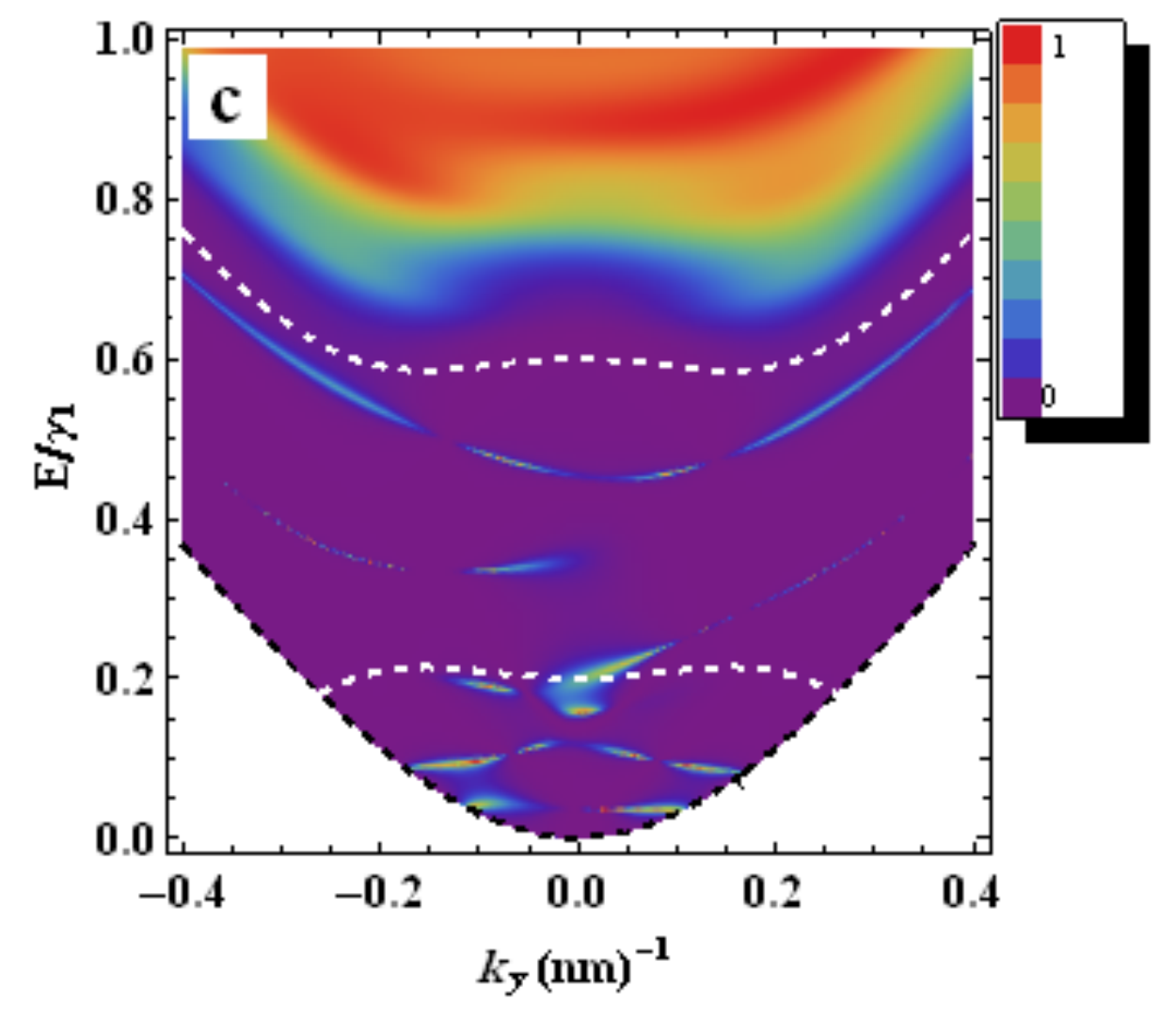}\ \
\includegraphics[width=2 in]{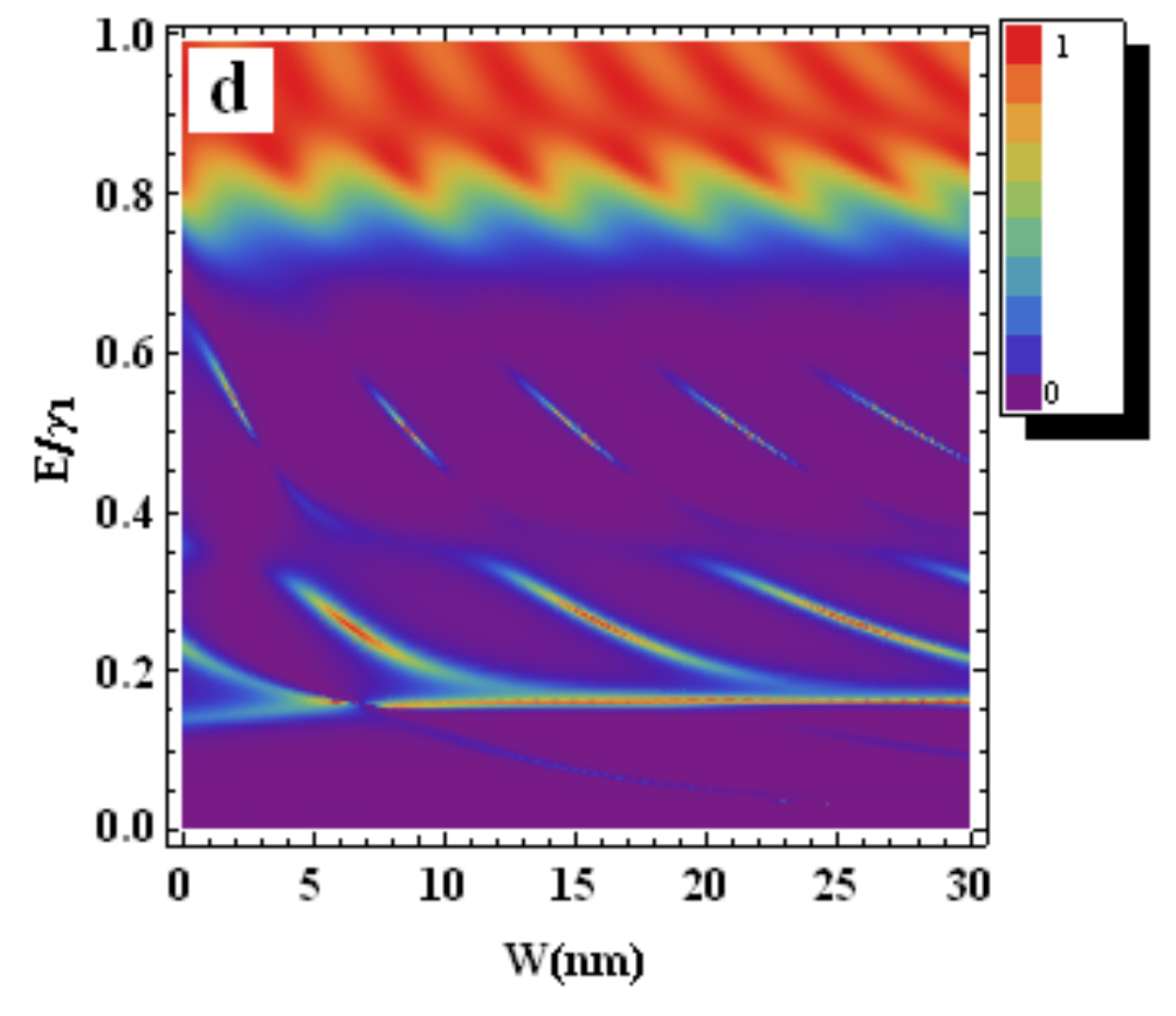}
\caption{Density plot of transmission probability, for
$\delta_2=\delta_4=0.2\ \gamma_1$, versus (a): $E$ and $k_y$ for
$U_2=U_4=0.4\ \gamma_1$, and $b_1=b_2=\Delta=10\ nm$, (b): $E$ and
$\Delta$ for the same parameters as in (a) but with $k_y=0$,
 (c,d): for the same parameters as in (a, b), respectively, but for
$U_2=0.4\ \gamma_1$, $U_4=0.6\ \gamma_1$. White and black dashed
lines represent the band inside and outside the first barrier,
respectively. }\label{fig010}
\end{figure}

\begin{figure}
\centering
\includegraphics[width=2 in]{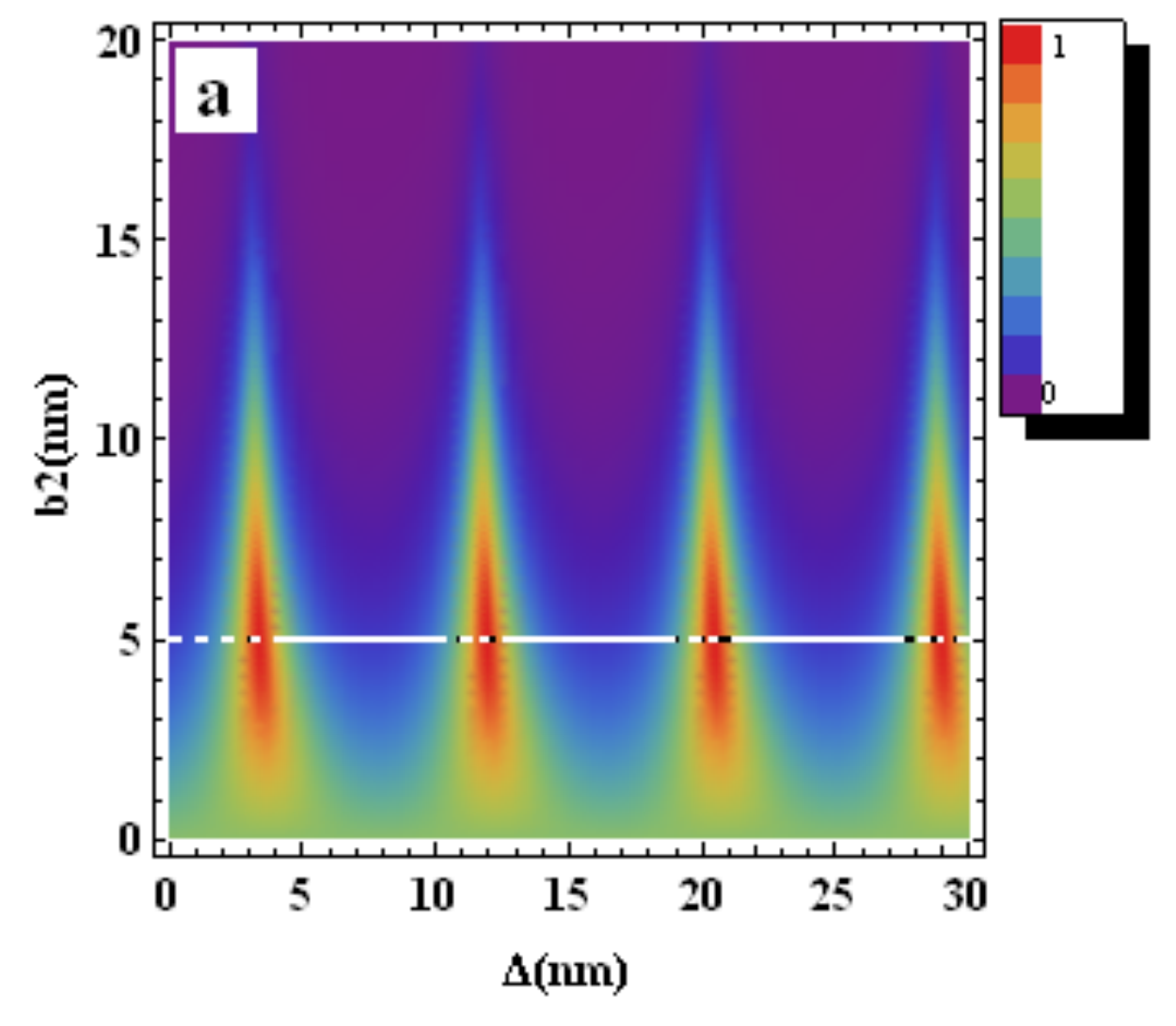}\ \ \
\includegraphics[width=2 in]{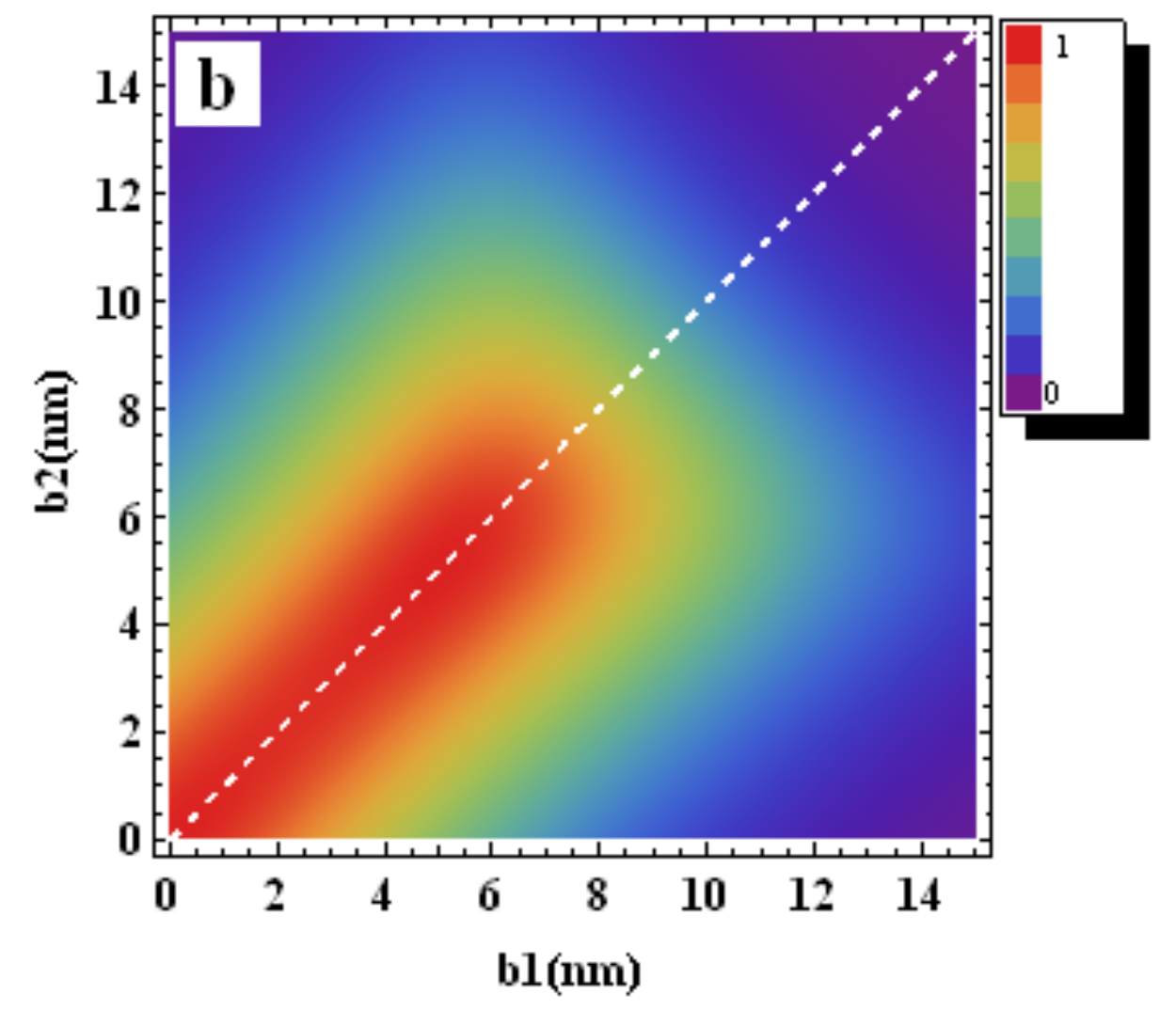}\\
\includegraphics[width=2 in]{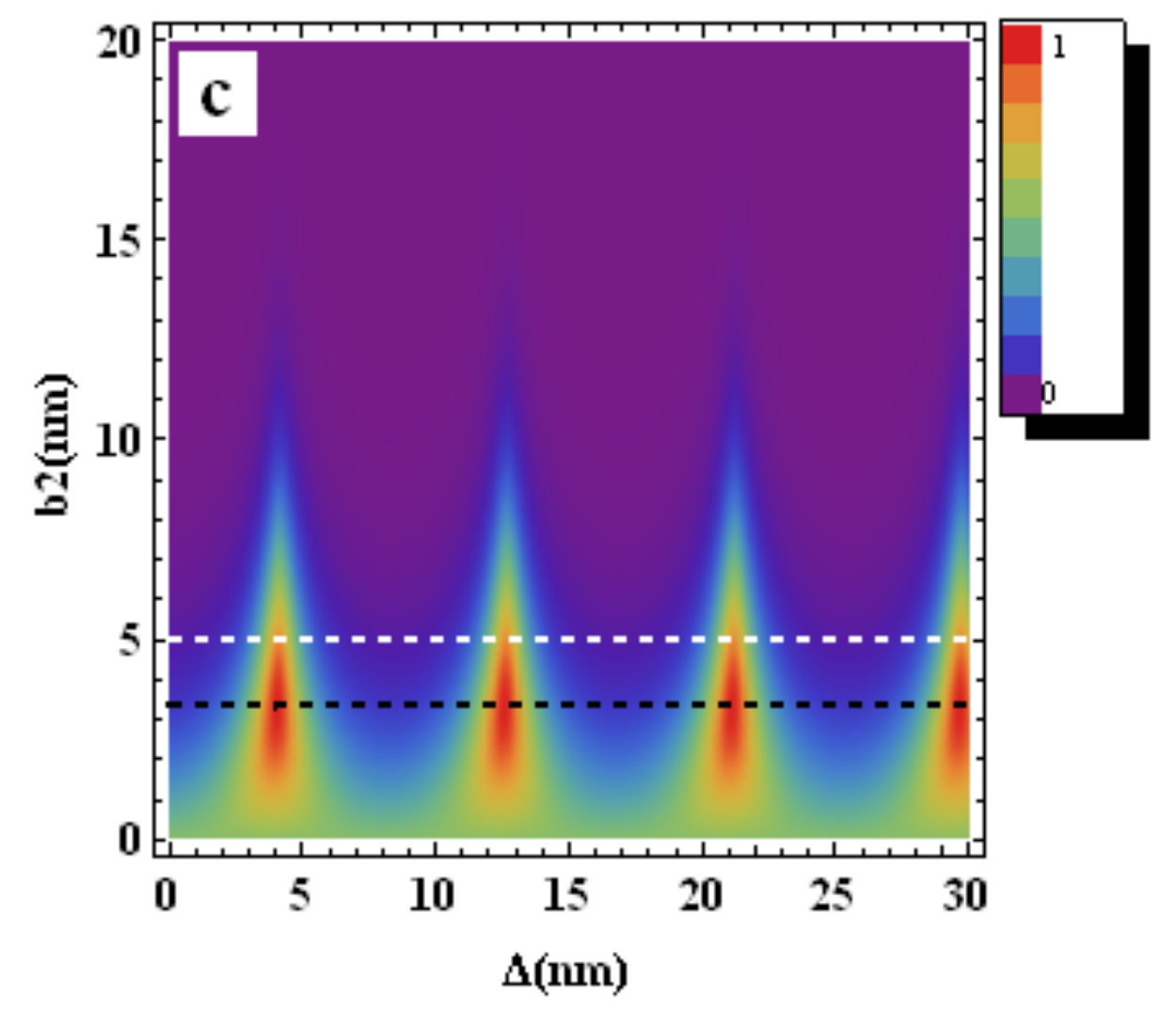}\ \ \
\includegraphics[width=2 in]{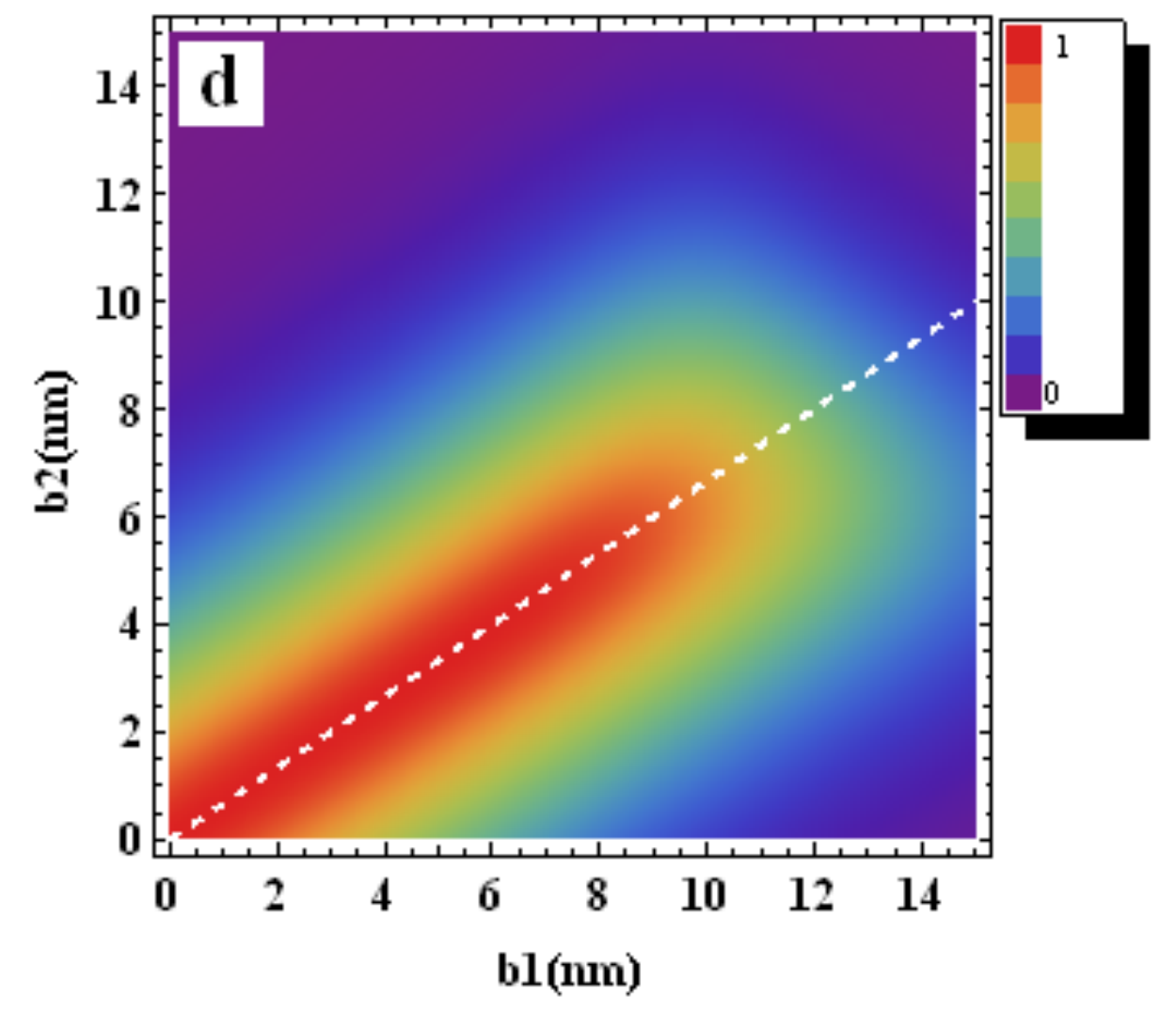}
\caption{ Density plot of transmission at normal incidence for
$E=\frac{4}{5}\ U_2$ and $\delta_2=\delta_4=0$. (a):
$U_{2}=U_{4}=0.4\ \gamma_{1}$, $b_1=5\ nm$. (b): $U_{2}=U_{4}=0.4\
\gamma_{1}$, $\Delta=3.36\ nm$. (c) $U_{2}=0.4\ \gamma_{1},
U_{4}=0.6\ \gamma_{1}$, $b_1=5\ nm$. (d): $U_{2}=0.4\ \gamma_{1},
U_{4}=0.6\ \gamma_{1}$, $\Delta=4\ nm$. The dashed white and black
lines in the left column represent the values of $b_1$ and $b_2$,
respectively, where the resonance occur.}\label{fig4}
\end{figure}
It is well-known that introducing an interlayer potential
difference induces an energy gap in the energy spectrum in bilayer
graphene. It is worth to see how this interlayer potential
difference will affect the transmission probability. To do so, we
extend the results presented in Figure \ref{fig3} to the case
$\delta_2=\delta_4=0.2\ \gamma_1$ to get Figure \ref{fig010}. In
agreement with \cite{27}, Figure \ref{fig010}a shows a full
transmission inside the gap in the energy spectrum, which
resulting from the available states in the well between the
barriers. In contrast to the single barrier case \cite{27, 30},
there are full transmission inside the energy gap. In Figure
\ref{fig010}b, we show the density plot of the transmission
probability as a function of $E$ and $\Delta$ for fixed thickness
of the tow barriers. We note that the resonances resulting from
the bound states in the well are highly influenced by the
interlayer potential difference where it removes part of them and
arises a full transmission at specific value of the energy
$E\approx 0.17\ \gamma_1$, which is absent in the case when there
is no interlayer potential difference ($\delta_2=\delta_4=0$ in
Figure \ref{fig3}b). Figures \ref{fig010}c,\ref{fig010}d show the
same result as in Figures \ref{fig010}a,\ref{fig010}b,
respectively, but with different heights of the barriers $U_2=0.4\
\gamma_1$ and $U_4=0.6\ \gamma_1$, which shows a decreasing in the
transmission probability as a results of the asymmetric structure
of the two barriers.

In Figure \ref{fig4} we observe how these resonances for normal
incidence are affected by the parameters of the barriers. In the
first row we fixed the thickness of the first barrier $b_1=5\ nm$
and set the height of the two barriers to be the same
$(U_{2}=U_{4}=0.4\ \gamma_{1})$, then we plot the transmission as
a function of $\Delta$ and the thickness of the second barrier
$b_2$ as depicted in Figure \ref{fig4}a. These resonances occur
frequently as $\Delta$ increases where $b_2$ (dashed black line)
is equal to $b_1$ (dashed wight line).
Picking up one of these resonances (i.e. at fixed distance between
barriers $\Delta=3.36\ nm$) and calculating the transmission as a
function of $b_1$ and $b_2$ as presented in Figure \ref{fig4}b, it
becomes clear that these resonances occur when $(b_1=b_2)$ for
fixed $\Delta$. In the second row, we show the transmission for
the same parameters as in the first row but with different heights
of the barriers $(U_{2}=0.4\ \gamma_{1}, U_{4}=0.6\ \gamma_{1})$.
Full transmission now occur for $b_1$ (dashed black line) $\neq
b_2$ (dashed black line) as shown in Figure \ref{fig4}c. It is
worth mentioning that for energies less than the strength of the
barriers, and for a fixed $\Delta$, full transmission resonances
occur always when $S_1=S_2$, $S_1$ and $S_2$ being the area of the
first and second barrier, respectively. Therefore, for fixed
$b_1$, the value of $b_2$ where the resonance occur is given by
$b_2=\frac{U_{2}}{U_{4}}\  b_1$ which is superimposed in Figure
\ref{fig4}b,\ref{fig4}d (the dashed white line). Moreover, the
cloak effect in the double barrier occur at non-normal incidence
for some states which is different from the single barrier case
\cite{34} that occur always at normal incidence.

\begin{figure}
\centering
\includegraphics[width=2 in]{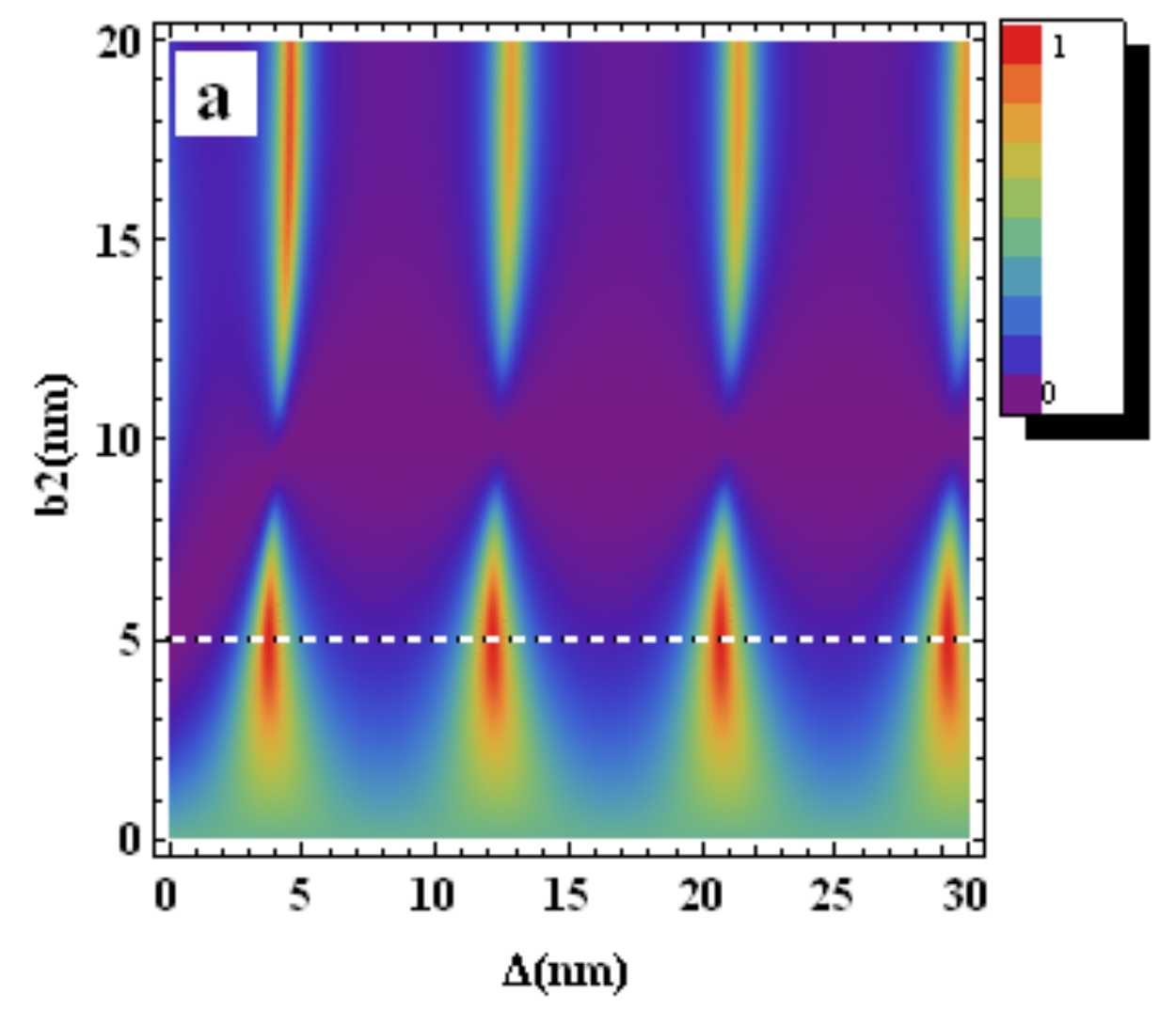}\ \ \
\includegraphics[width=2 in]{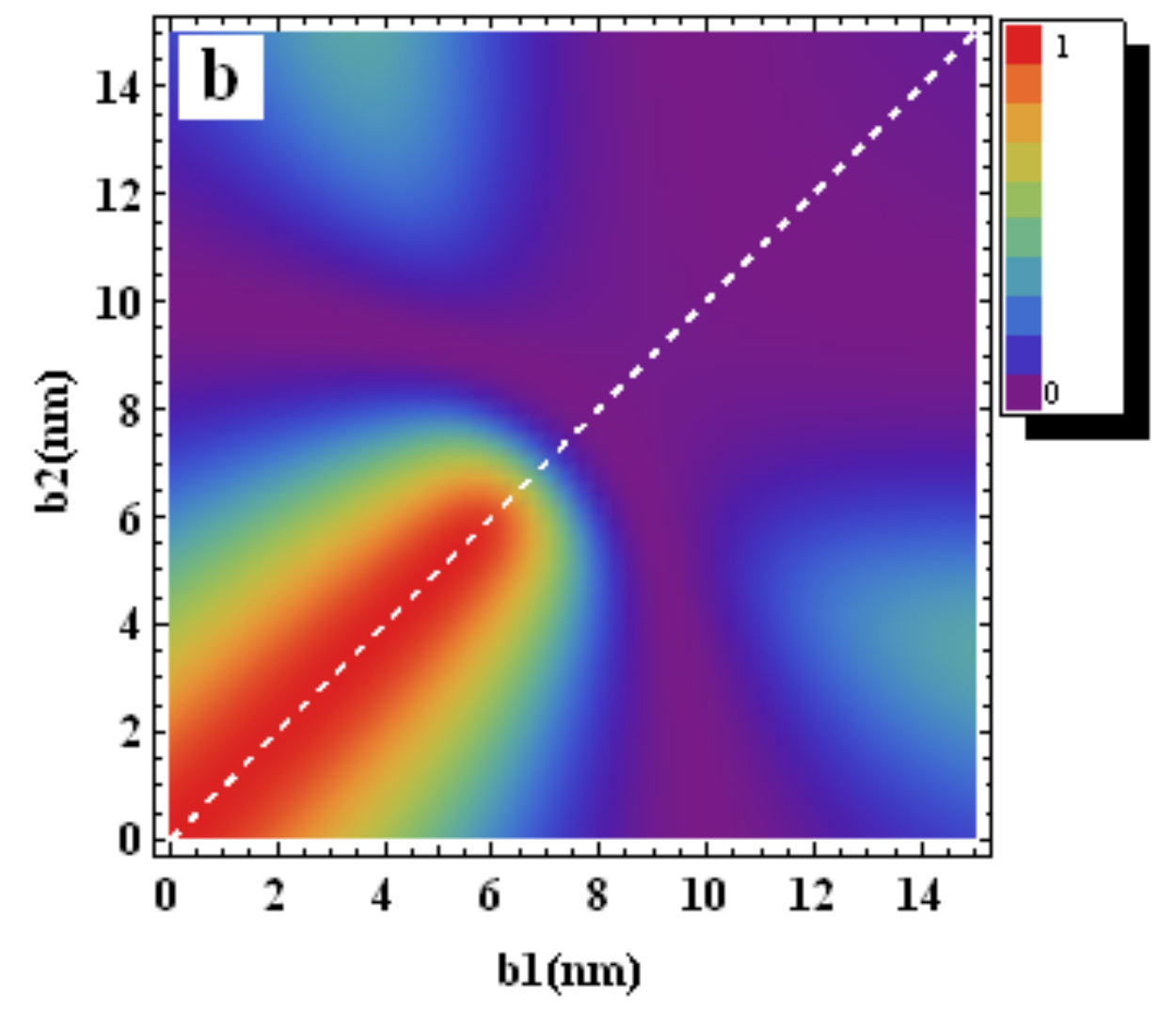}\\
\includegraphics[width=2 in]{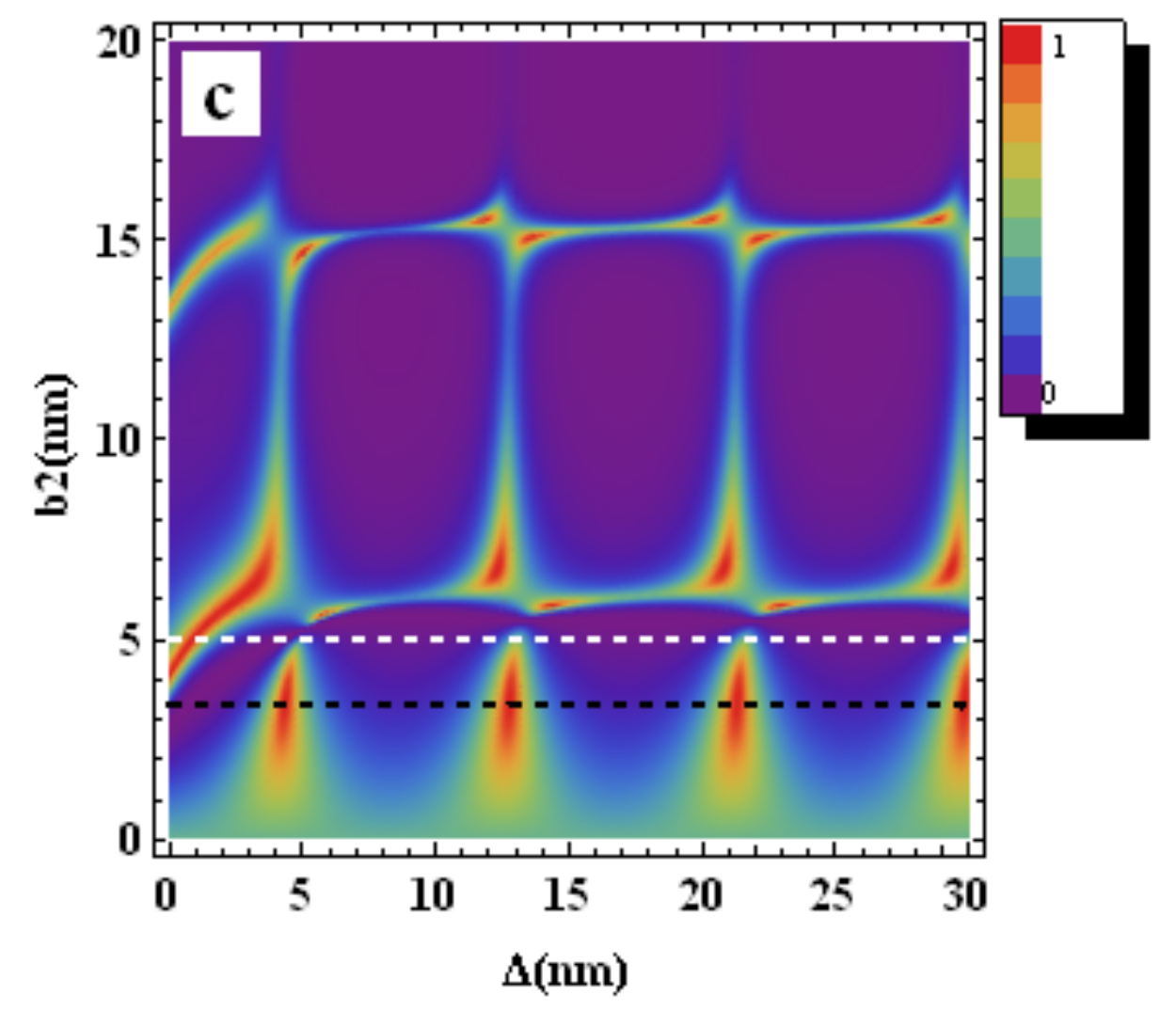}\ \ \
\includegraphics[width=2 in]{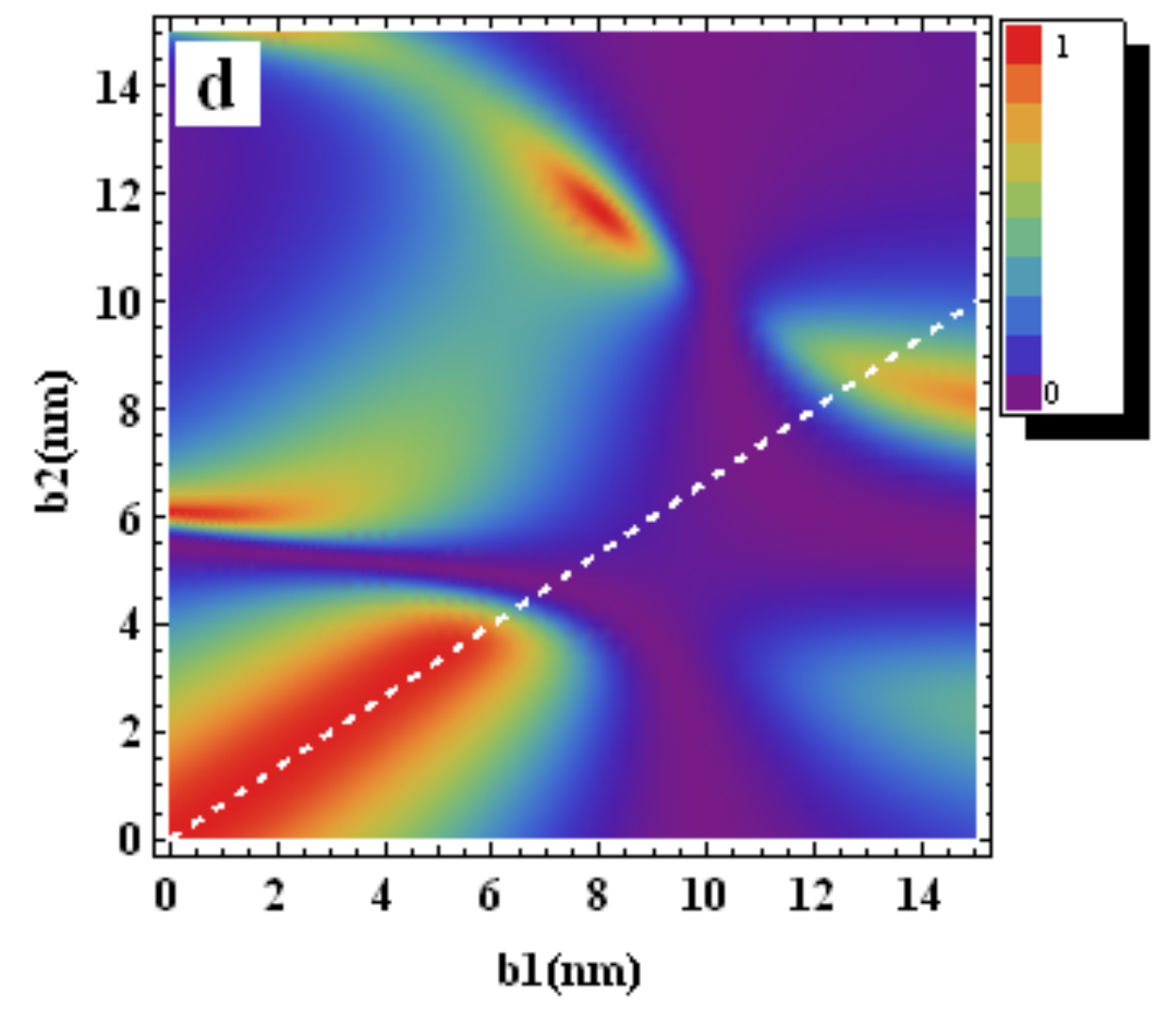}
\caption{ Density plot of transmission at normal incidence for
$E=\frac{4}{5}\ U_2$ and $\delta_2=\delta_4=0.1\ \gamma_1$. (a):
$U_{2}=U_{4}=0.4\ \gamma_{1}$, $b_1=5\ nm$. (b): $U_{2}=U_{4}=0.4\
\gamma_{1}$, $\Delta=3.7\ nm$. (c): $U_{2}=0.4\ \gamma_{1},
U_{4}=0.6\ \gamma_{1}$, $b_1=5\ nm$. (d) $U_{2}=0.4\ \gamma_{1},
U_{4}=0.6\ \gamma_{1}$, $\Delta=4.3\ nm$. The dashed white and
black lines in the left column represent the values of $b_1$ and
$b_2$, respectively, where the resonance occur.}\label{fig114}
\end{figure}

In Figure
\ref{fig114} we extend the results in Figure \ref{fig4} but with
interlayer potential difference ($\delta_2=\delta_4=0.1\
\gamma_1$) for the same other parameters. As we note the total
transmission probability is decreasing and some of the original
resonances are splitting as a sequence of the induced energy gap.
Let us now see how the transmission probability is affected by the
double barrier parameters.

\begin{figure}
\centering
\includegraphics[width=2 in]{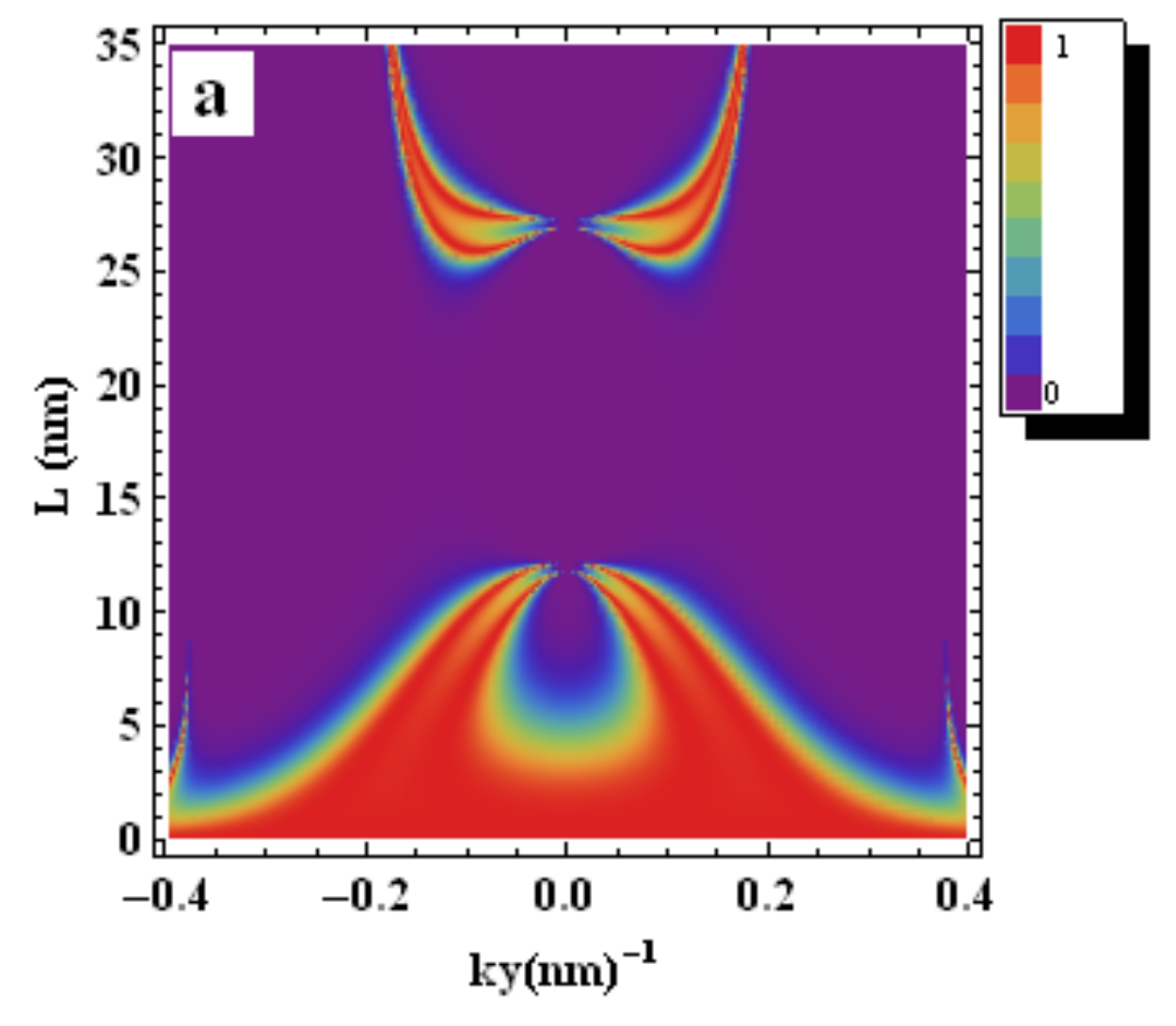}\ \ \
\includegraphics[width=2 in]{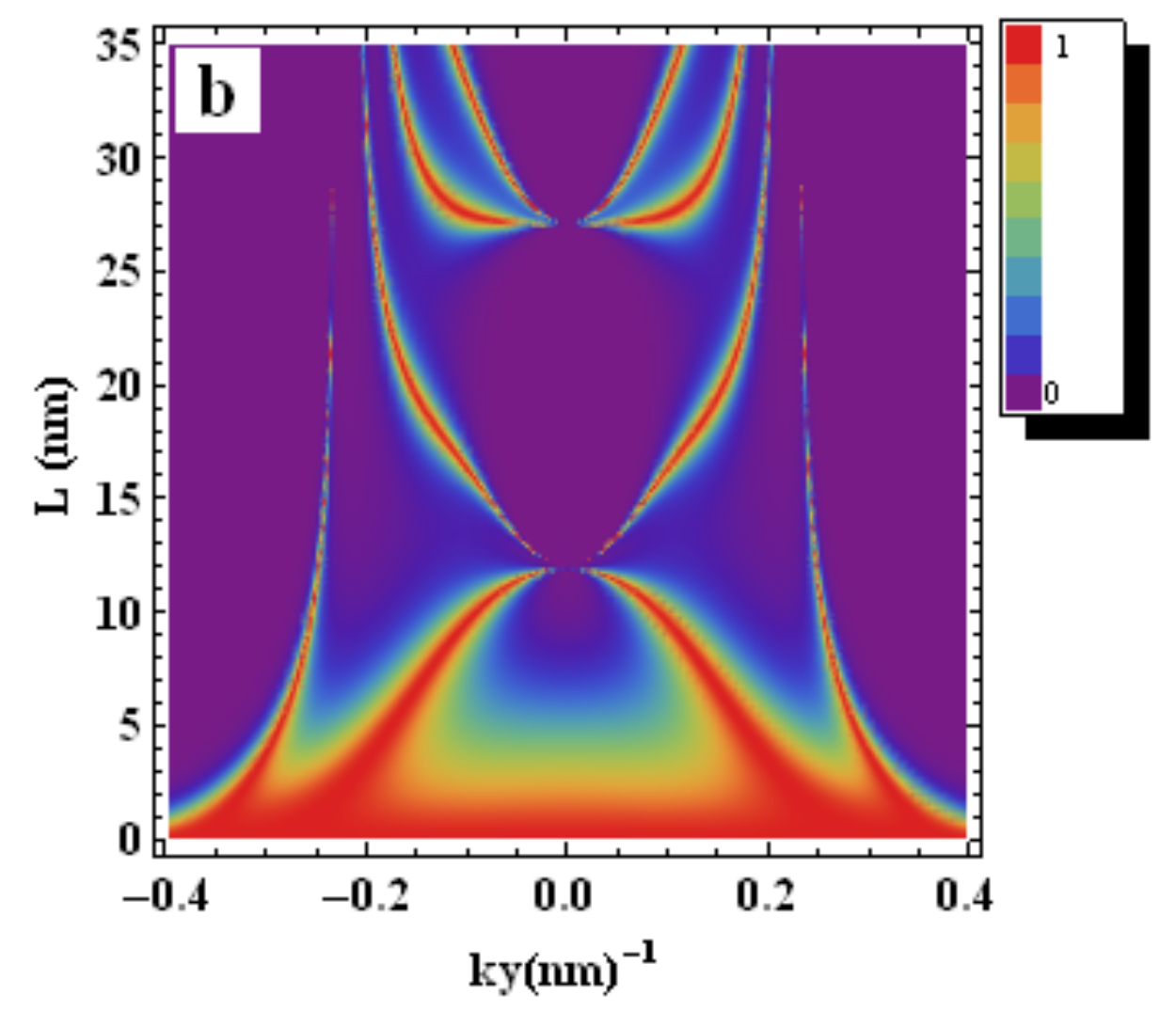}\\
\includegraphics[width=2 in]{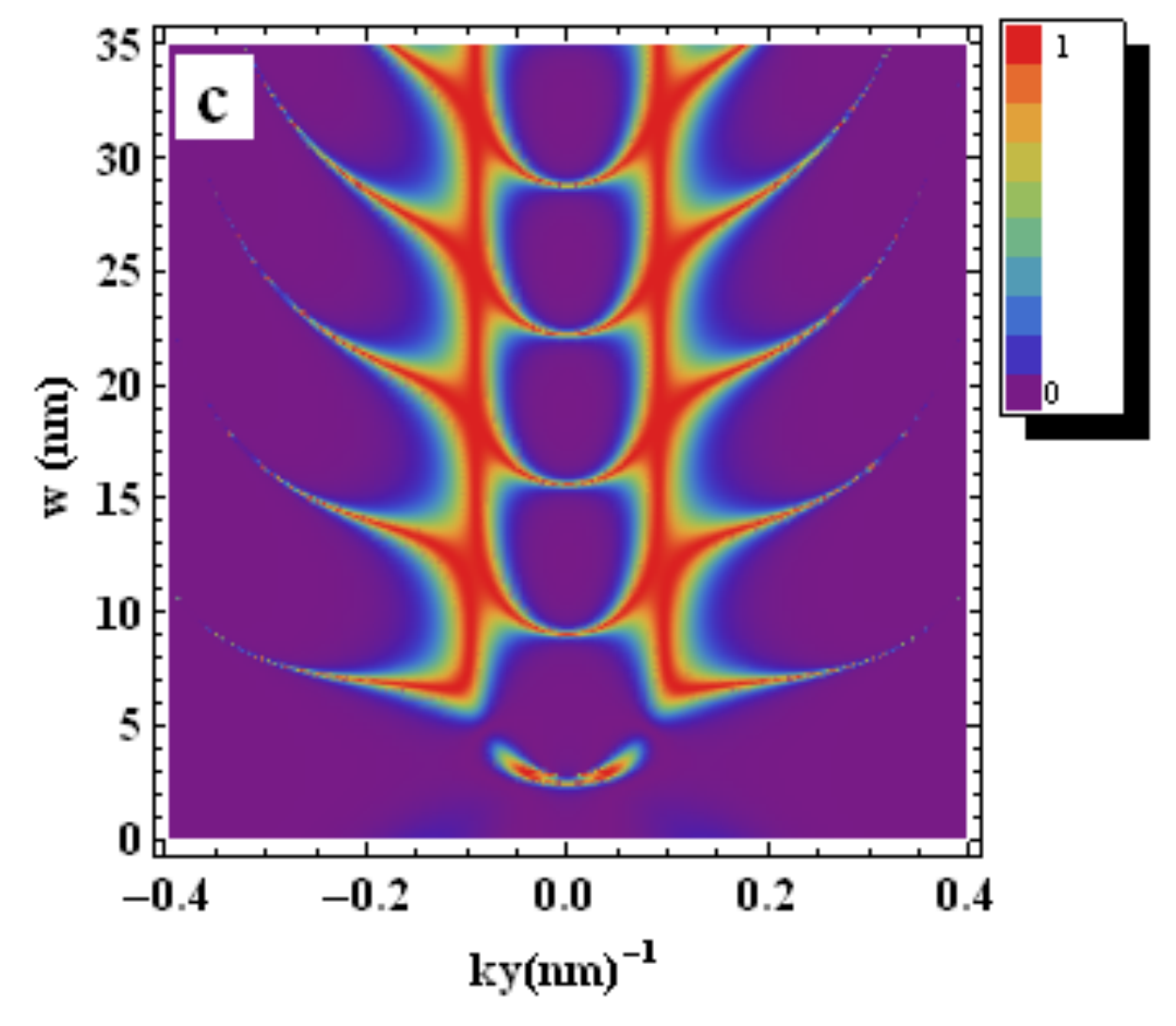}\ \ \
\includegraphics[width=2 in]{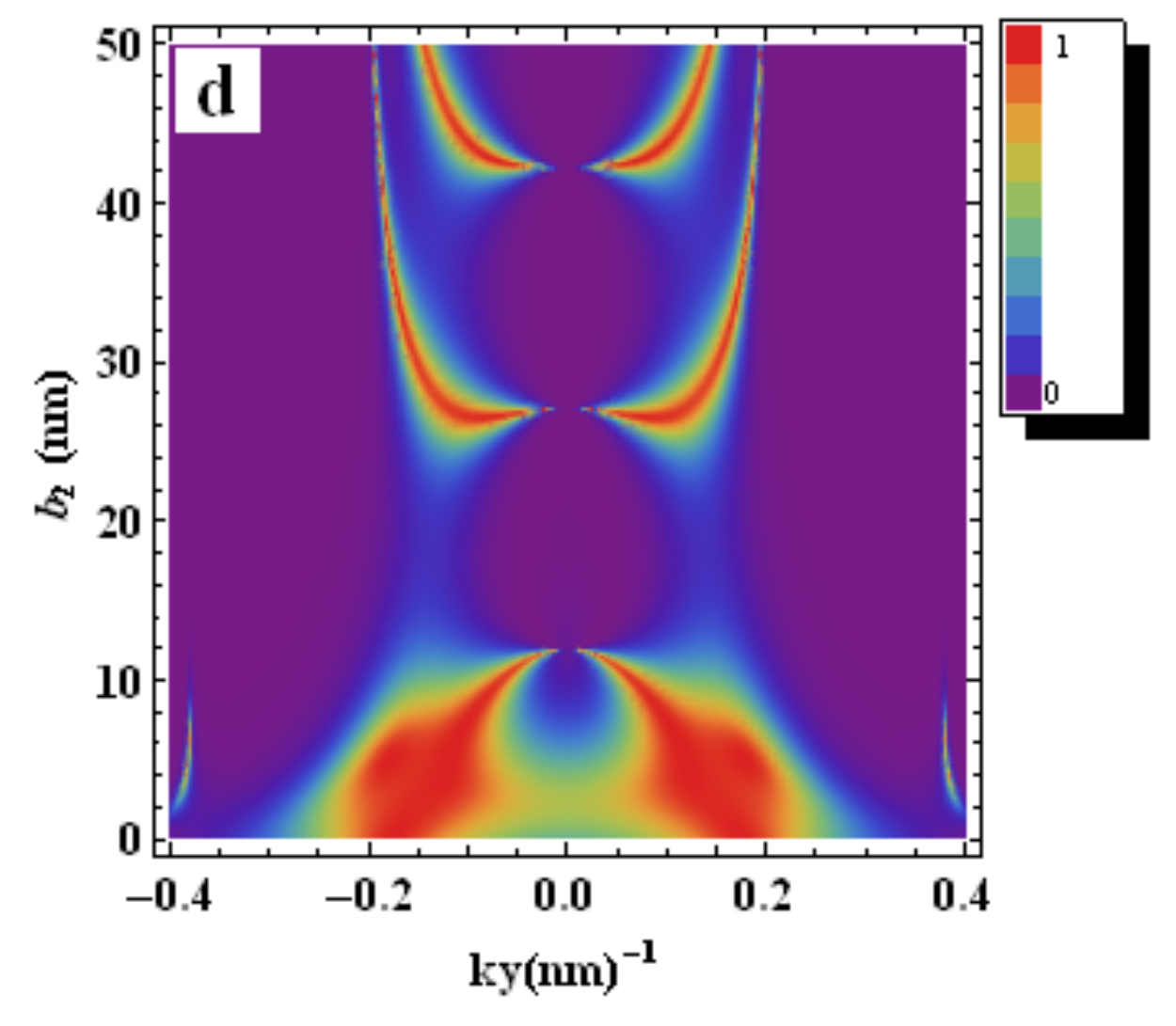}
\caption{Density plot of the transmission probability versus (a,b): $k_y$ and the width of the two barriers $(b_1=b_2=L)$ for
$U_2=U_4=0.6\ \gamma_1$, $E=\frac{4}{5}\ U_4$ and $ \Delta=10\ nm,
15\ nm$, respectively. (c): $k_y$ and $\Delta$ for the same
parameters as in (a) and for $b_1=b_2=10\ nm$. (d): $ky$ and $b_2$
with $b_1=5\ nm$ and $\Delta=10\ nm$.}\label{fig009}
\end{figure}

\begin{figure}
\centering
\includegraphics[width=2 in]{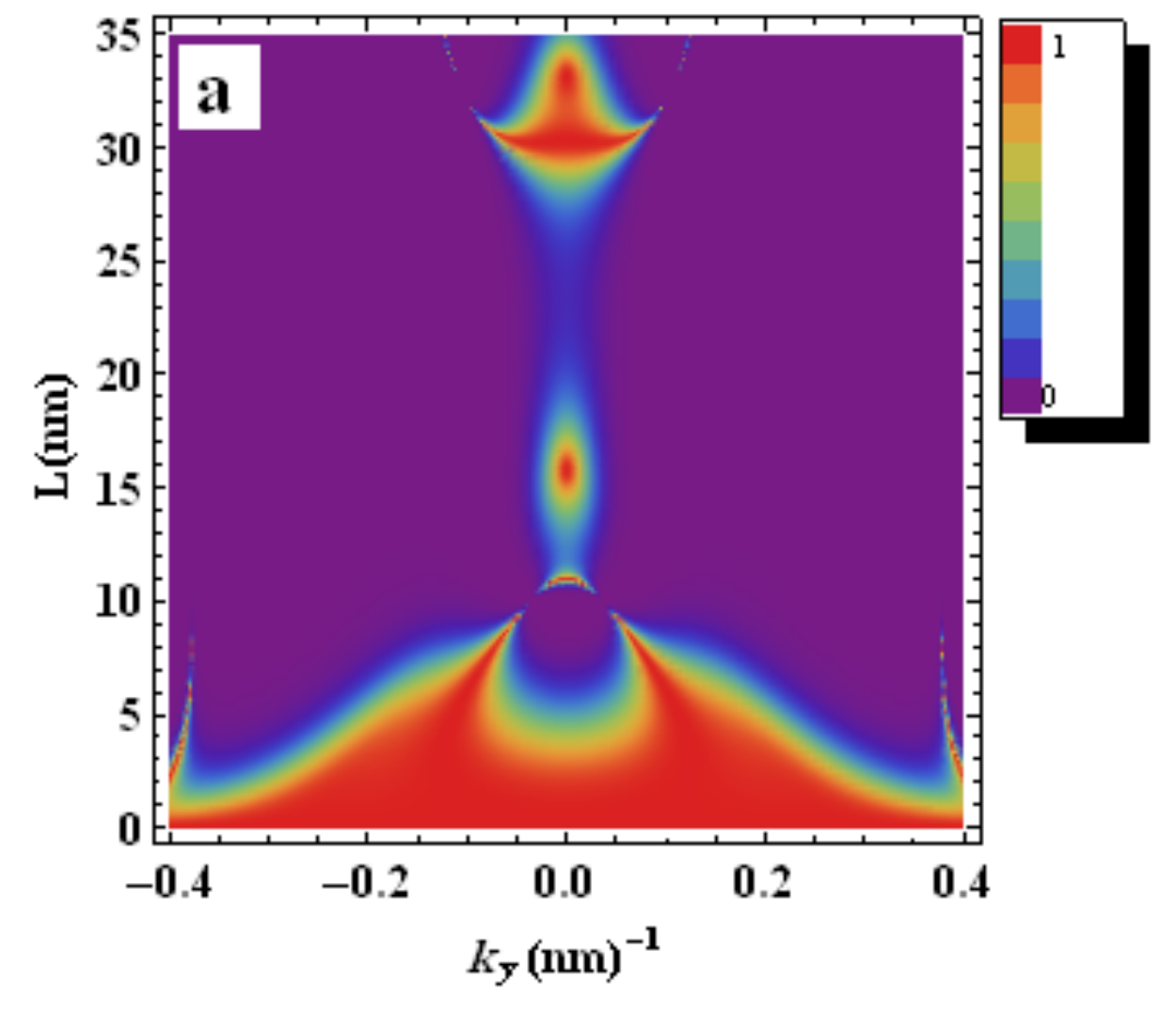}\ \ \
\includegraphics[width=2 in]{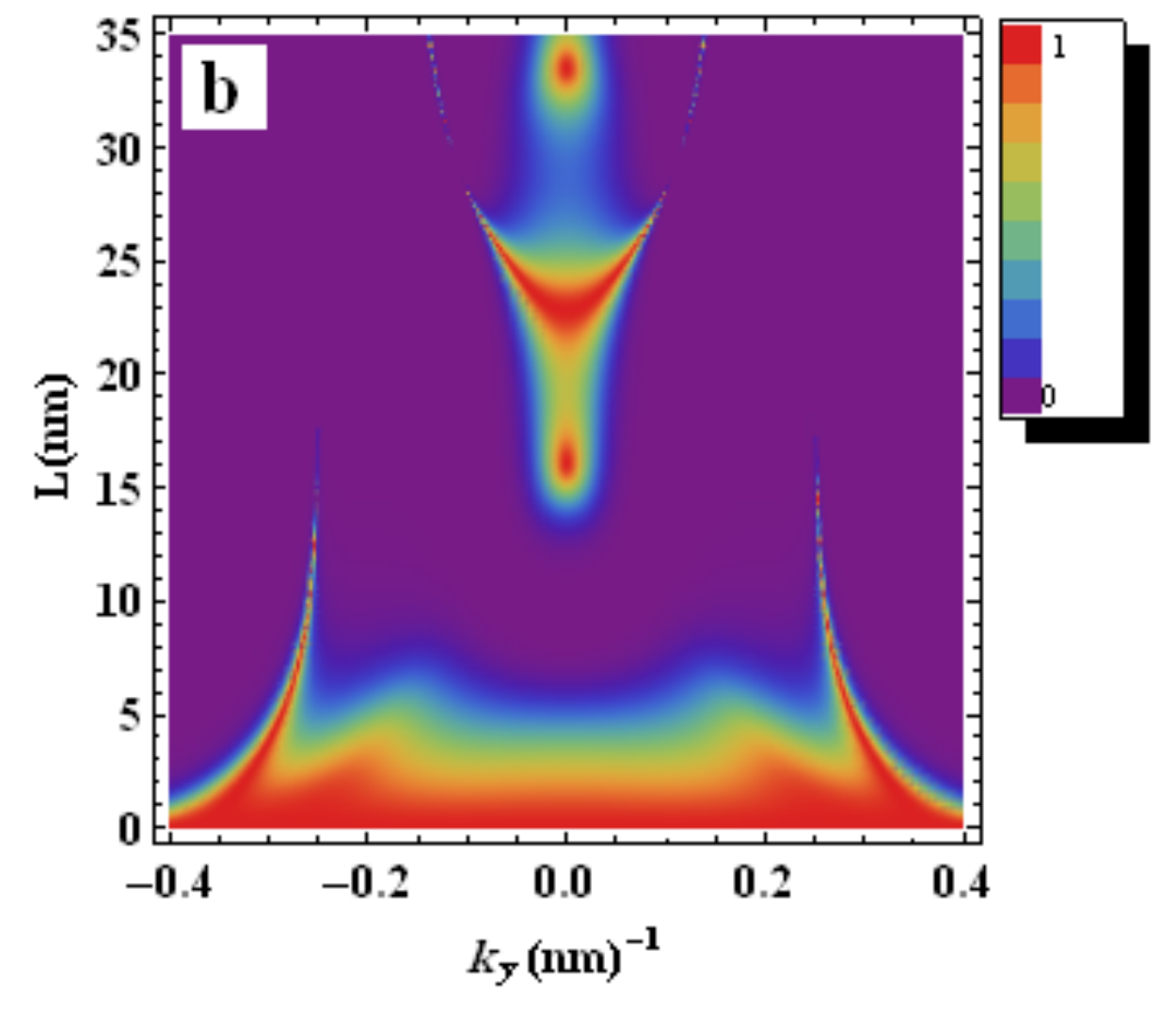}\\
\includegraphics[width=2 in]{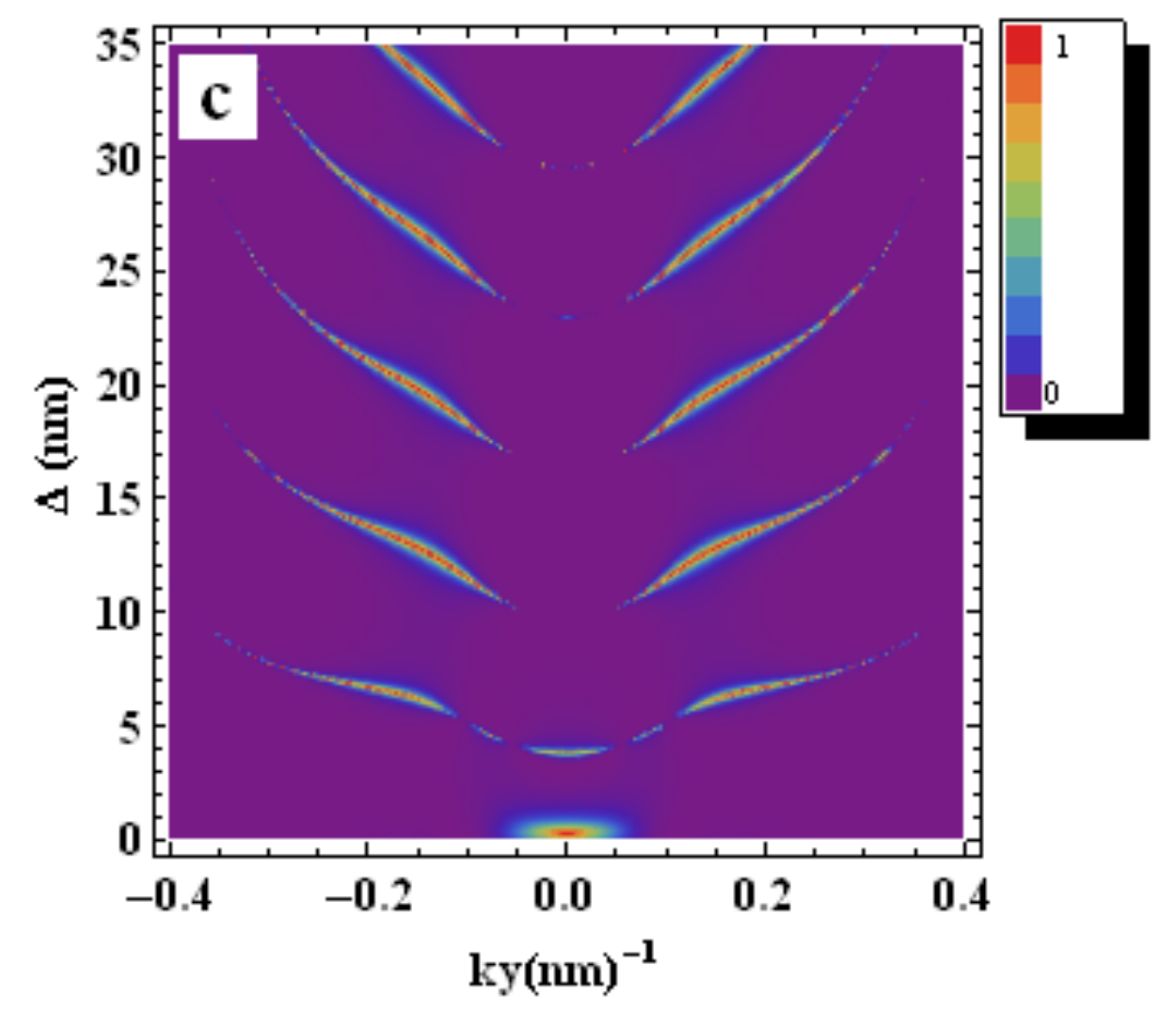}\ \ \
\includegraphics[width=2 in]{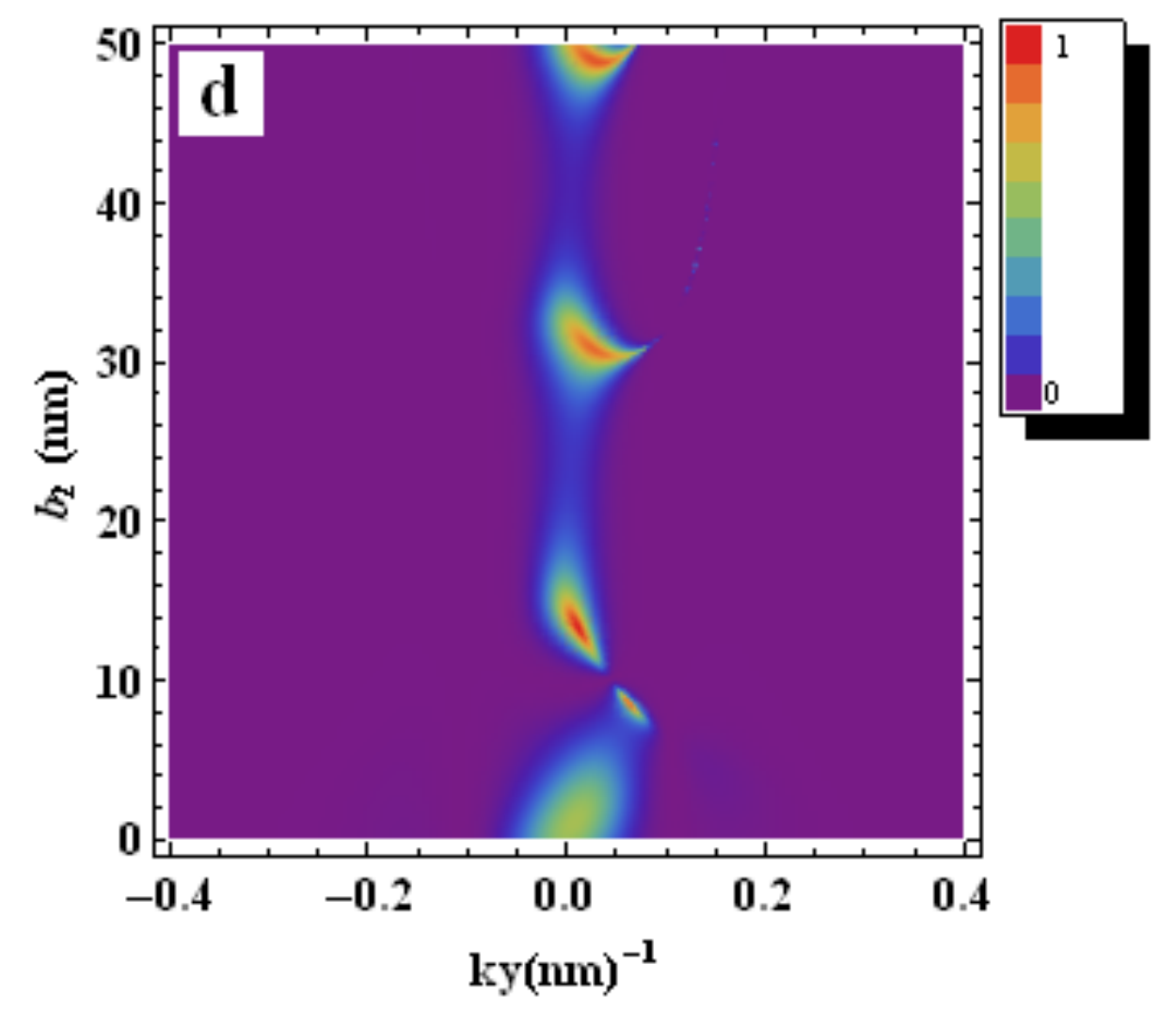}
\caption{Density plot of the transmission probability versus (a,b): $k_y$ and the width of the two barriers $(b_1=b_2=L)$ for
$U_2=U_4=0.6\ \gamma_1$, $E=\frac{4}{5}\ U_4$,
$\delta_2=\delta_4=0.1\ \gamma_1$ and $ \Delta=10\ nm, 15\ nm$,
respectively. (c): $k_y$ and the width of the well $\Delta$ for the
same parameters as in (a) and for $b_1=b_2=10\ nm$. (d): $ky$ and
$b_2$ with $b_1=5\ nm$ and $\Delta=10\ nm$.}\label{fig0059}
\end{figure}

In Figures \ref{fig009}a,\ref{fig009}b
we show the density plot of the transmission probability for
$U_2=U_4$, $E<U_2=U_4$ and different values of $\Delta$, as a
function of $k_y$ and the thickness of the two barriers $L$ (i.e.
with changing the width of the two barriers simultaneously by
setting $b_1=b_2=L$). For $\Delta=10\ nm$ and for small $L$ we
have a full transmission for wide range of $k_y$, with increasing
$L$, transmission probability dramatically decreases however, some
resonances still show up as depicted in Figure \ref{fig009}a. In
contrast, for $\Delta=15\ nm$ the transmission probability is
completely different where the position and number of resonant
peaks change as depicted in Figure \ref{fig009}b. This stress that
the crucial parameters that determine the number of resonant peaks
and their position is the width of the well $\Delta$ not the
thickness of the two barriers $b_1$ and $b_2$\cite{27,133}.
$\Delta$ dependance of the transmission probability is shown in
Figure \ref{fig009}c, we note a full transmission frequently occur
for normal incidence. Moreover, after certain value of $\Delta$ we
start getting a full transmission for specific value of $k_y$ and
for all values of $\Delta$. In Figure \ref{fig009}d we show how
the transmission probability changes with $b_2$ and $k_y$  for
fixed $\Delta$ and $b_1$.

The effect of the interlayer potential
difference on the transmission probability with respect to the
geometry of the barriers is depicted in Figure \ref{fig0059} for
the same parameters as in Figure \ref{fig009} but for
$\delta_2=\delta_4=0.1\ \gamma_1$, we note that most of the
resonances disappeared as one can conclude from Figure
\ref{fig009} due to the gap in the spectrum resulting from the
induced electric field.

%%%%%%%%%%%%%%%%%%%%%%%%%%%%%%%%%%%%%%%%%%%%%%%%%%%
\section{Four band tunneling} % ($E>\gamma_{1}$)}
%%%%%%%%%%%%%%%%%%%%%%%%%%%%%%%%%%%%%%%%%%%%%%%%%%%

\begin{figure}[h]
\centering
\includegraphics[width=2 in,height=1.7 in]{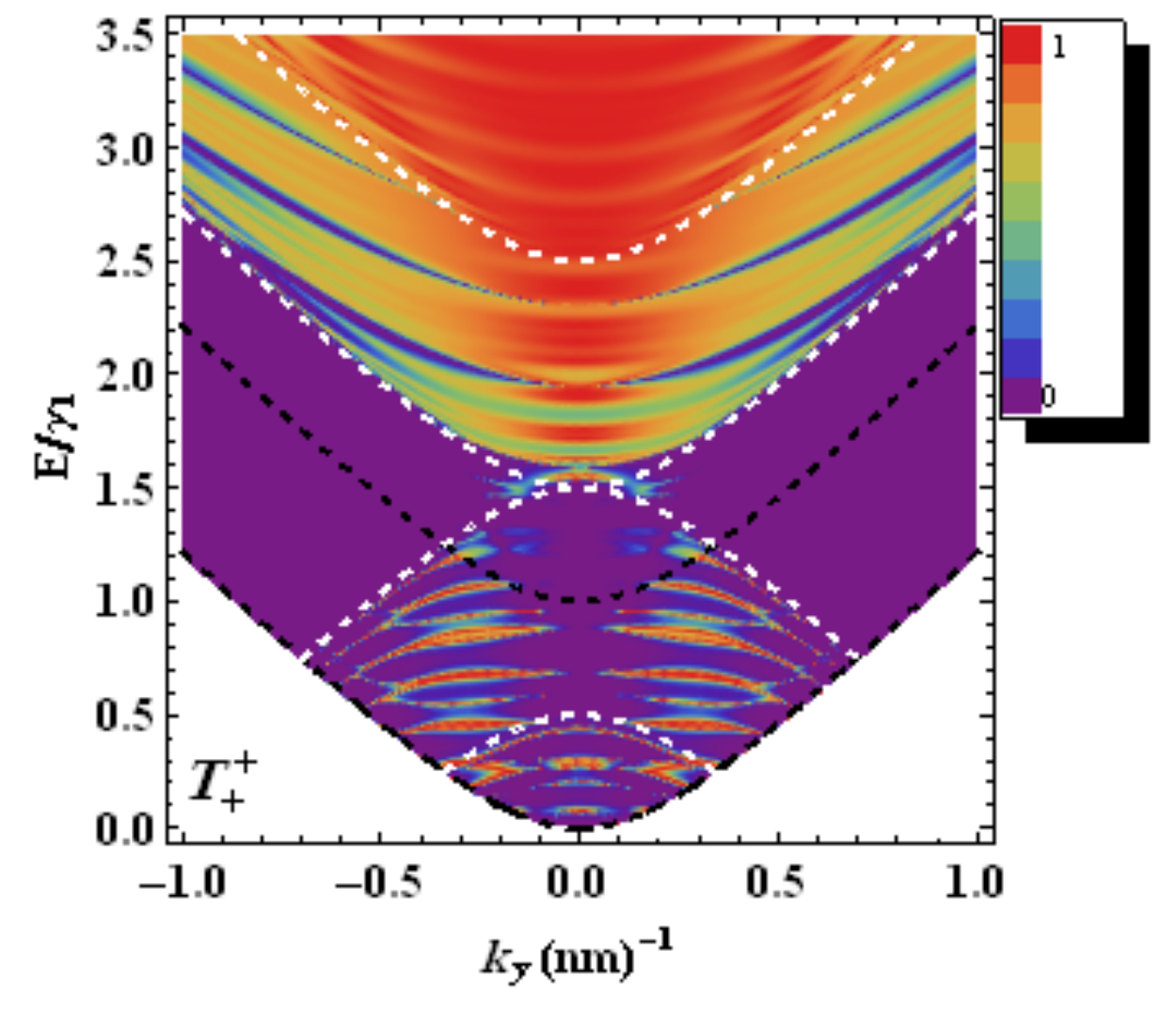}
\includegraphics[width=2 in,height=1.7 in]{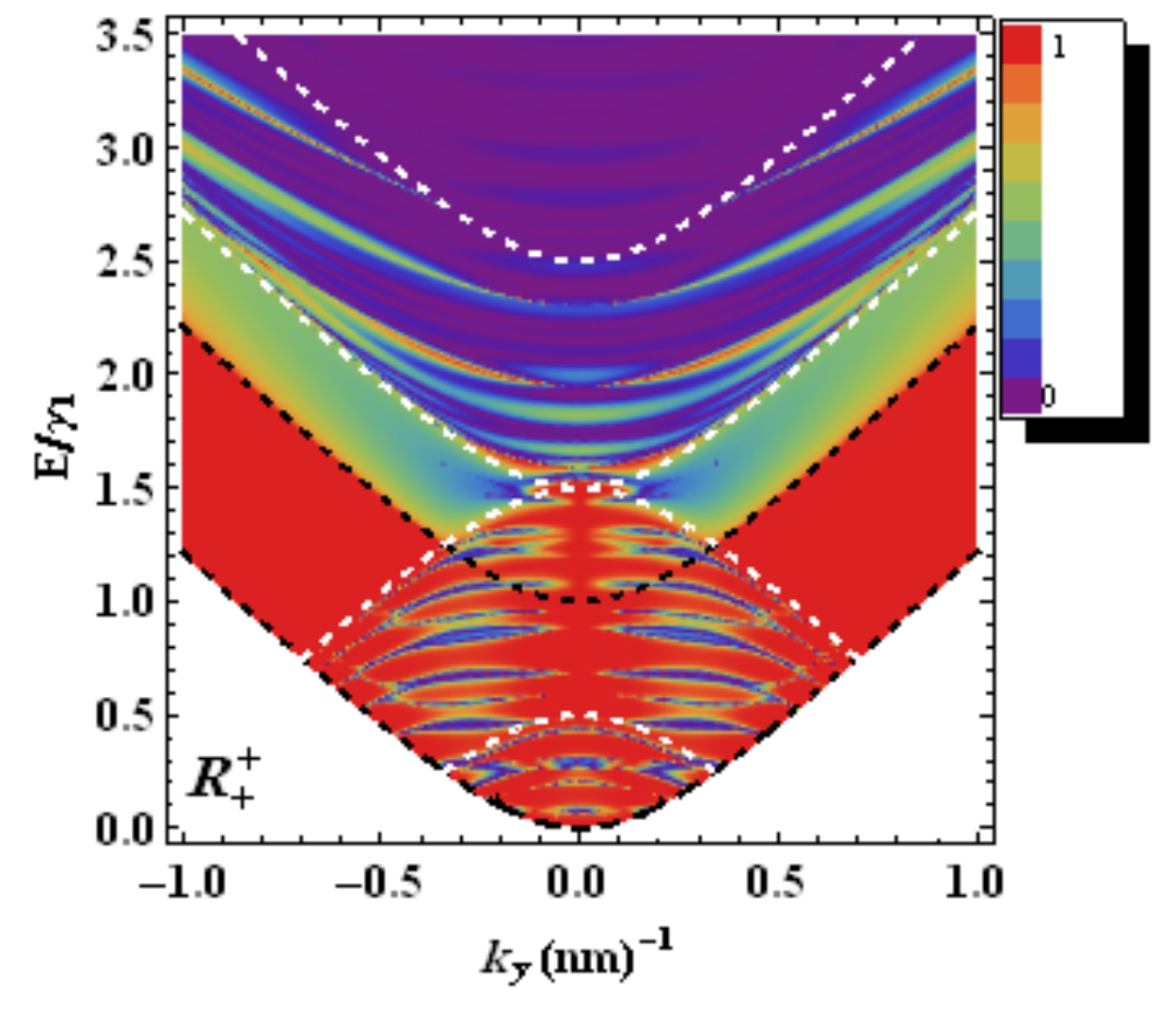}
\includegraphics[width=2 in,height=1.7 in]{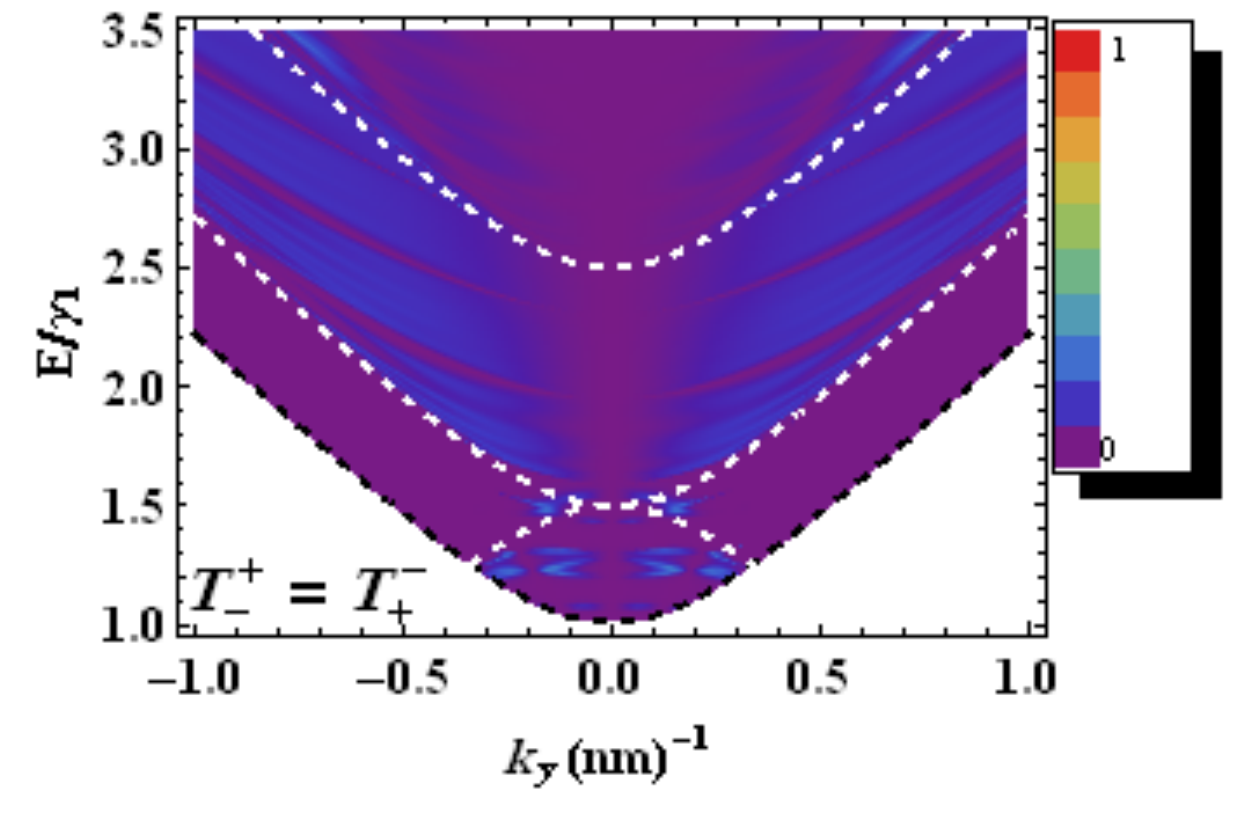}\\
\includegraphics[width=2 in,height=1.7 in]{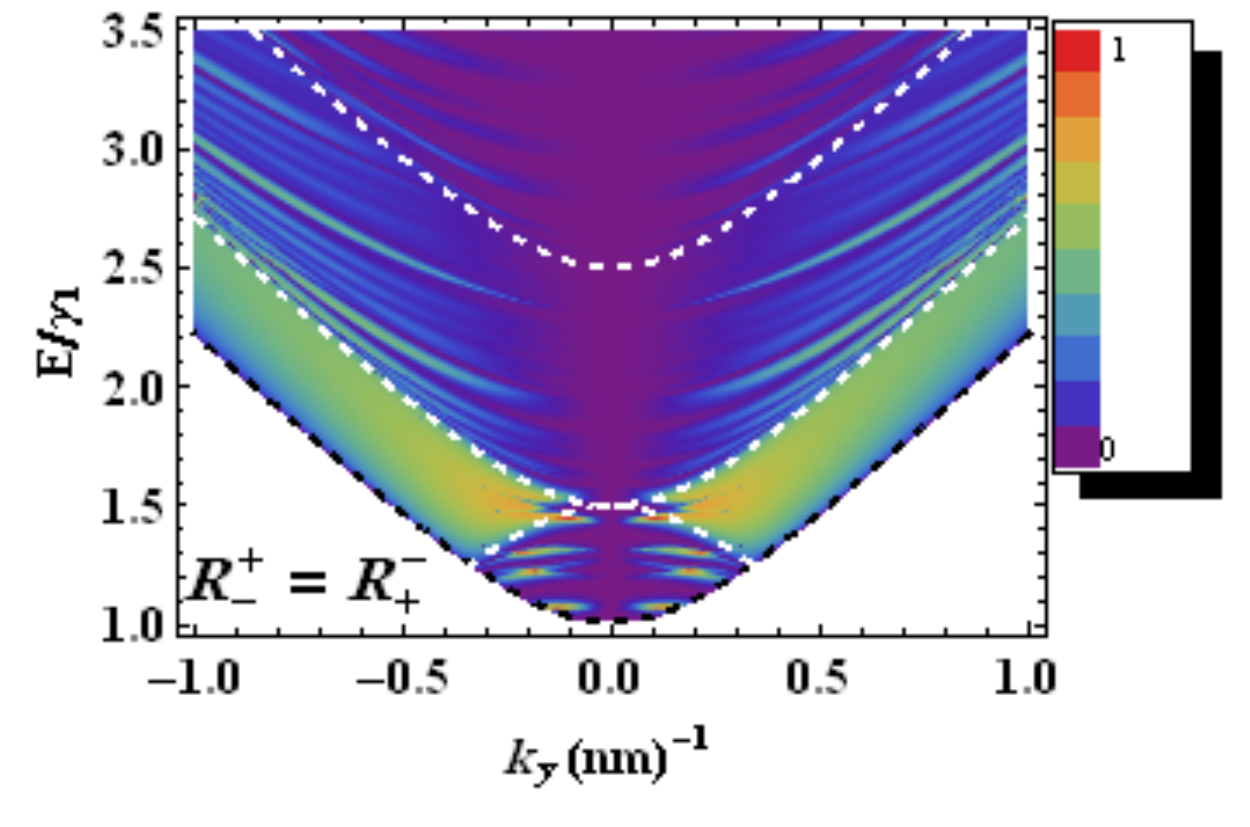}
\includegraphics[width=2 in,height=1.7 in]{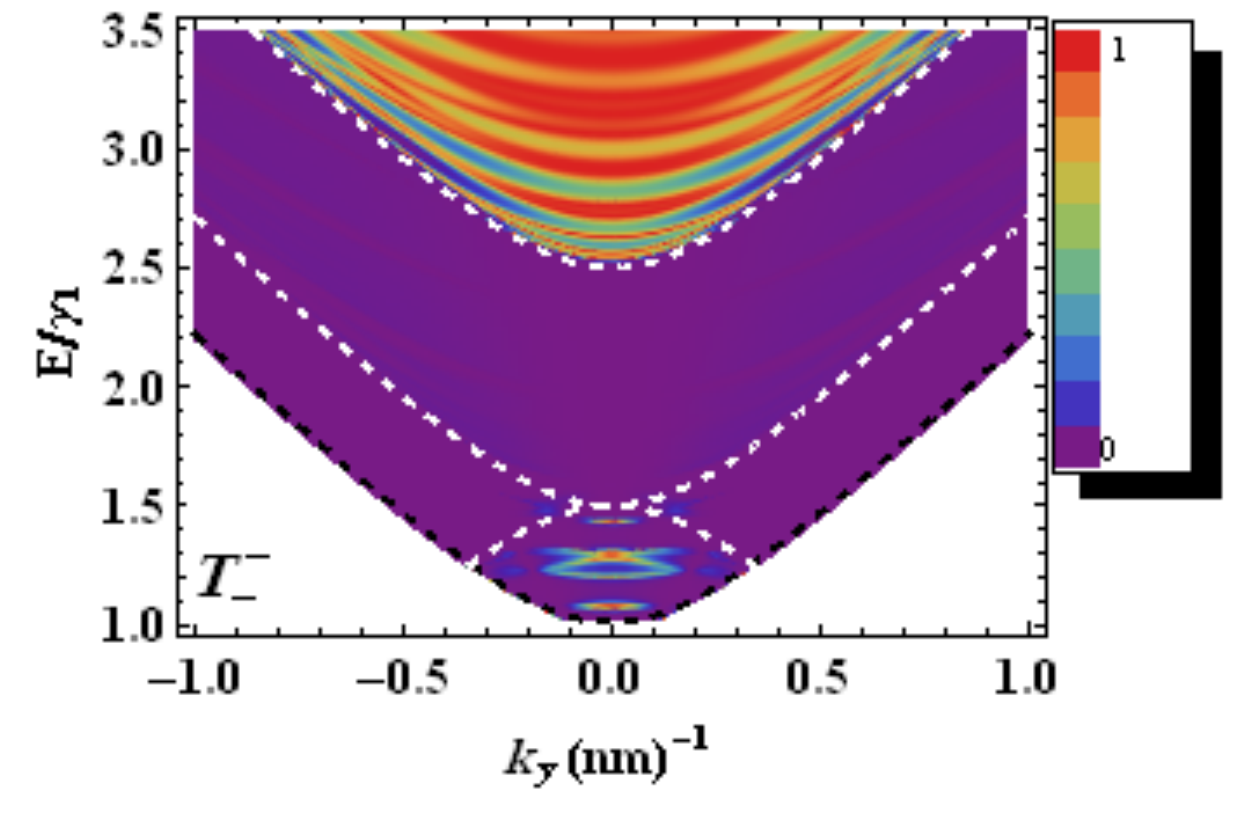}
\includegraphics[width=2 in,height=1.7 in]{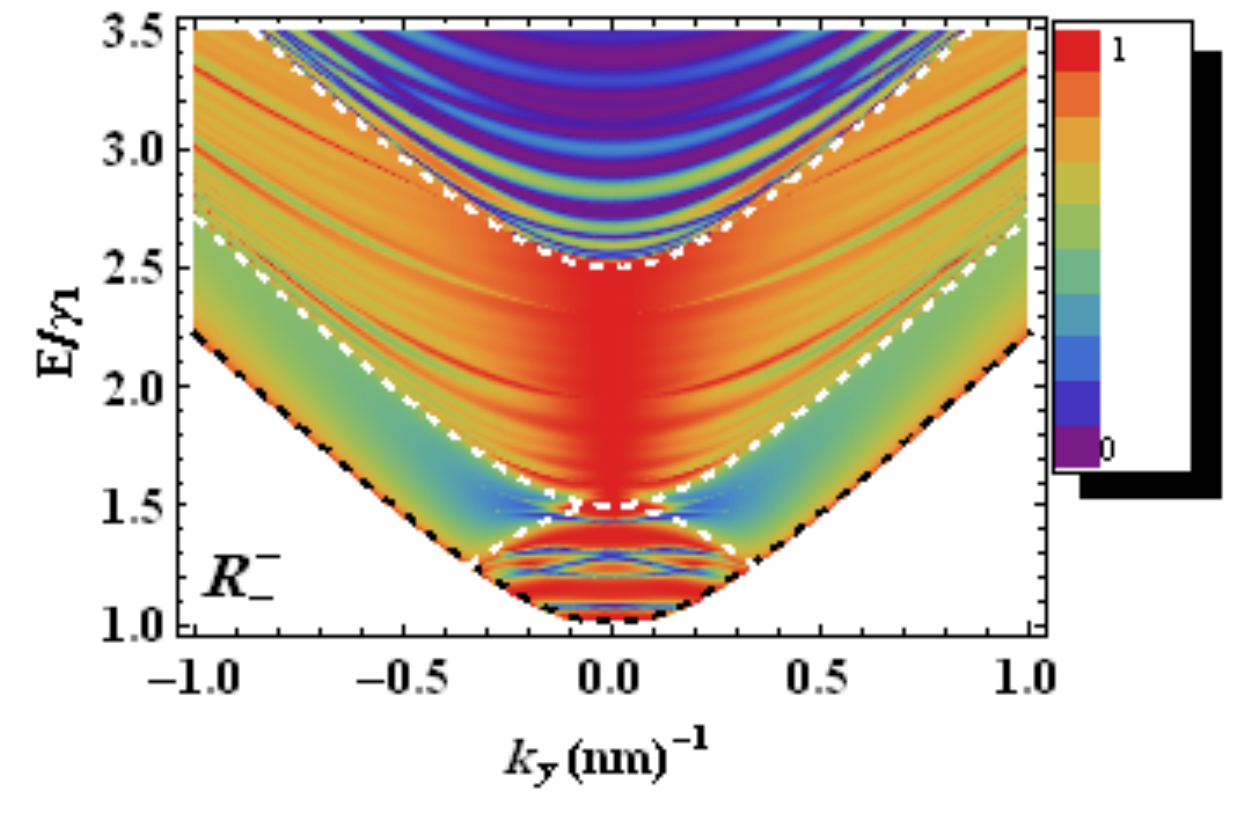}
\caption{Density plot of transmission and reflection probabilities
with $U_{2}=U_{4}=1.5\ \gamma_{1}$, $b_1=b_2=20\ nm$ and
$\Delta=10\ nm$. The dashed white and black lines represent the
band inside and outside the barrier, respectively.}\label{fig5}
\end{figure}

For energies larger than $\gamma_1$, the particles can use the two
conduction band for propagation which gives rise to four channels
of transmission and four for reflection. In Figure \ref{fig5} we
present these reflection and transmission probabilities for a
double barrier structure as a function of $k_y$ and $E$. The
potential barriers heights are set to be $U_2=U_4=\frac{3}{2}\
\gamma_1$ and the interlayer potential difference is zero.
Different regions are shown up in the spectrum ($E,k_y$) which
appeared as a result of the different propagating modes inside and
outside the barriers. The superimposed dashed curves in the
density plot in Figure \ref{fig5} indicates the borders between
these different regions \cite{30}. In the double  barrier, the
cloak effect \cite{34} in $T^{+}_+$ and $T^{+}_-$ $(T^{-}_+)$
occurs in the region $U_{2}-\gamma_1<E<U_{2}$ for nearly normal
incidence $k_y\approx0$ where the two modes $k^+$ and $k^-$ are
decoupled and therefore no scattering occurs between them
\cite{30}. However, this effect also exist for some states for
non-normal incidence as a result of the available electrons states
in the well as mentioned in the previous section. For non-normal
incidence the two modes $k^+$ and $k^-$ are coupled and hence the
electrons can be scattered between them, so that the transmission
$T^{+}_+$ and $T^{+}_-$ $(T^{-}_+)$ in the same region are not
zero for non-normal incidence. For energies less than
$U_{2}-\gamma_1$ electrons propagate via $k^+$ mode inside the
barriers which give the resonances in $T^{+}_+$ in this region.
Increasing (decreasing) the area of the barriers or the well
between them will increase (decrease) the number of these
resonances.

\begin{figure}[h]
\centering
\includegraphics[width=2 in,height=1.7 in]{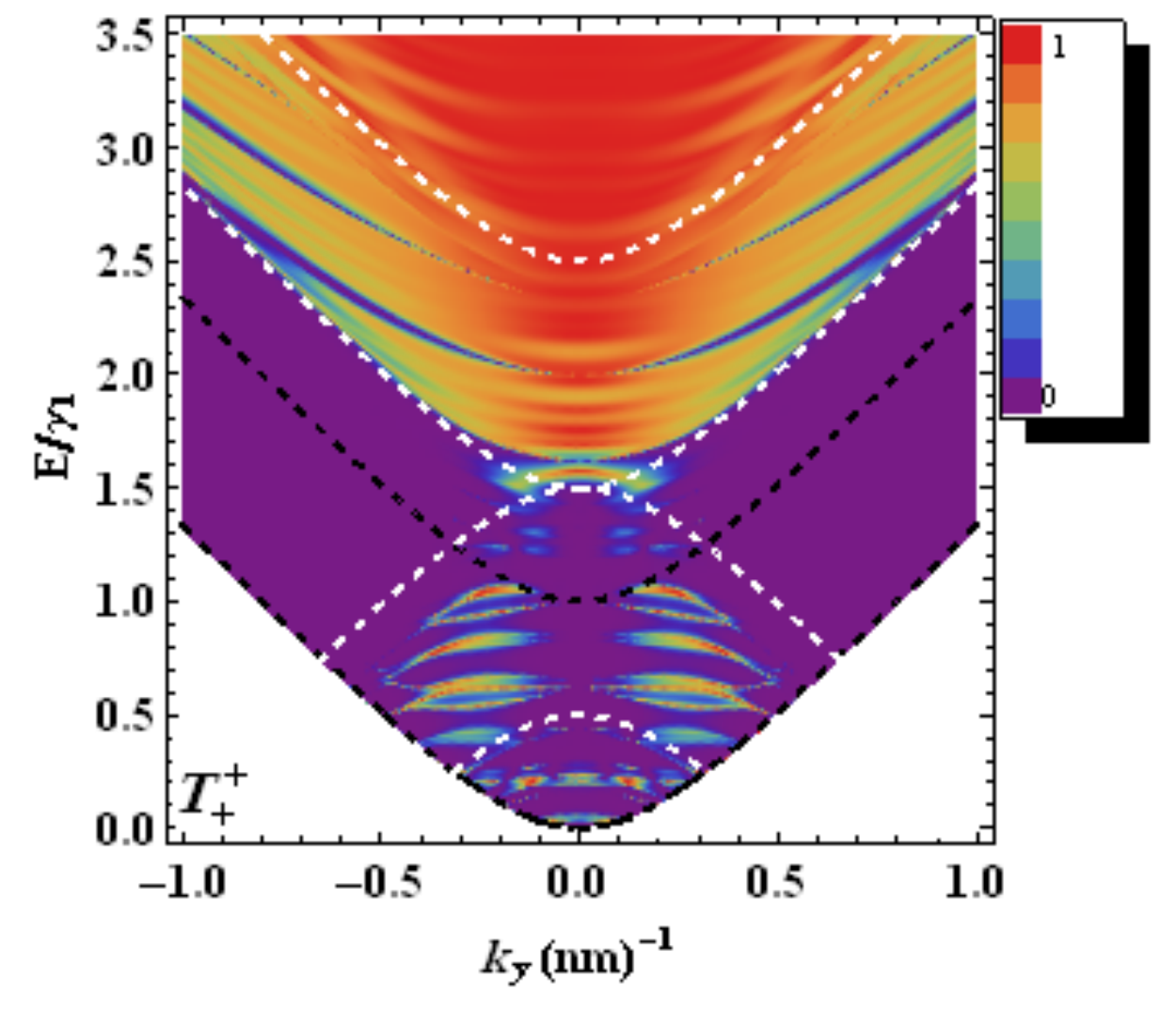}
\includegraphics[width=2 in,height=1.7 in]{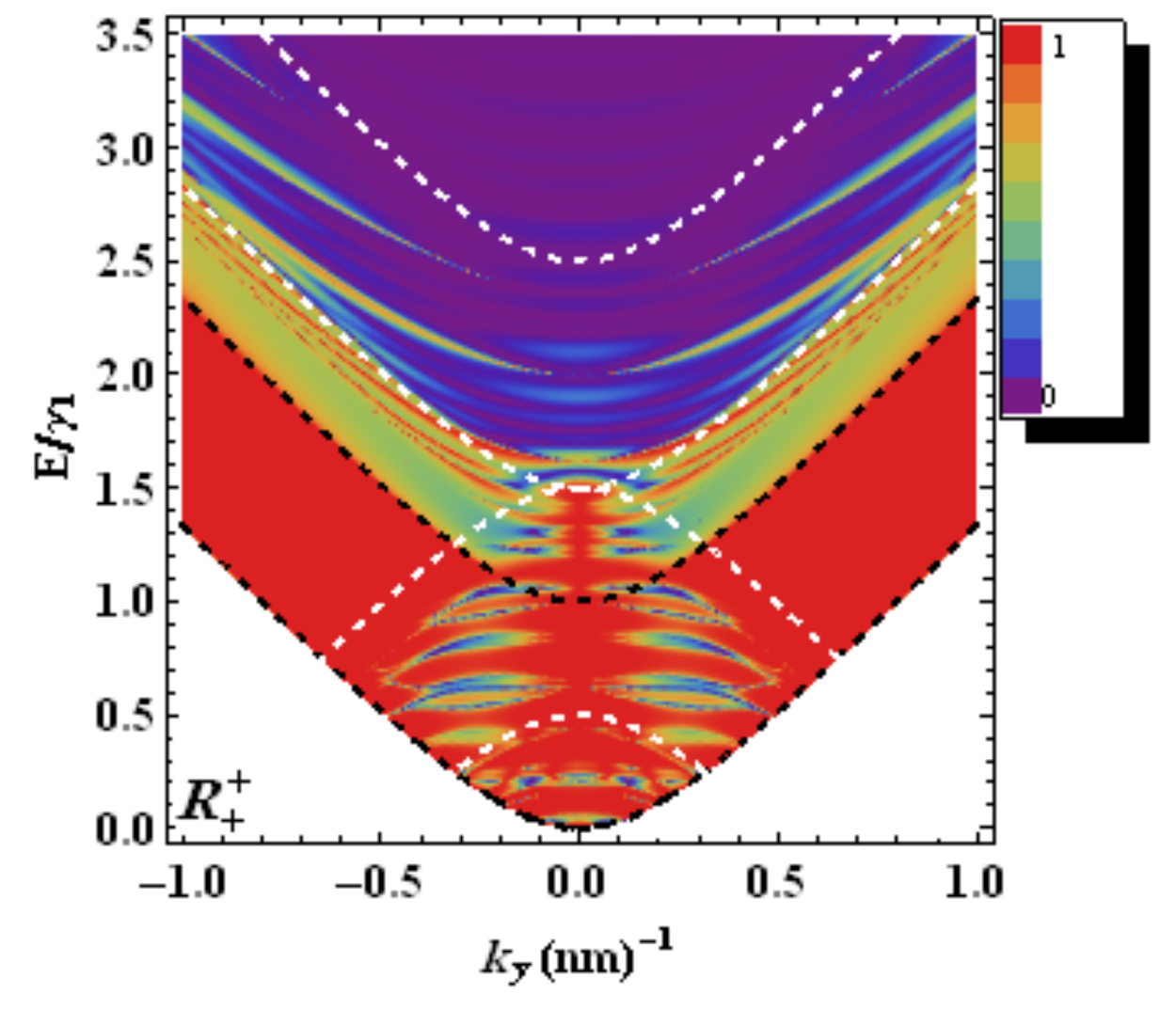}
\includegraphics[width=2 in,height=1.7 in]{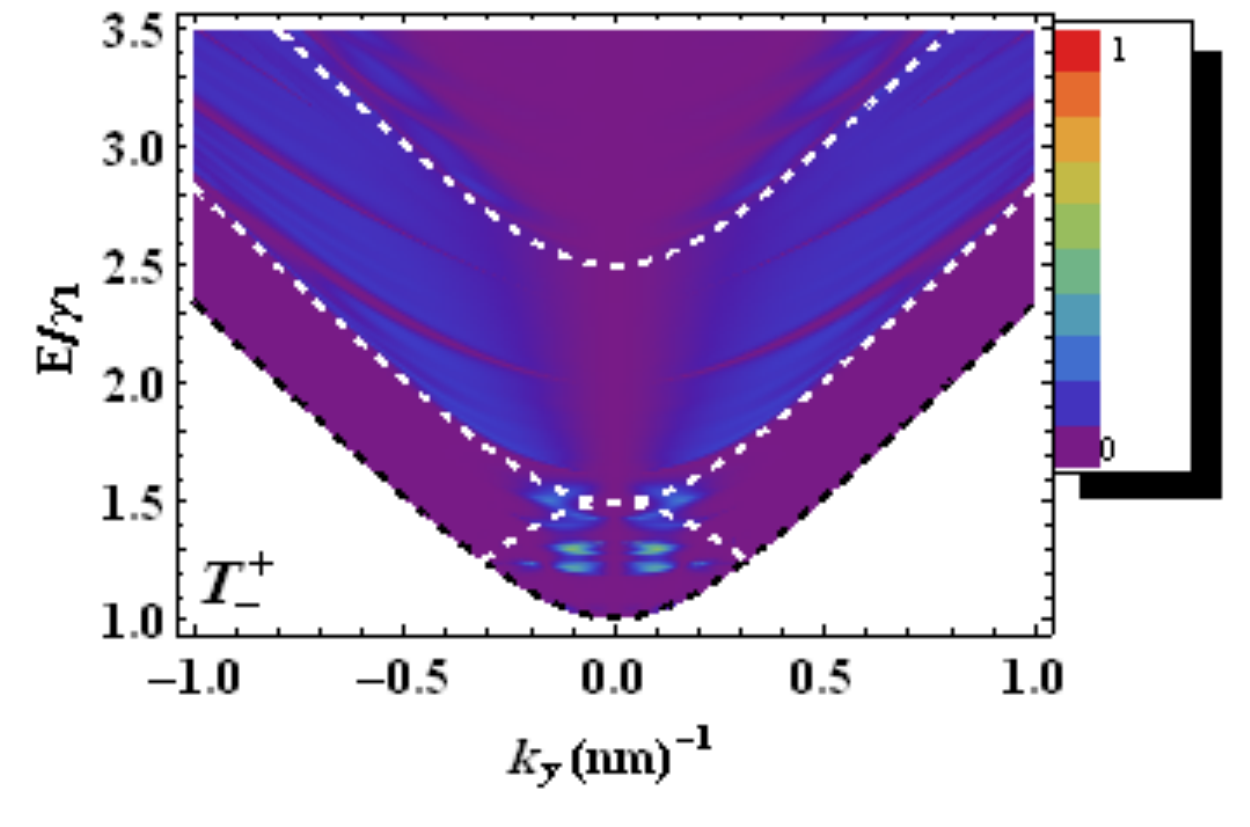}\\
\includegraphics[width=2 in,height=1.7 in]{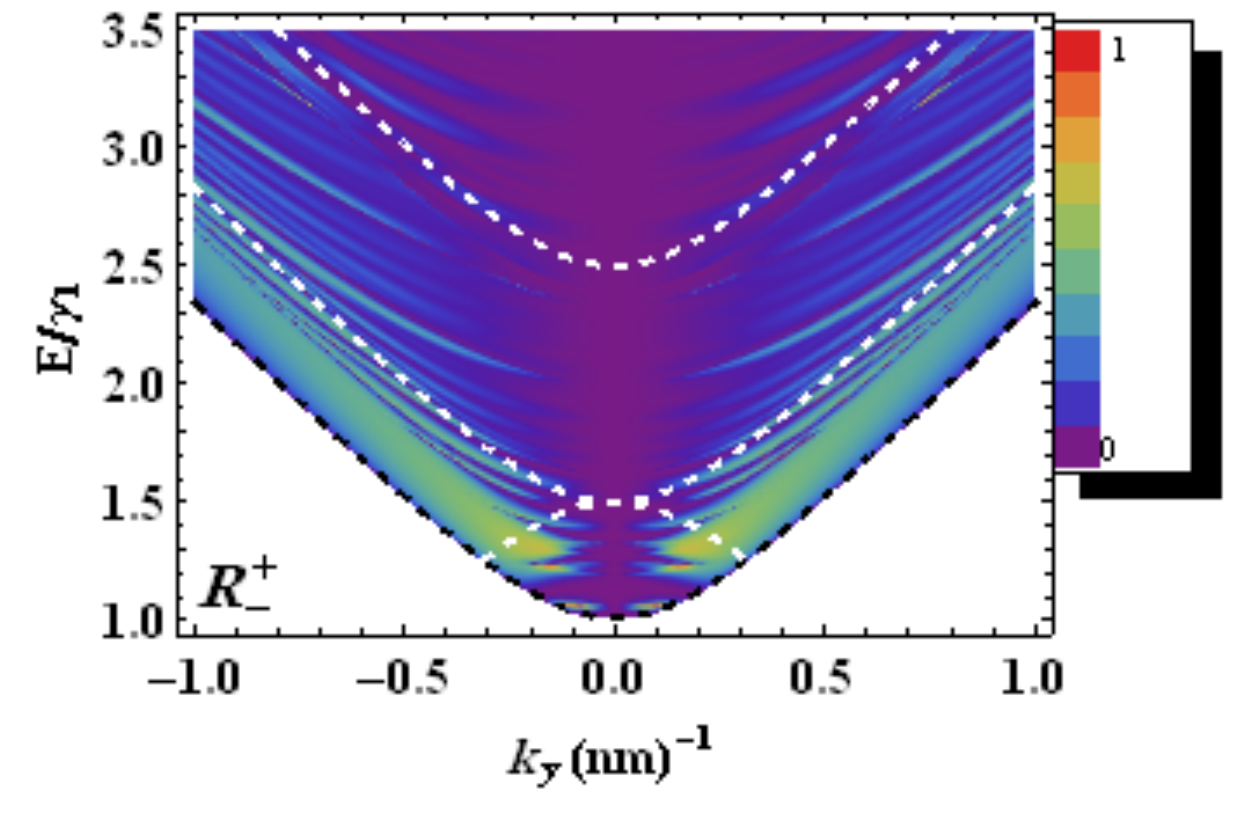}
\includegraphics[width=2 in,height=1.7 in]{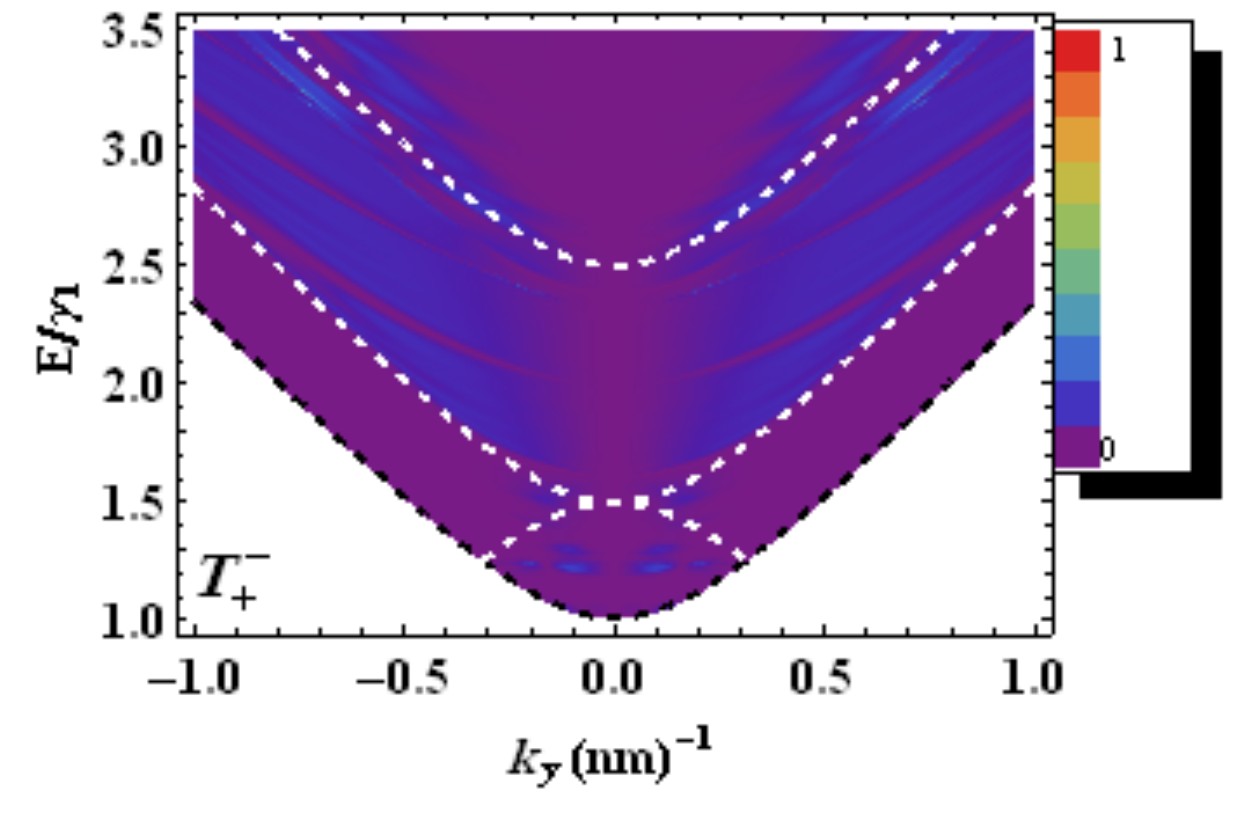}
\includegraphics[width=2 in,height=1.7 in]{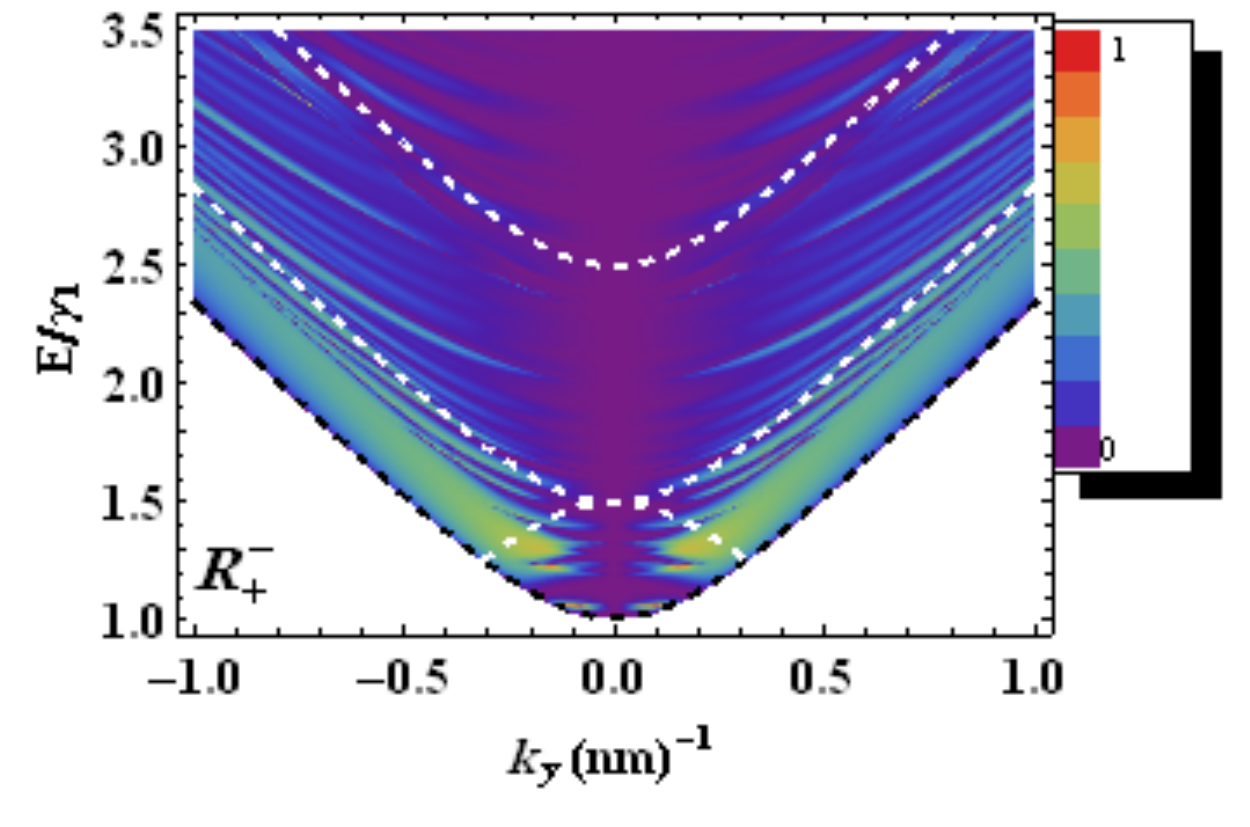}\\
\includegraphics[width=2 in,height=1.7 in]{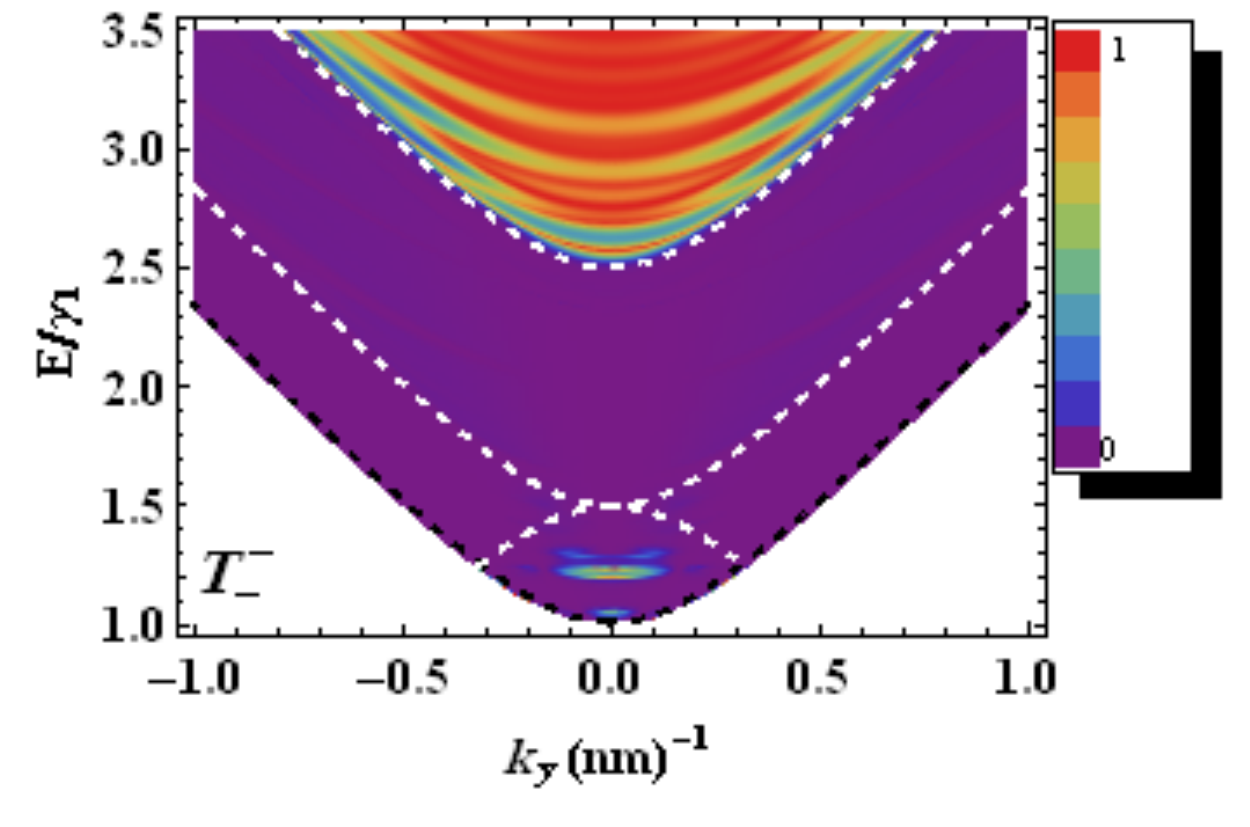}
\includegraphics[width=2 in,height=1.7 in]{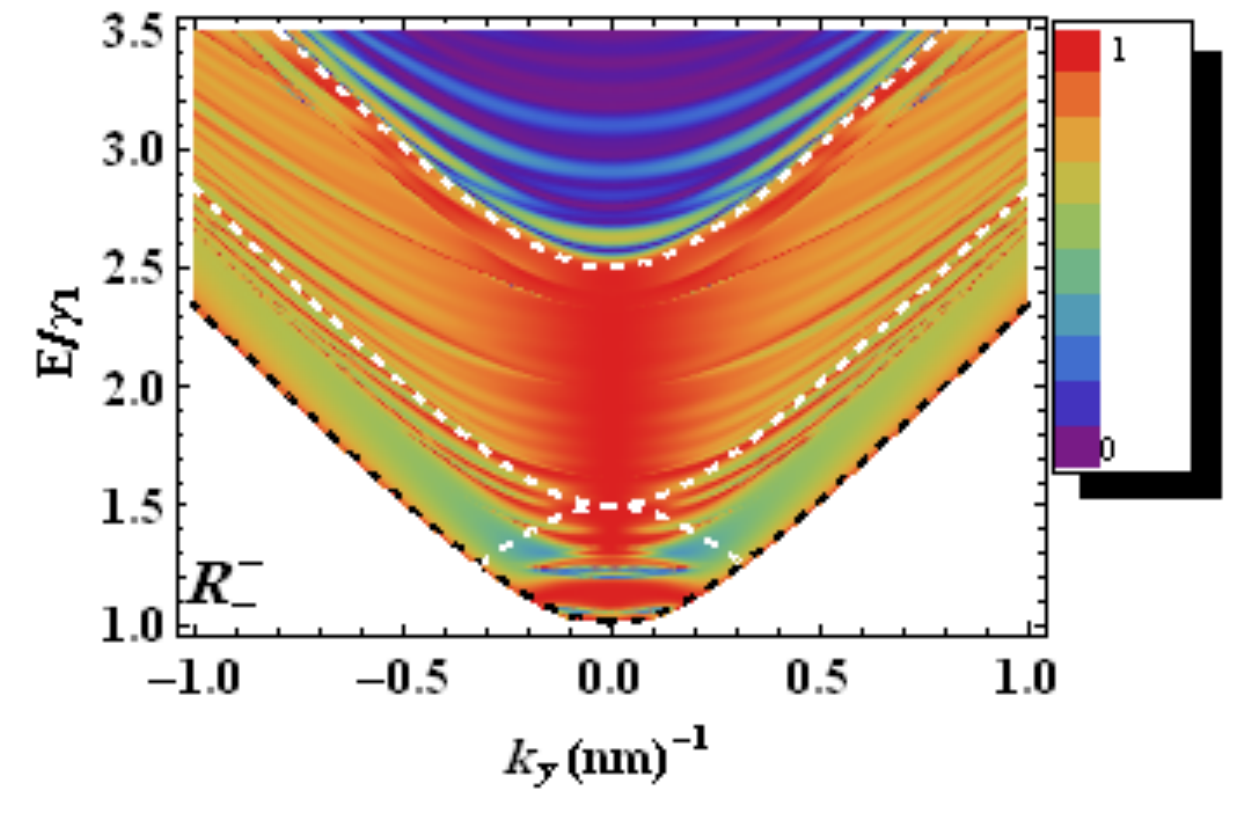}
\caption{Density plot of transmission and reflection probabilities
with $U_{2}=1.3\ \gamma_1$, $U_{4}=1.5\ \gamma_{1}$, $b_1=b_2=20\
nm$ and $\Delta=10\ nm$. The dashed white and black lines
represent the band inside and outside the second barrier,
respectively.}\label{fig0233}\nonumber
\end{figure}
For $T^{-}_-$ electrons propagate via $k^-$ mode which is absent
inside the barriers so that the transmission is suppressed in this
region and this is equivalent to the cloak effect \cite{30}. The
transmission probabilities $T^{+}_-$ and $T^{-}_+$ are the same
just when the time reversal symmetry holds (in this case when
$\delta_j=0,\ U_2=U_4$) which means that electrons moving in
opposite direction (moving from left to right and scattering from
$k^+\rightarrow k^-$ in the vicinity of the first valley or moving
from right to left and scattering from $k^- \rightarrow k^+$ in
the vicinity of the second valley) are the same because of the
valley equivalence \cite{30}.

Introducing asymmetric double
barrier structure with $U_2=1.3\ \gamma_1$, $U_4=1.5\ \gamma_1$
and without interlayer potential difference will break this
equivalence symmetry such that $T^{+}_- \neq T^{-}_+$ as depicted
in Figure \ref{fig0233}. In contrast, the reflection probabilities
$R^{+}_-$ and $R^{-}_+$ stay the same because the incident
electrons return again in an electron states \cite{30}. In
addition, the resonant peaks in $T^{+}_+$ are less intens
comparing to $T^{+}_+$ with $U_2=U_4$ in Figure \ref{fig5}.

\begin{figure}[h]
\centering
\includegraphics[width=2 in,height=1.7 in]{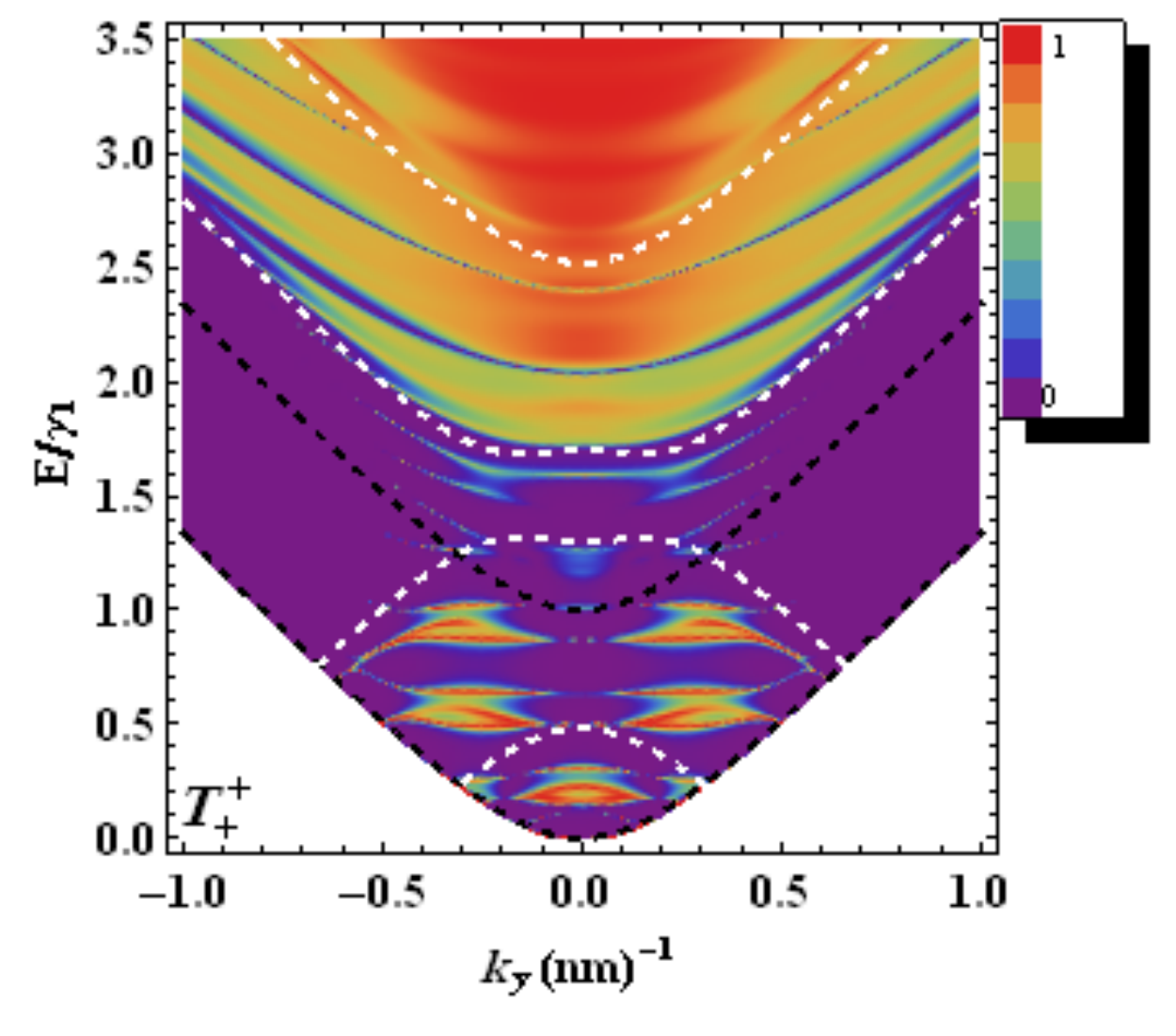}
\includegraphics[width=2 in,height=1.7 in]{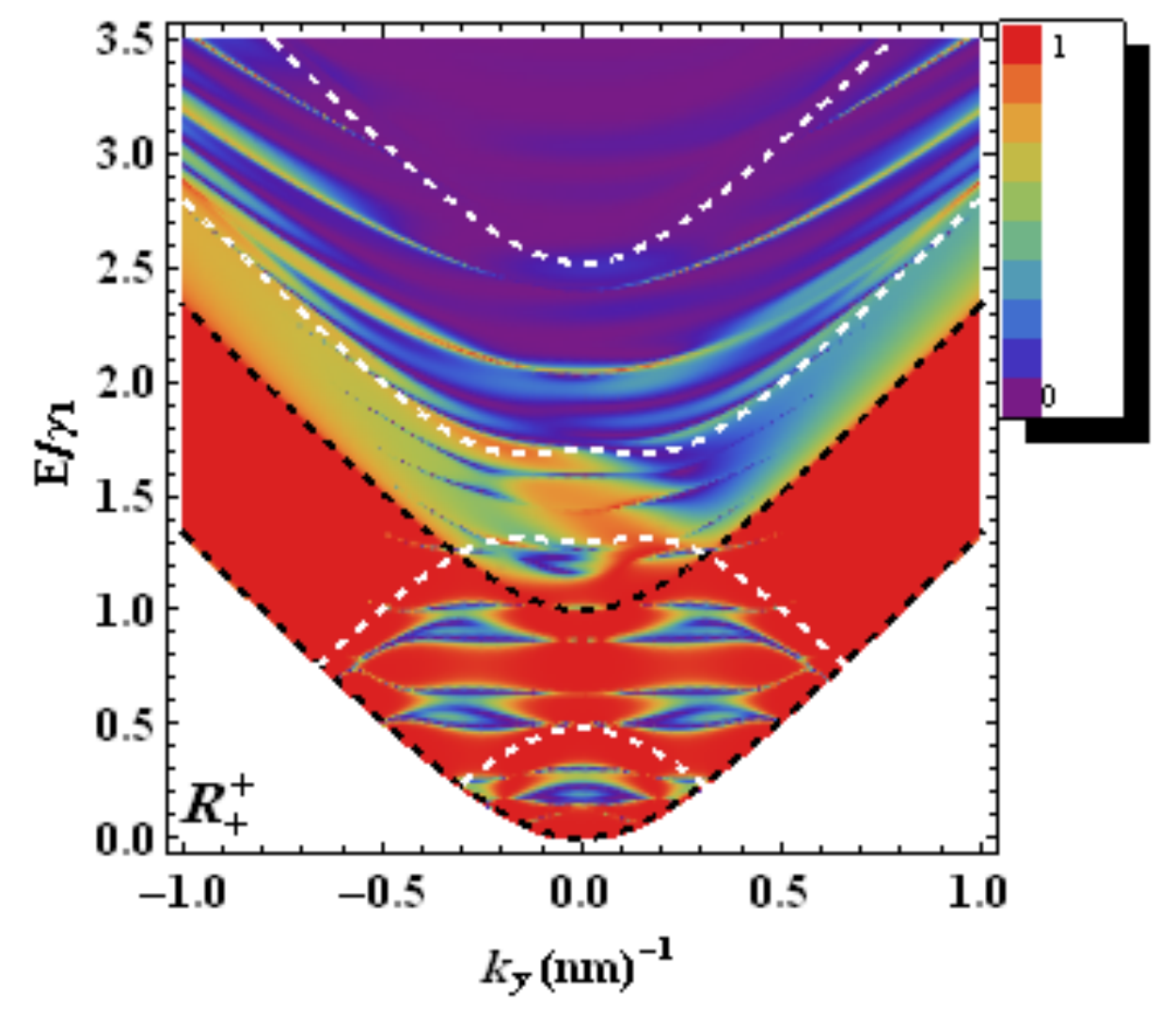}
\includegraphics[width=2 in,height=1.7 in]{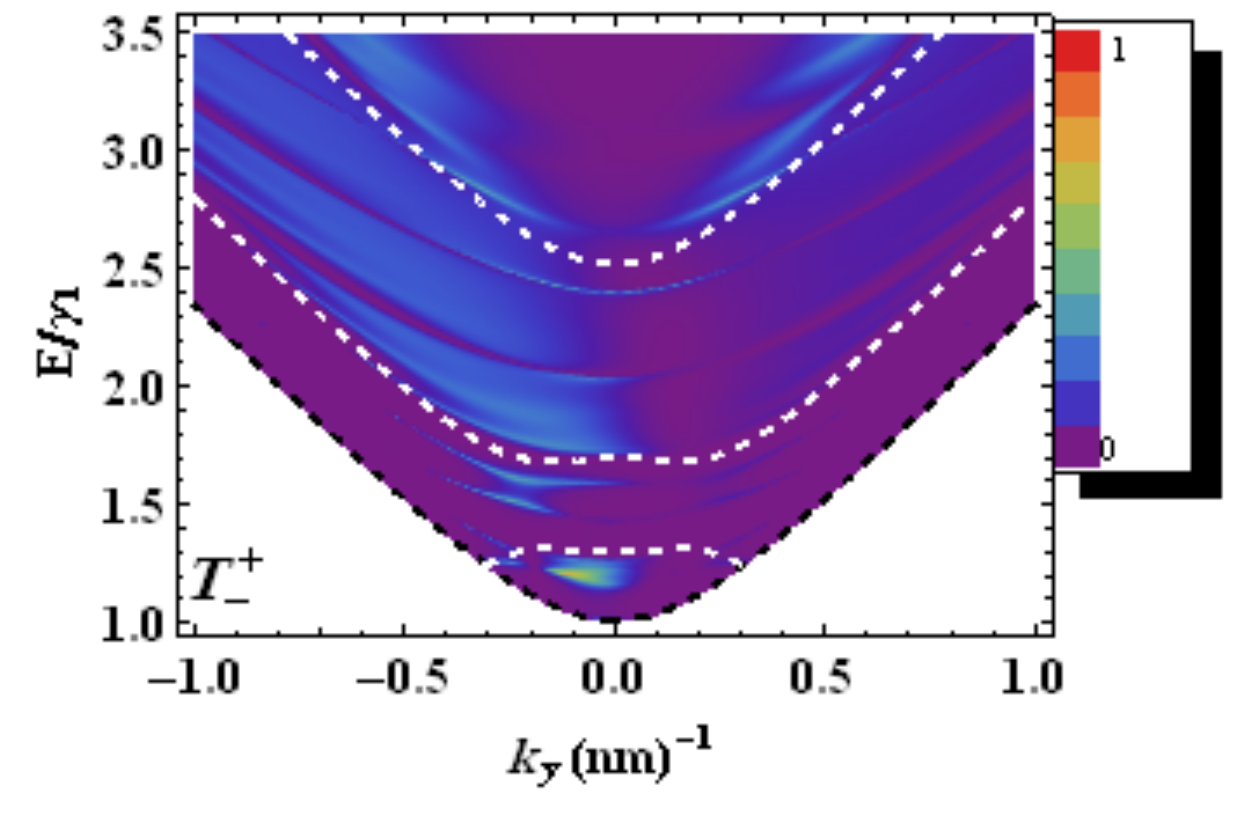}\\
\includegraphics[width=2 in,height=1.7 in]{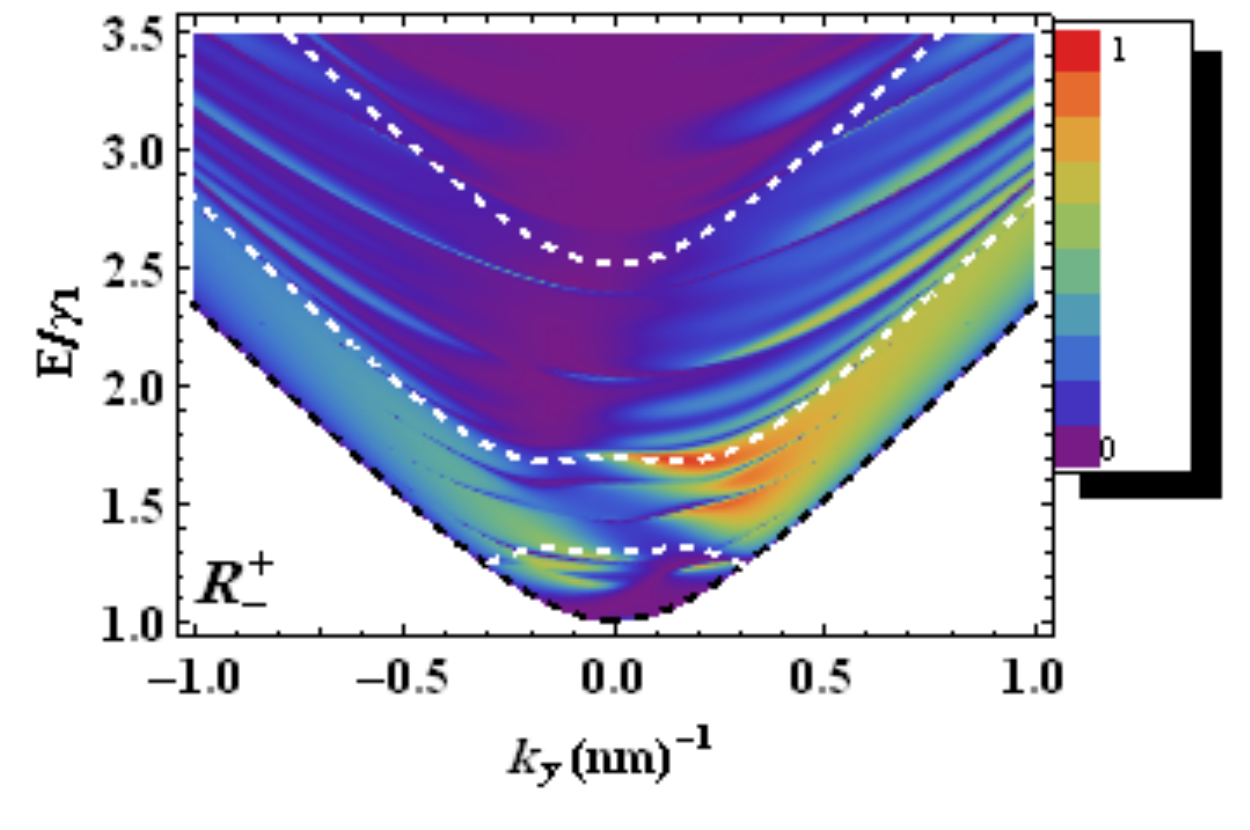}
\includegraphics[width=2 in,height=1.7 in]{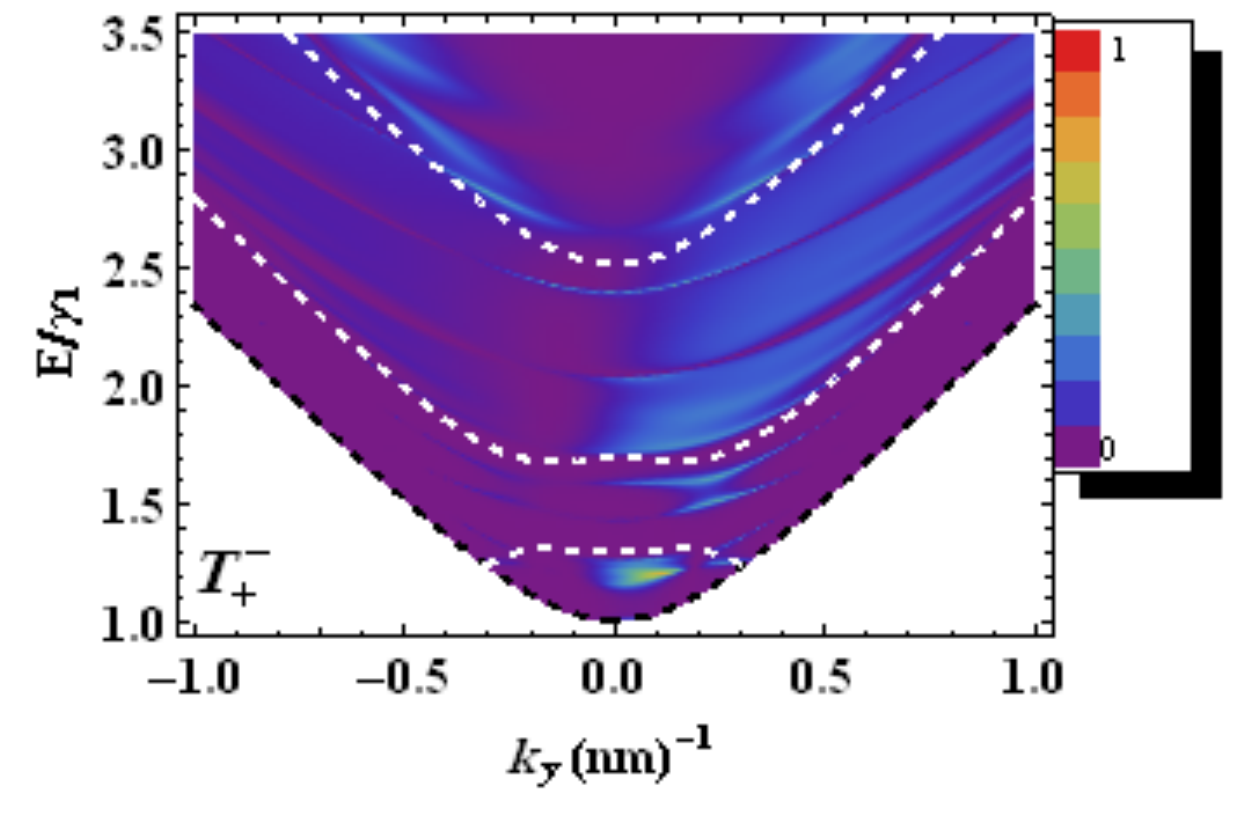}
\includegraphics[width=2 in,height=1.7 in]{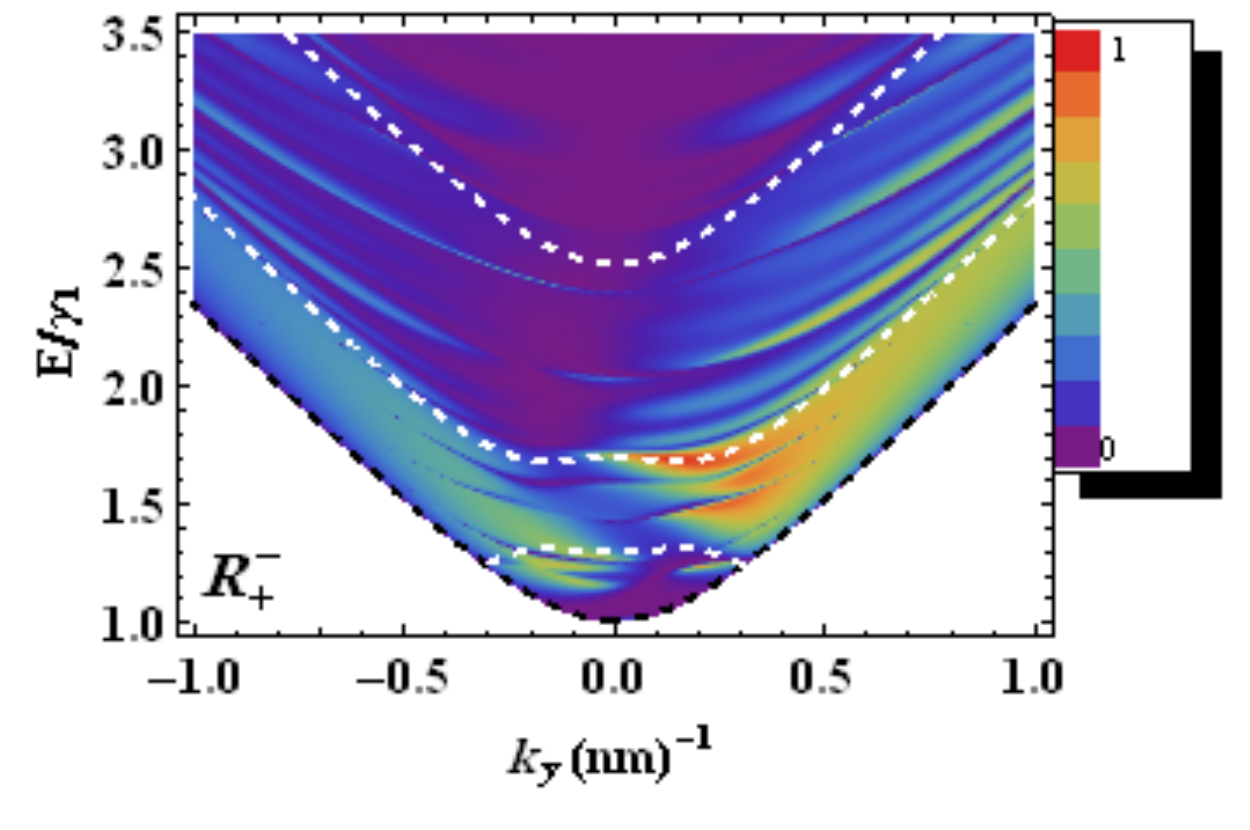}\\
\includegraphics[width=2 in,height=1.7 in]{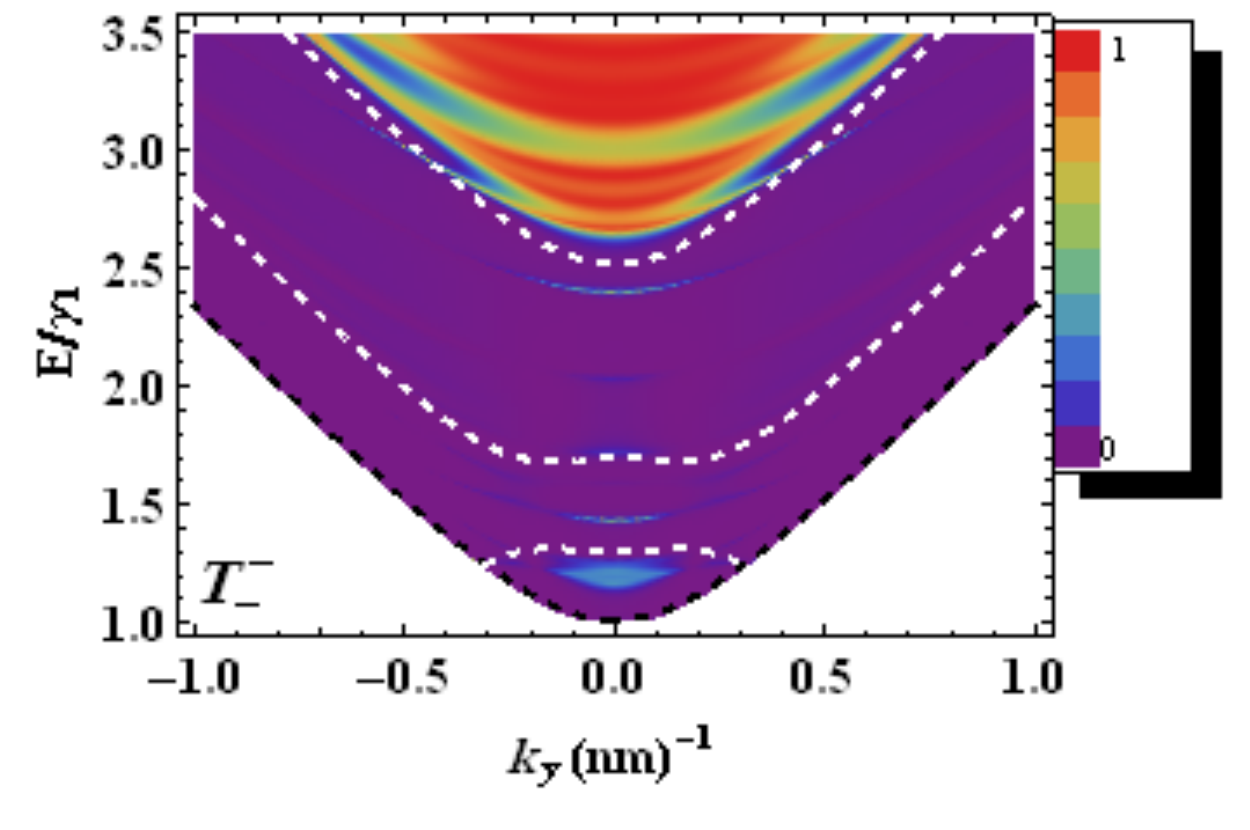}
\includegraphics[width=2 in,height=1.7 in]{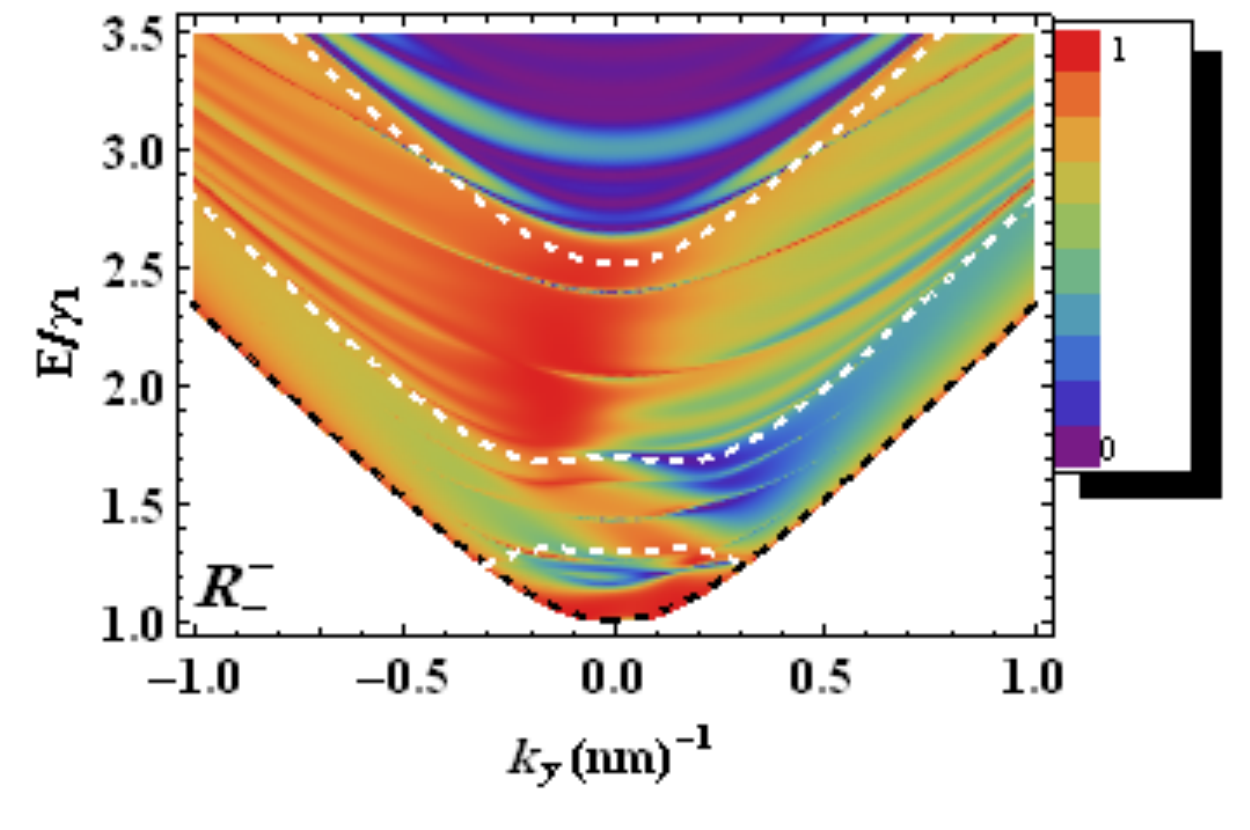}
\caption{Density plot of transmission and reflection probabilities
with $U_{2}=U_{4}=1.5\ \gamma_{1}$, $\delta_2=\delta_4=0.2\
\gamma_1$ and $b_1=b_2=\Delta=10\ nm$. The dashed white and black
lines represent the band inside and outside the second barrier,
respectively.}\label{fig015}
\end{figure}

Now let see how the interlayer potential difference will affect
the different channels of transmission and reflection. Figure
\ref{fig015} reveals the probabilities of the different
transmission and reflection channels  as a function of $k_y$ and
$E$ for $U_2=U_4=1.5\ \gamma_1$ and $\delta_2=\delta_4=0.2\
\gamma_1$. The general behavior of these different channels
resemble the single barrier case \cite{30} with some major
differences, such as observing extra resonances in the energy
region $0<E<U_j$ due to these bounded states in the well. In
addition, the induced gap does not completely suppressed the
transmission in the energy region $U_j\pm \delta_j$ as it the case
in the single barrier \cite{30} and this is also attributed to
these bounded states.

\begin{figure}[h]
\centering
\includegraphics[width=2 in,height=1.7 in]{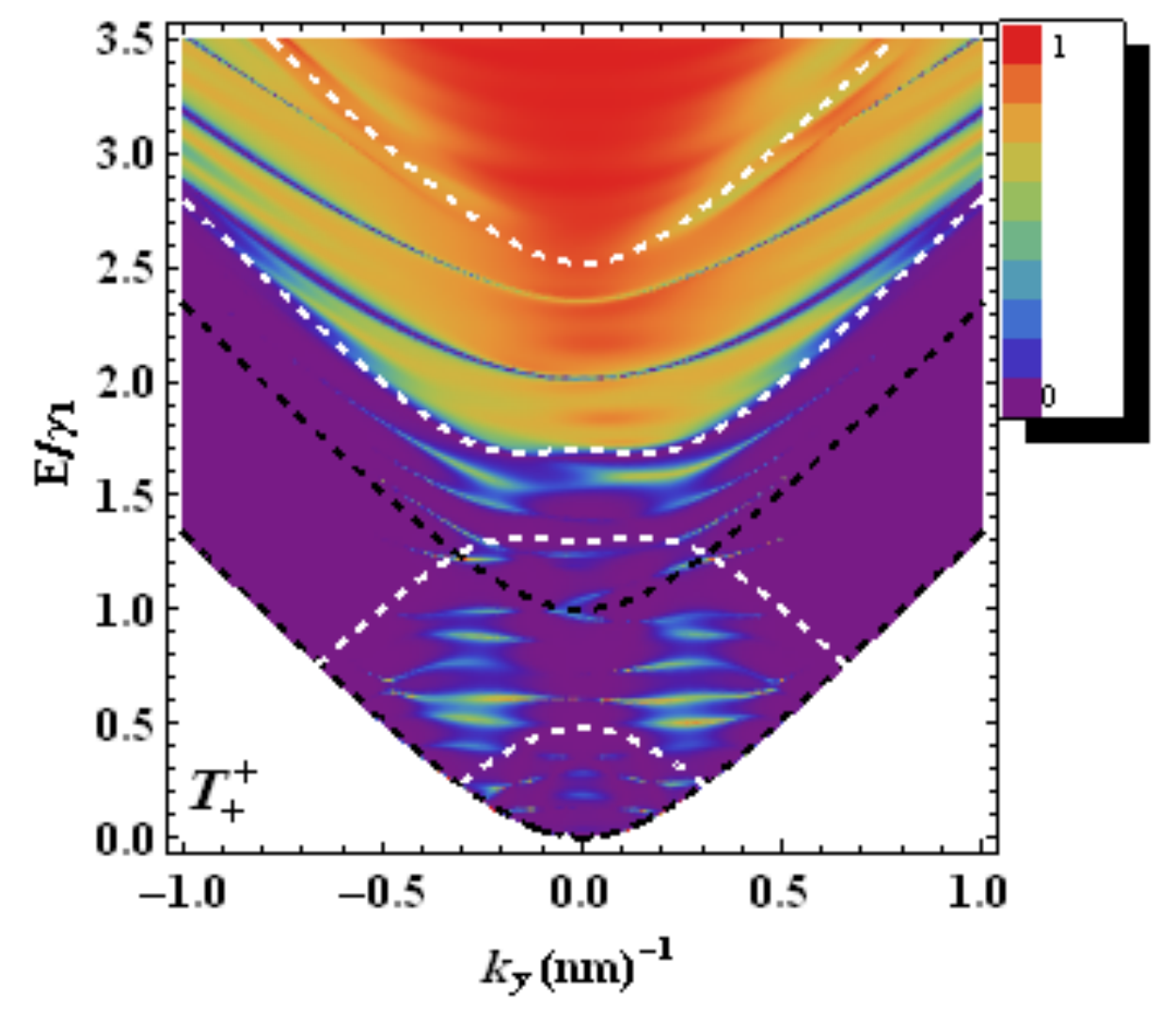}
\includegraphics[width=2 in,height=1.7 in]{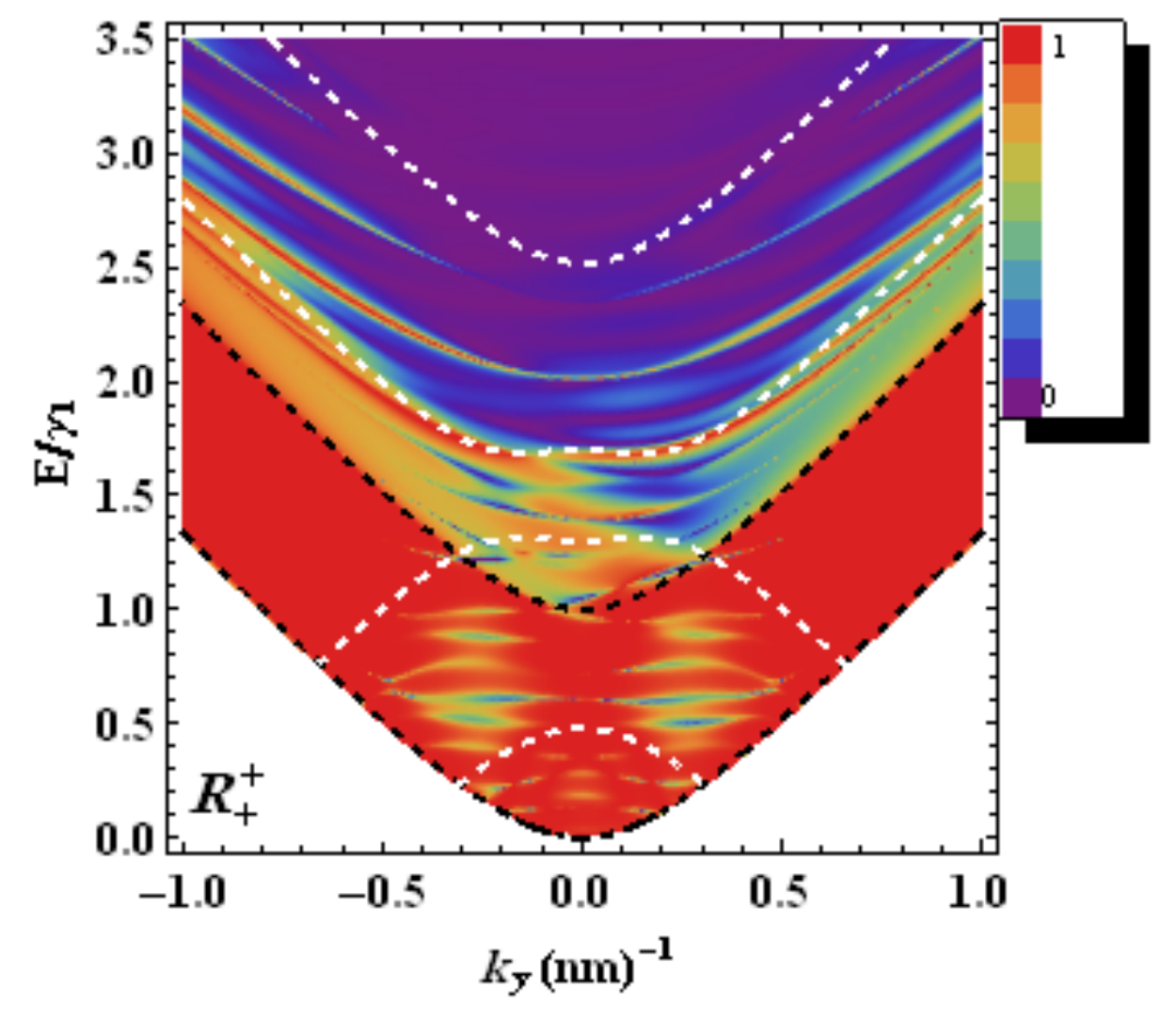}
\includegraphics[width=2 in,height=1.7 in]{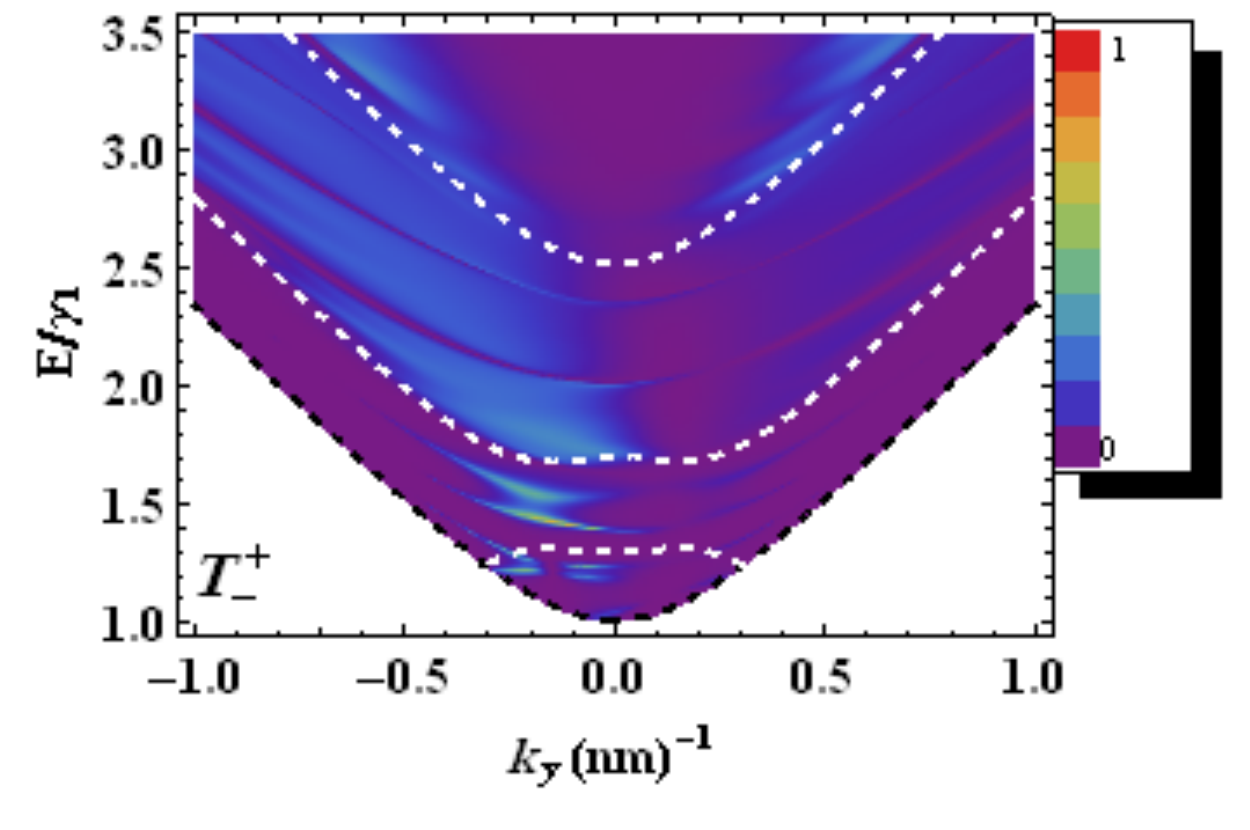}\\
\includegraphics[width=2 in,height=1.7 in]{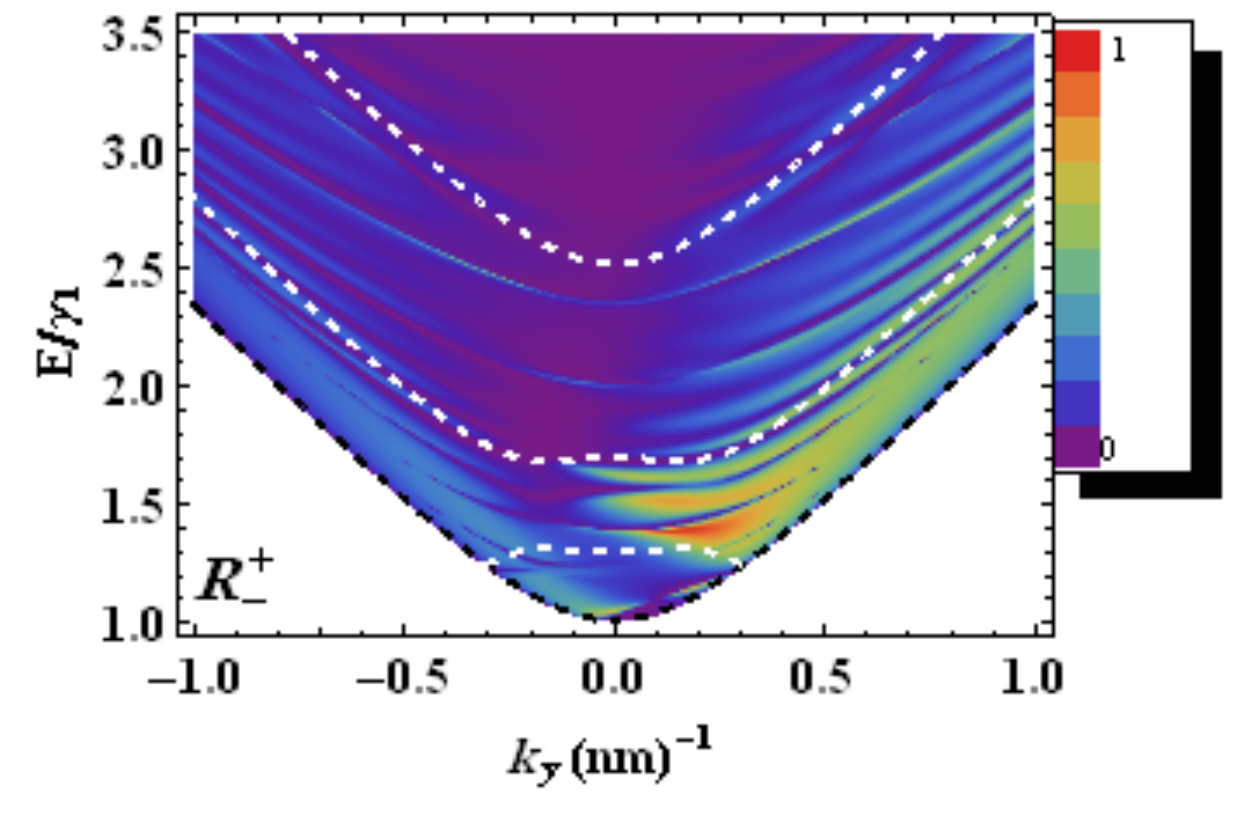}
\includegraphics[width=2 in,height=1.7 in]{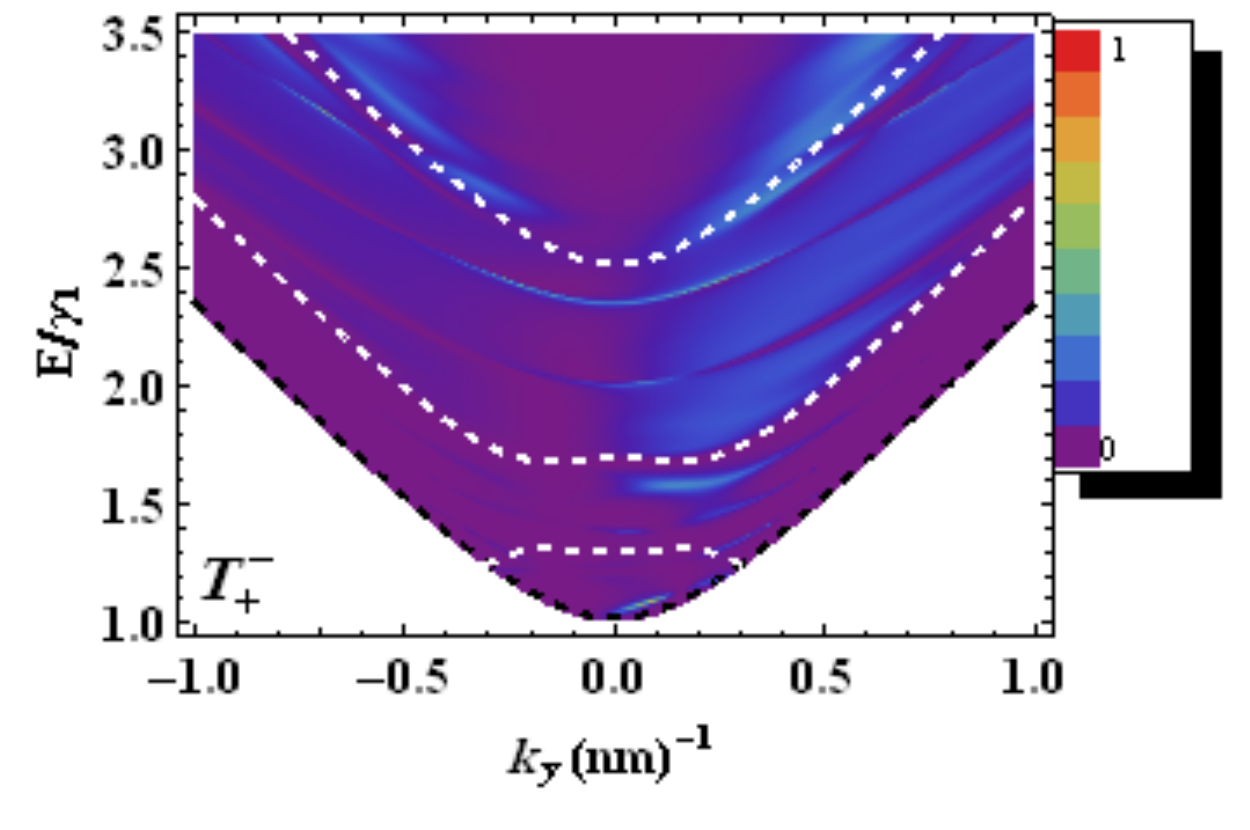}
\includegraphics[width=2 in,height=1.7 in]{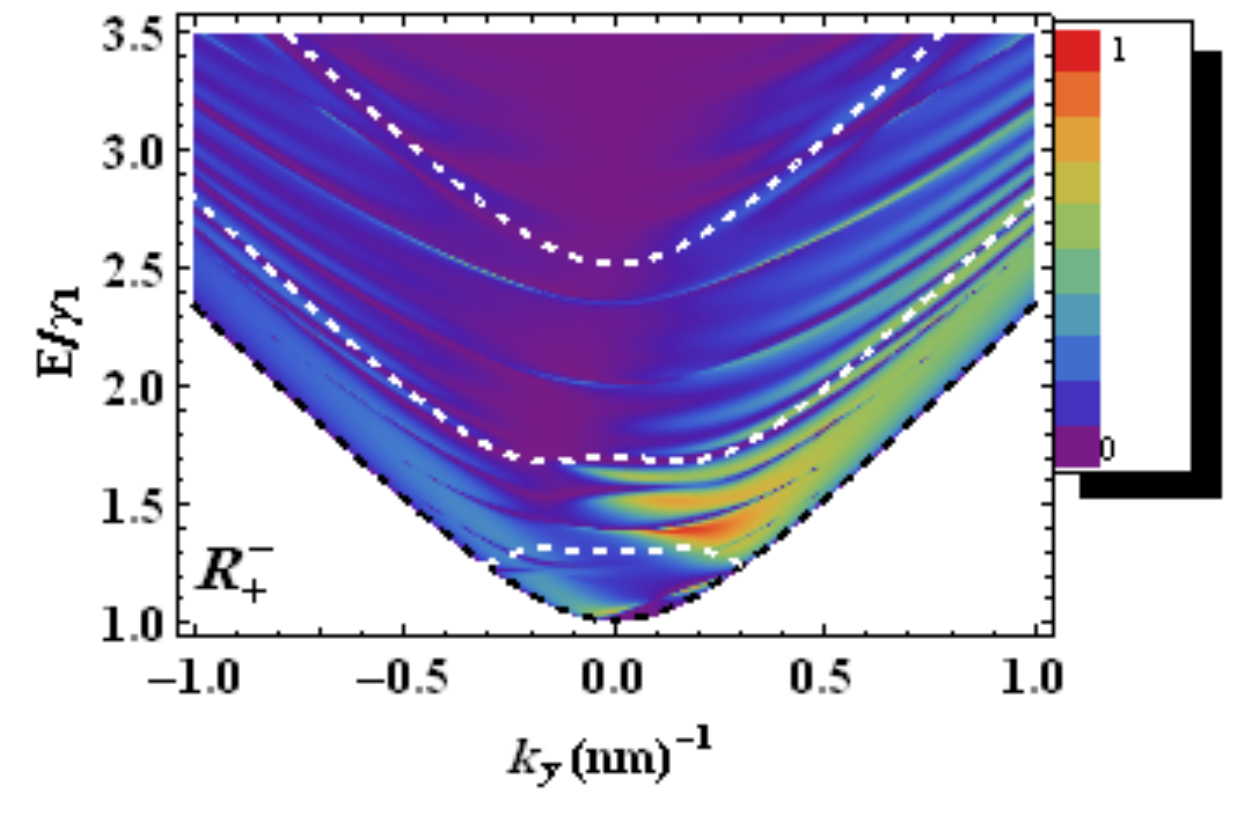}\\
\includegraphics[width=2 in,height=1.7 in]{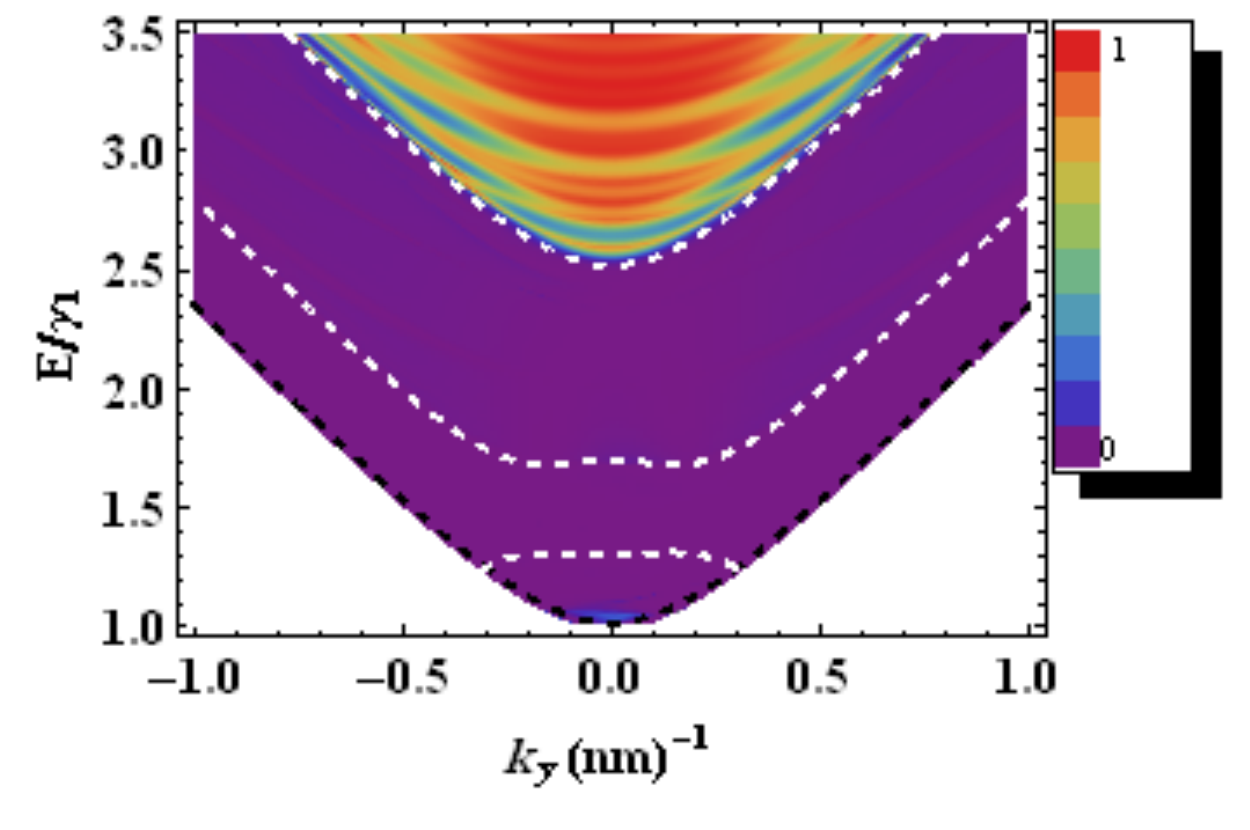}
\includegraphics[width=2 in,height=1.7 in]{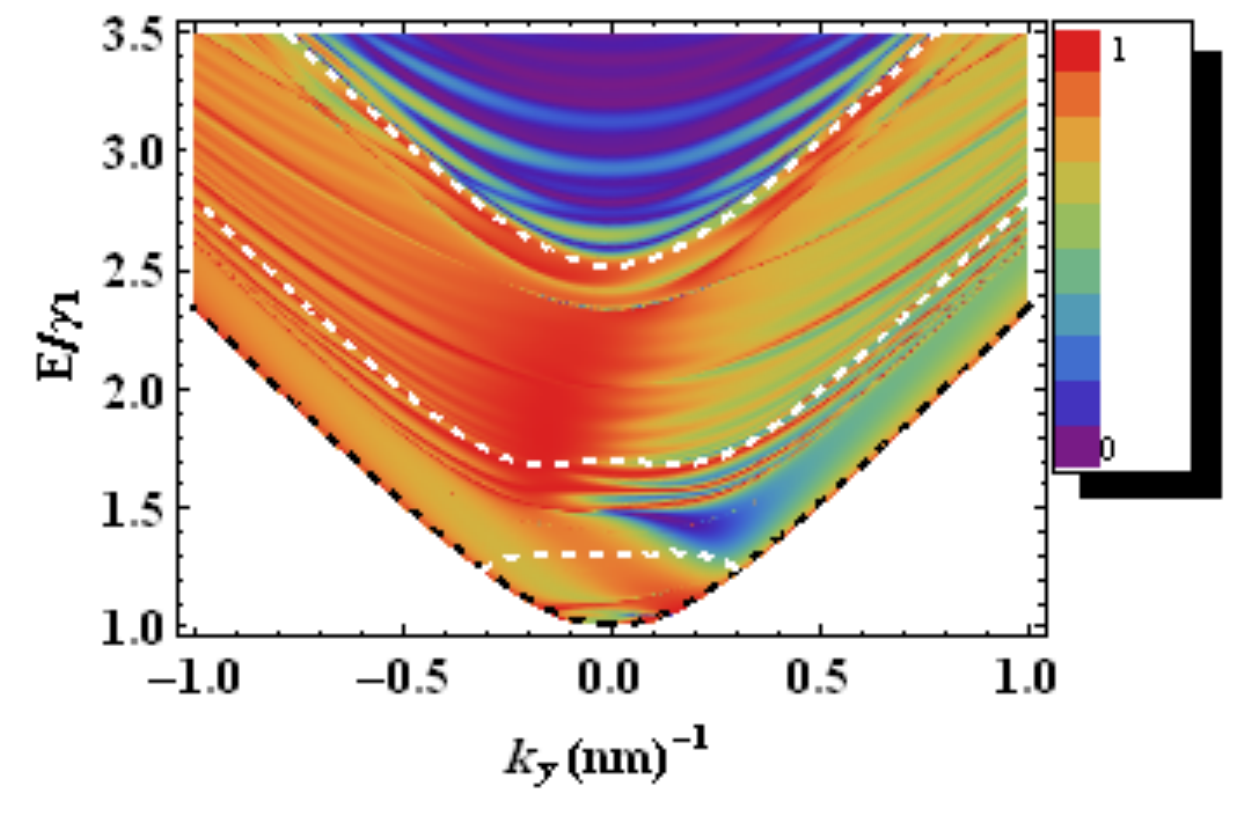}
\caption{Density plot of transmission and reflection probabilities
with $U_{2}=1.3\ \gamma_1$, $U_{4}=1.5\ \gamma_{1}$,
$\delta_2=\delta_4=0.2\ \gamma_1$ and $b_1=b_2=\Delta=10\ nm$. The
dashed white and black lines represent the band inside and outside
the second barrier, respectively.}\label{fig02331}
\end{figure}

With the interlayer potential difference and
different height of the barriers for $U_2=1.3\ \gamma_1$,
$U_4=1.5\ \gamma_1$ and $\delta_2=\delta_4=0.2\ \gamma_1$ we show
the different channels of transmission and reflection
probabilities in Figure \ref{fig02331}. In the same manner, the
effect of this different height of the barriers is reducing the
transmission probabilities. However, we note that it becomes more
intens inside the gap and this is because the available states
outside the first barrier which are in the same energy zone of the
gap on the second barrier.

%%%%%%%%%%%%%%%%%%%%%%%%%%%%%%%%%%%%%%%%%%%%%%%%%%%%%%%%%%%%%%%%%%%%%%%%%%%%%%%%%%%
\section{Conductance}
%%%%%%%%%%%%%%%%%%%%%%%%%%%%%%%%%%%%%%%%%

\begin{figure}[h]
\centering
\includegraphics[width=3 in]{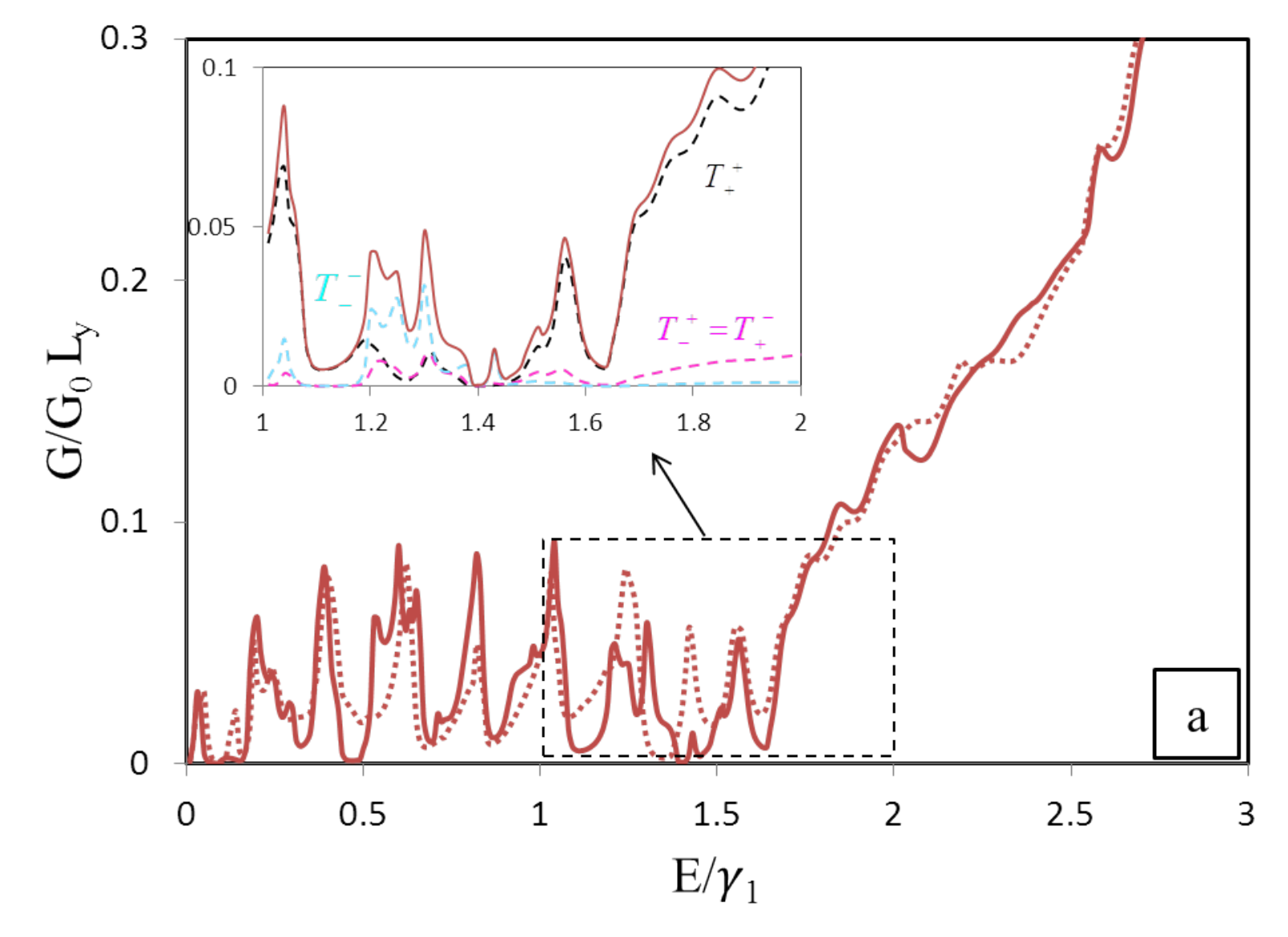}\ \ \
\includegraphics[width=3in]{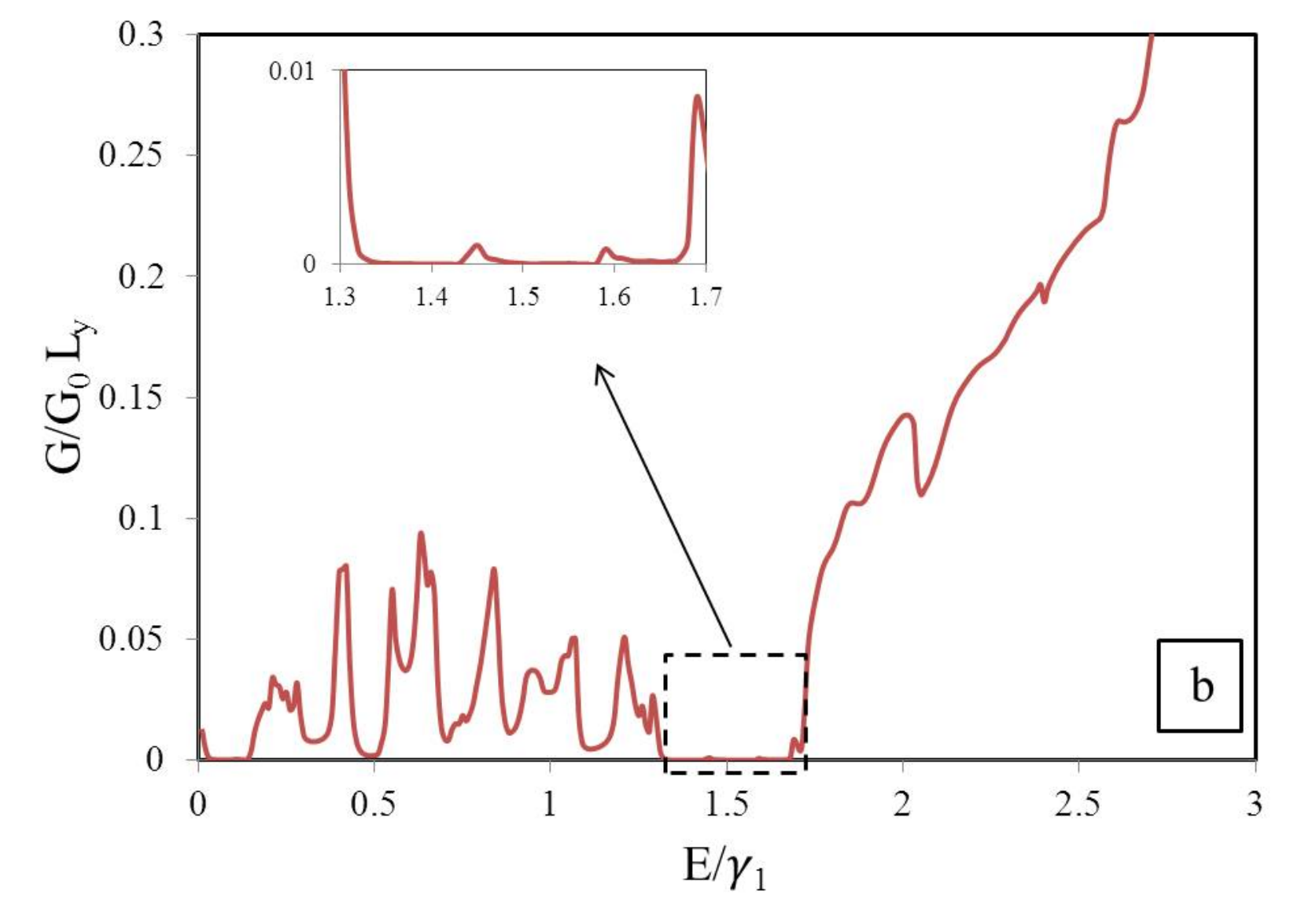}\\
\includegraphics[width=3 in]{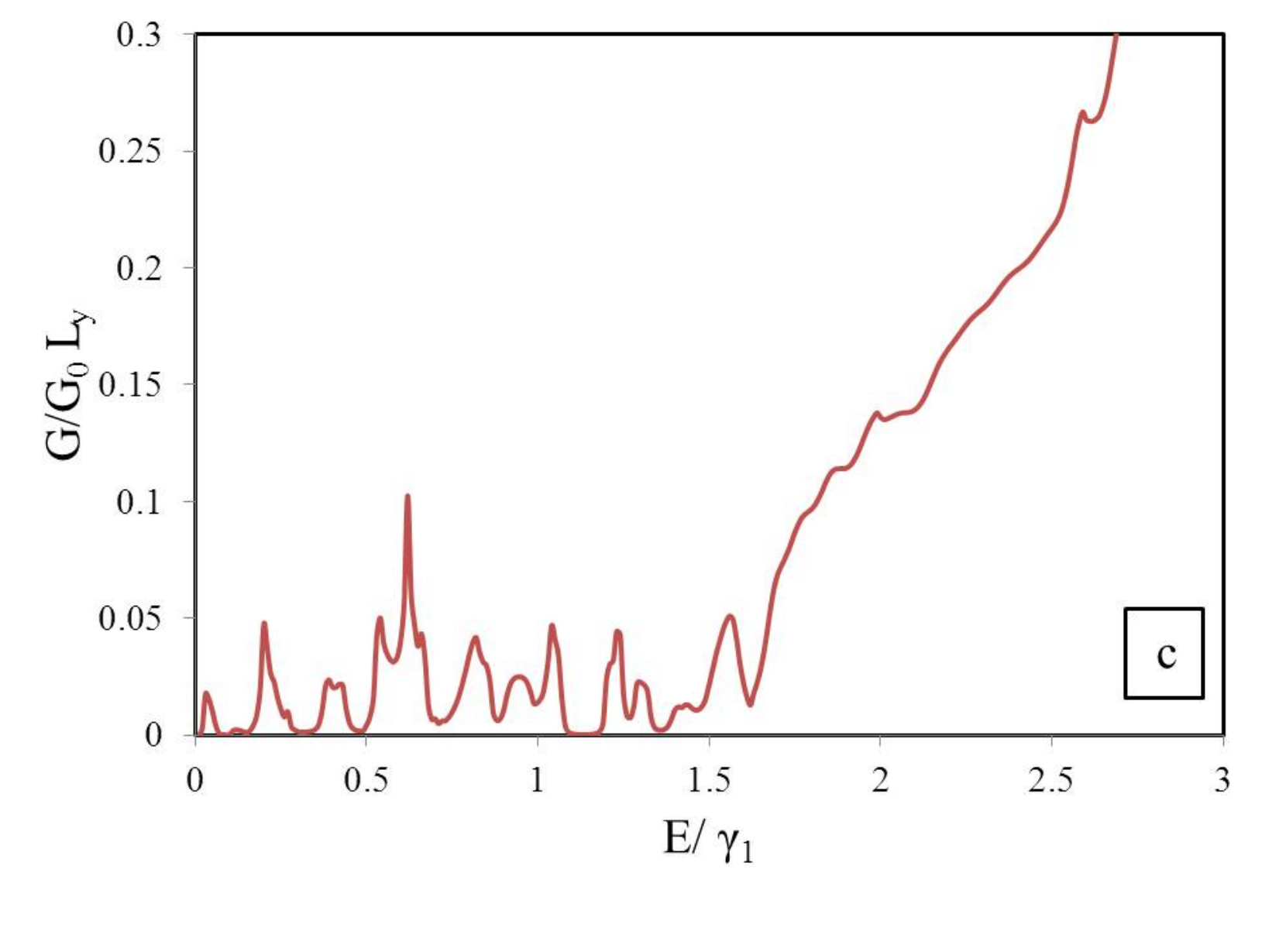}\ \ \
\includegraphics[width=3 in]{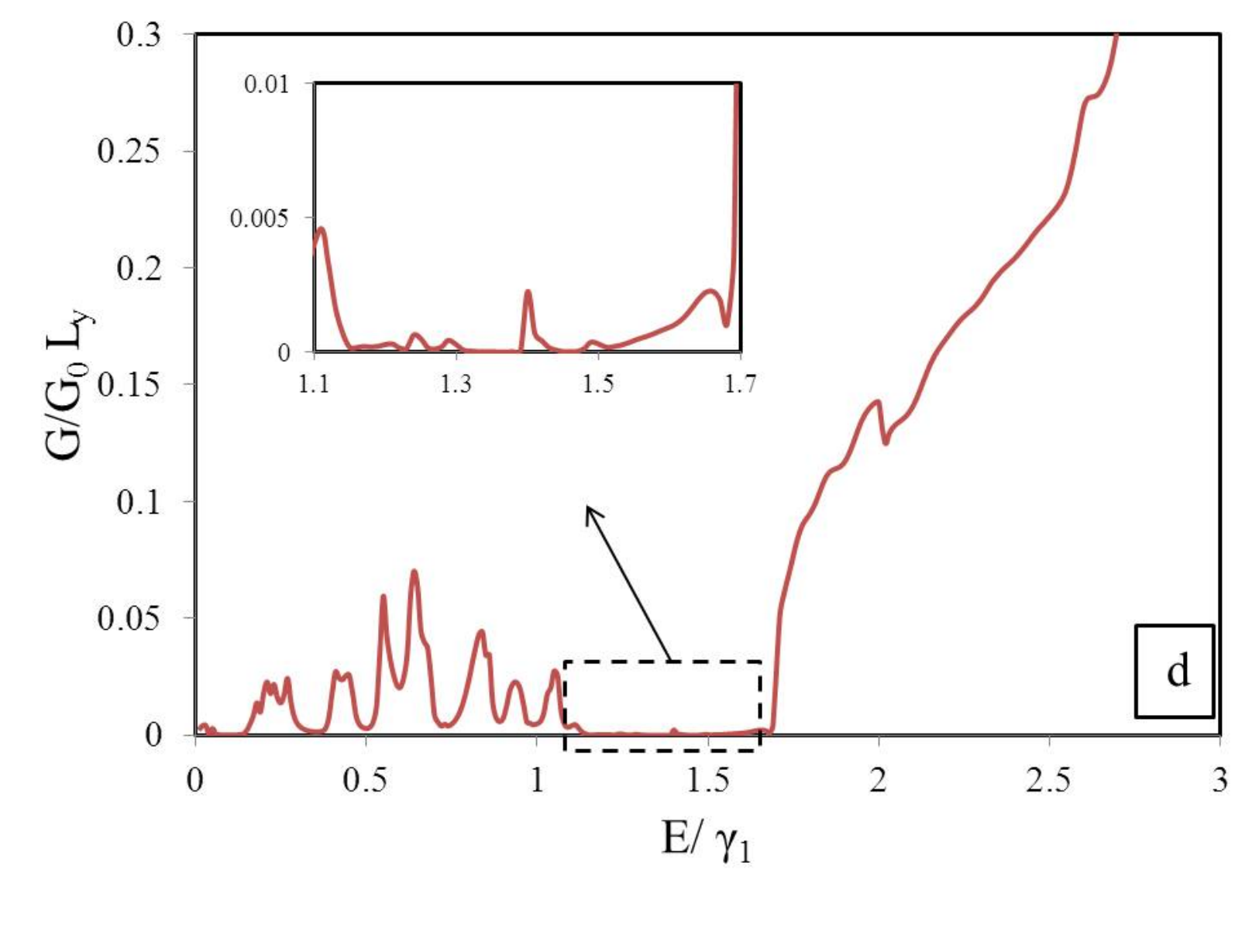}
\caption{Conductance of the double barrier structure as a function
of energy for $b_1=b_2=20\ nm$. (a)  $U_{2}=U_{4}=1.5\ \gamma_1$,
 $\Delta= 10\ nm$ (solid), $\Delta= 5\ nm$ (dotted) and
$\delta_2=\delta_4=0$. (b) $U_2=U_4=1.5\ \gamma_1$ and
$\delta_2=\delta_4=0.2\ \gamma_1$. (c) $U_{2}=1.3\ \gamma_1$ ,
$U_{4}=1.5\ \gamma_1$,  $\Delta= 10\ nm$ and $
\delta_2=\delta_4=0$. (d) $U_{2}=1.3\ \gamma_1$, $U_{4}=1.5\
\gamma_1$, $\Delta= 10\ nm$ and $\delta_2=\delta_4=0.2\ \gamma_1$.
}\label{fig6}
\end{figure}

In Figure \ref{fig6} we show the conductance as a function of the
energy $E$. Figure \ref{fig6}a shows the conductance of the double
barrier structure for $U_2=U_4=1.5\ \gamma_1$,
$\delta_2=\delta_4=0$ for $\Delta=5\ nm$ (dotted curve) and
$\Delta=10\ nm$ (solid curve). The peaks in the conductance of the
double barrier have extra shoulders as a results of the resonances
in the transmission probabilities due to the existence of the
bound electron states in the well. These resonances show up as
convex curves, which were absent for the single barrier, in
$T^{+}_{+}$ in the region $0<E<U_{2}=U_{4}$ and in $T^{+}_{-},
T^{-}_{+}$ and $T^{-}_{-}$ in the region $\gamma_1<E<U_{2}=U_{4}$
as depicted in figure \ref{fig5}.
For energies larger than $U_{2}+\gamma_1$ the channel $T^{-}_{-}$
is not suppressed (cloaked) anymore so that we notice these very
pronounced peaks in the conductance in this regime. The inset of
Figure \ref{fig6}a show the contribution of each channel to the
conductance for $\Delta=10\ nm$ in the region $
\gamma_1<E<2\gamma_1$. For energies between the interlayer
coupling and the barriers's height all channel contribute to the
conductance, but for energies larger than the barrier's height the
contribution of $T^{-}_{-}$ is zero due to the cloak effect which
is clarified in the inset of Figure \ref{fig6}a. In Figure
\ref{fig6}b we show the conductance of the double barrier with the
interlayer potential difference $\delta_2=\delta_4=0.2\ \gamma_1$
and for the same height of the two barriers $U_2=U_4=1.5\
\gamma_1$. As a result of the none zero transmission inside the
gap (see Figure \ref{fig015}) we also have none zero conductance
inside the gap as clarified in the inset of Figure \ref{fig6}b. In
Figure \ref{fig6}c we represent the result in Figure \ref{fig6}a
but with asymmetric double barrier structure such that $U_2=1.3\
\gamma_1$ and $U_4=1.5\ \gamma_1$ for $\delta_2=\delta_4=0$, we
see that the asymmetric structure of the double barrier reduces
the conductance and even removing some shoulders of the peaks. The
effect of the asymmetric double barrier structure together with
the interlayer potential difference is presented in Figure
\ref{fig6}d for $U_2=1.3\ \gamma_1$, $U_4=1.5\ \gamma_1$,
$\delta_2=\delta_4=0.2\ \gamma_1$. Similarly to the previous case,
the conductance here also decreases and some of the main peaks are
removed as a consequence of this asymmetric structure of the
double barriers and the induced gap in the spectrum. Although the
interlayer potential difference is the same on the both barriers,
the gap in the conductance is not anymore $2\ \delta_2=2\
\delta_4=0.4\ \gamma_1$ as the case in Figure \ref{fig6}c instead
it becomes $3\ \delta_2=3\ \delta_4=0.6\ \gamma_1$ as depicted in
the inset of Figure \ref{fig6}. Moreover, although at
$E=U_{2}=U_{4}=V$ there are no available states, the conductance
is not zero (in the single and double barrier) and this is due to
the presence of resonant evanescent modes which are responsible
for the pseudo-diffusive transport at the Dirac point \cite{26}.

\begin{figure}
\centering
\includegraphics[width=2.6 in]{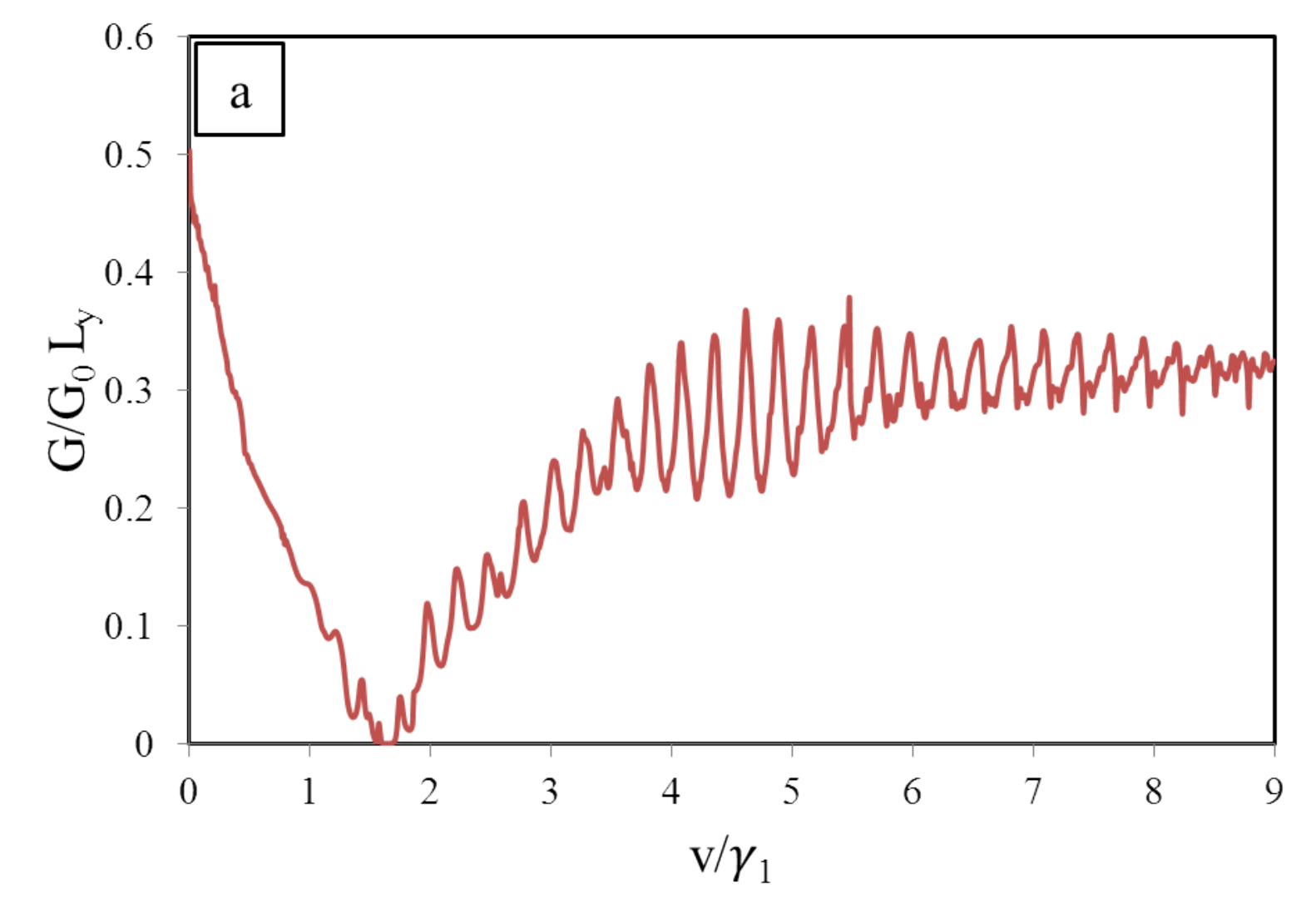}\ \ \
\includegraphics[width=2.7 in]{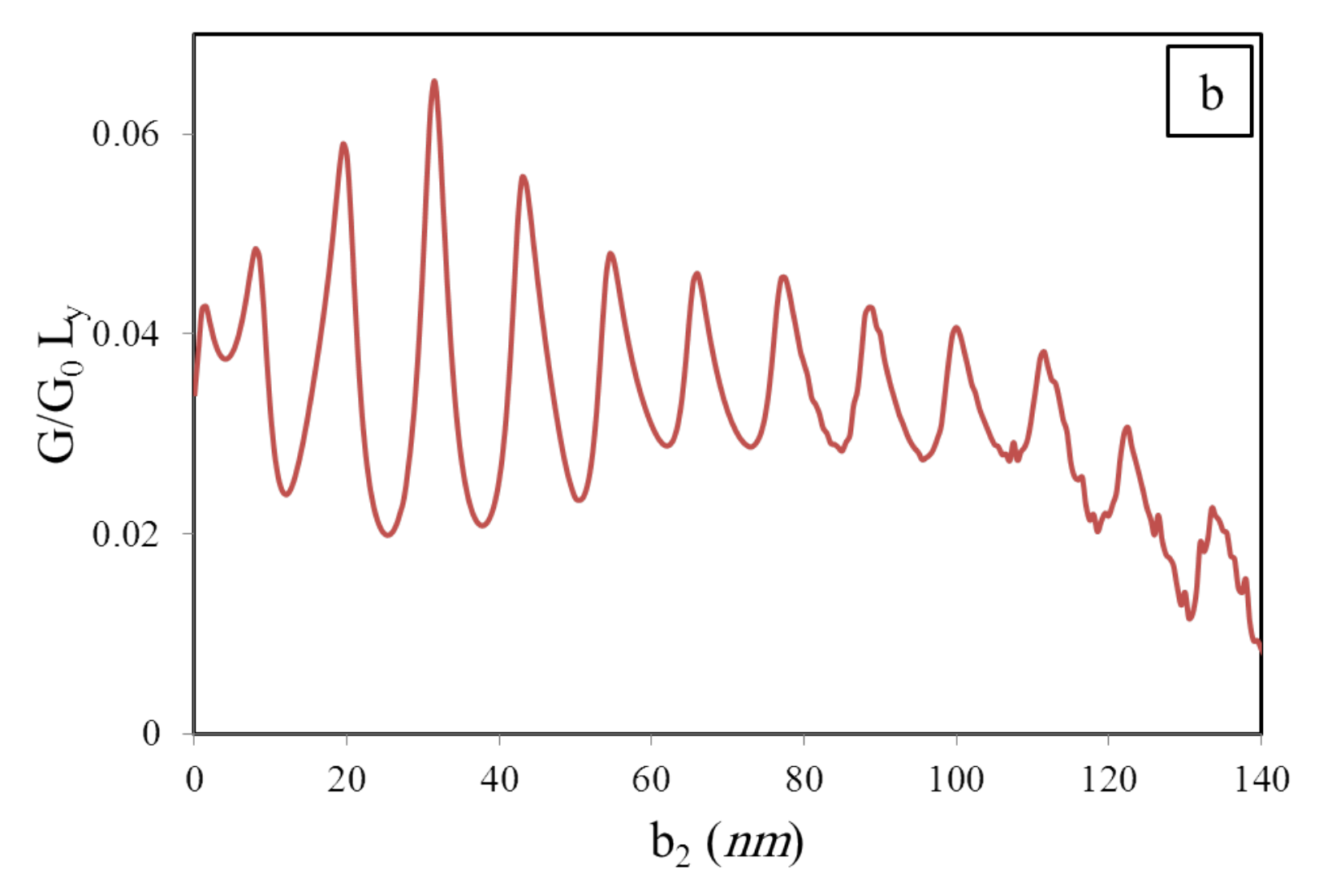}\\
\includegraphics[width=2.7 in]{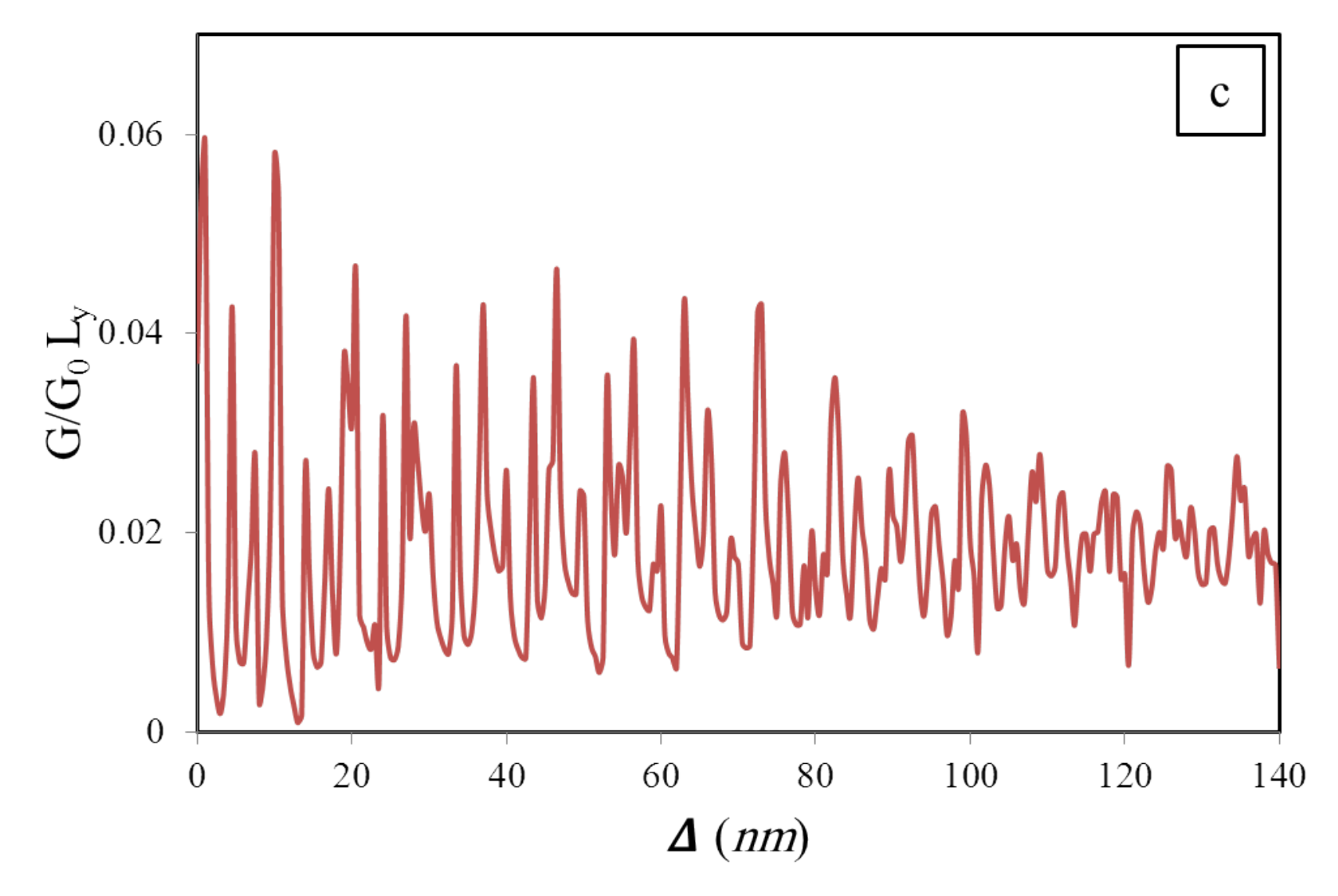}
\caption{The conductance of the double barrier as a function of:
(a) barriers's height V ($U_{2}=U_{4}=$V) for $E=1.5\ \gamma_1$,
$b_1=b_2=20\ nm$  and $\Delta=10\ nm$. (b) $b_2$ for V$= 1.5\
\gamma_1$, $E=1.3\ \gamma_1$, $b_1=20\ nm$ and $\Delta=10\ nm$.
(c) $\Delta$ for V$= 1.5\ \gamma_1$, $E=1.3\ \gamma_1$ and
$b_1=b_2=20\ nm$.}\label{fig7}
\end{figure}

The conductance dependance on the double barriers parameters is
shown in Figure \ref{fig7}. For $E=1.5\ \gamma_1$ and
$\delta_2=\delta_4=0$  we show the conductance as a function of
the height of the barriers V ($U_2=U_4=V$) in Figure \ref{fig7}a.
In the region $E>$V$>0$ the conductance decreases with increasing
$V$, whereas in the region $V>E$ it increases with increasing $V$
till it reaches a Plato constant value which is an odd behavior.
This behavior is attributed to the resonance in the region $E<V$
since the conductance is minimum at the Dirac point (in this case
$E=V$) leading to an increase of the conductance on the both sides
of the Dirac point ( $E>V$ and $E<V$) \cite{26}. In contrast,
increasing $b_2$ for fixed other parameters decreases the
conductance as depicted in Figure \ref{fig7}b and the number of
resonances appearing in the conductance remains the same with
increasing $b_2$. Finally, in Figure \ref{fig7}c we plot the
conductance versus $\Delta$. The conductance is seen to oscillate
with increasing width of the well and then reaches to a constant
asymptotic value.

\begin{figure}
\centering
\includegraphics[width=2.5 in]{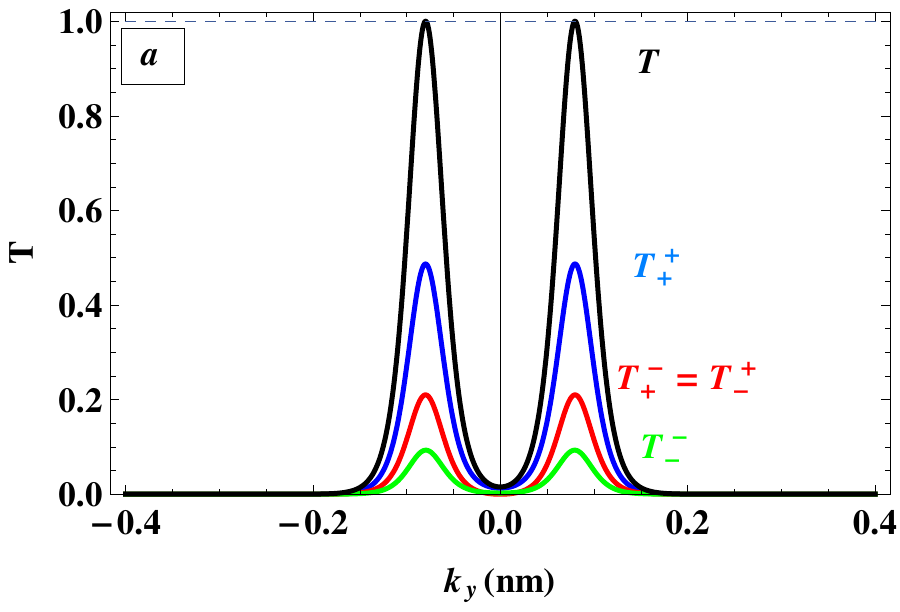}\ \ \
\includegraphics[width=2.5 in]{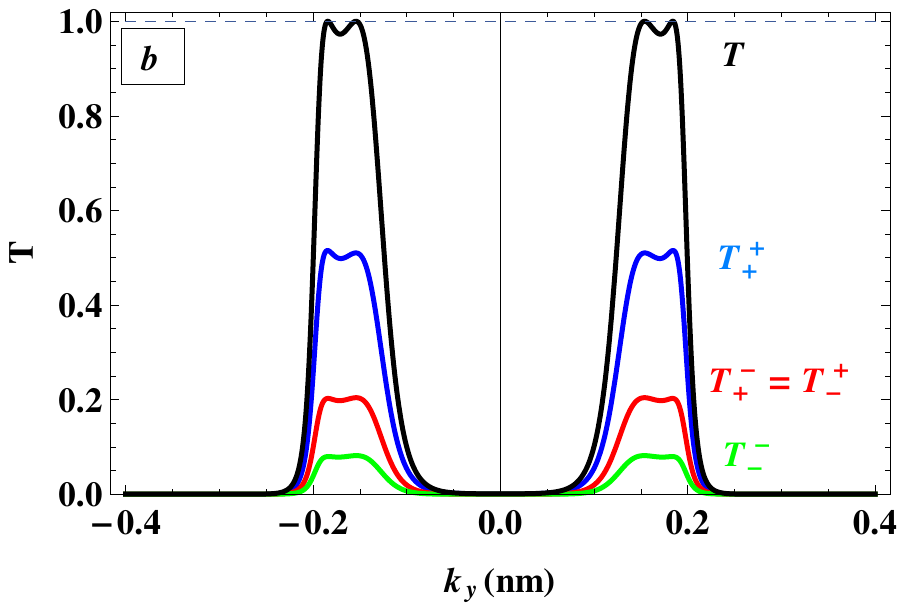}\\
\includegraphics[width=2.5 in]{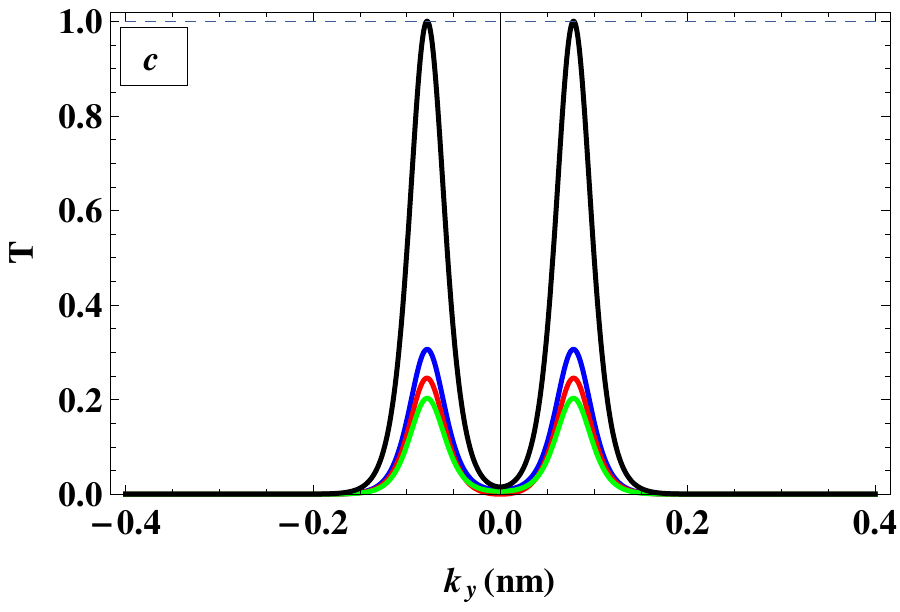}\ \ \
\includegraphics[width=2.5 in]{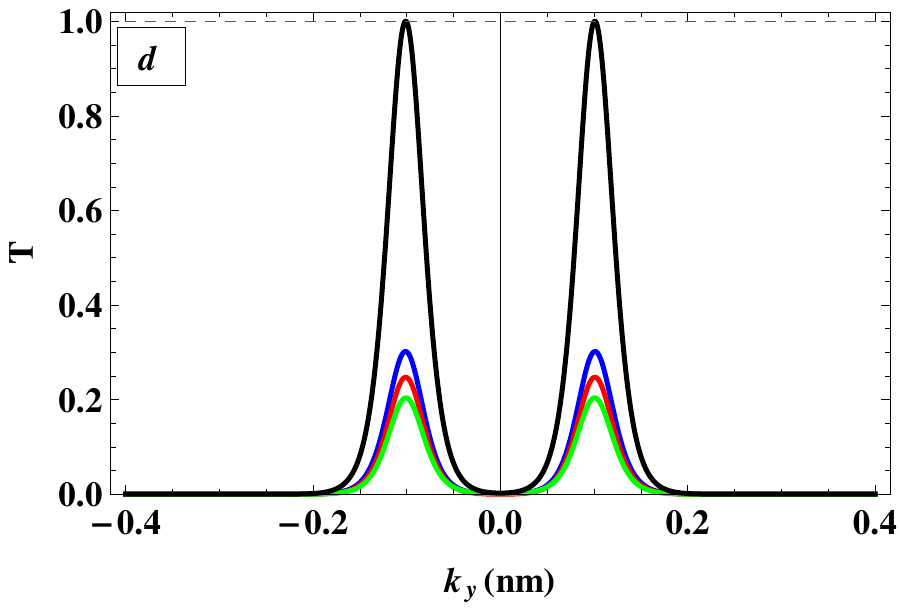}
\caption{The transmissions coefficients around the Dirac point for
$(E=V=1.5\ \gamma_1)$ and $b_1=b_2=20\ nm$. (a) single barrier
with $\Delta=0$. (b) double barrier with $\Delta=10\ nm$. (c, d)
single and double barrier transmission for the same parameters as
in (a, b), respectively, but for ($E=V=5\ \gamma_1$). Where
$T=\sum_{s,n=\pm}(T^{s}_{n})$ }\label{fig8}
\end{figure}

The transmissions coefficients of these evanescent modes are shown
in Figure \ref{fig8}a,\ref{fig8}b for a single and double barrier,
respectively. At high potential strength $(\ U_2=U_4=V\gg
\gamma_1)$ and for $\delta_2=\delta_4=0$, the four channels at
$E=V$ will give almost identical contributions,
$T_{+}^{+}=T_{+}^{-}=T_{-}^{+}=T_{-}^{-}$, for single and double
barrier because the electrons now can not differentiate between
the two modes, see Figure \ref{fig8}c,\ref{fig8}d.

%%%%%%%%%%%%%%%%%%%%%%%%%%%%%%%%%%%%%%%%%%%%%%%%%%%%%%%
\section{Conclusion}
%%%%%%%%%%%%%%%%%%%%%%%%%%%%%%%%%%%%%%%%%%%%%%%%%%%%%%%%%%%%
In conclusion, we have evaluated the reflection and transmission
probabilities of electrons through symmetric and asymmetric double
barrier potential in a bilayer graphene system. Based on the four
band model we found the solution in each potential region and by
matching them at the interface of each region and obtained the
different transmission and reflection coefficients. Subsequently,
the transmission of electrons through symmetric and asymmetric
double barrier structure for various barriers parameters was
investigated for energy ranges $E<\gamma_1$ and $E>\gamma_1$ where
there occurs one and two propagating mode, respectively.

We compared our results with previous work \cite{34} (For
$E<\gamma_1$) and showed that the cloak effect may occur for
non-normal incidence and exhibits a sequence of the resonances in
the transmission in the region $E<V$ due to bounded electrons in
the well between the two barriers. Furthermore, for normal
incidence we found that these resonances, which were absent for
the single barrier, always occur for fixed energy ($E<U_j$) when
$S_1=S_2$ where $S_1, S_2$, that it requires equality of the areas
of the first and second barrier. We also found that the most
important parameter that control the position and the number of
these resonances, in both cases $E<\gamma_1$ or $E>\gamma_1$, is
the well width between the tow barriers not the thickness of the
barriers in agreement with \cite{27,133}.

Introducing the interlayer potential difference open a gap in the
density plot of the transmission probabilities where it is not
completely suppressed as it the case in the single barrier
\cite{30}. This is a consequence of the bound states in the well
between the two barriers. The asymmetric structure of the double
potential barrier reduces the transmission probabilities and
removes the sharp resonant peaks. We observed that the resulting
conductance for the double barrier was different from that of the
single barrier. This difference manifests itself through the
presence of many extra resonances which are associated with the
bound electron states in the well.

The effect of the interlayer potential difference on the
transmission probabilities is reflected on the conductance where
we obtain a gap with non zero conductance. Moreover, the
asymmetric structure of the double barrier reduces the conductance
and removes the shoulders of main peaks. Finally, we studied the
conductance dependance on the double barrier parameters. The
conductance as a function of the height of the barrier showed a
region where it increases with increasing the potential height,
this is an odd behavior which can be correlated to the minimum
conductance around the Dirac point.

%%%%%%%%%%%%%%%%%%%%%%%%%%%%%%%%%%%%%%%%%%%%
\section*{Acknowledgments}
%%%%%%%%%%%%%%%%%%%%%%%%%%%%%%%%%%%%%%%%%
The generous support provided by the Saudi Center for Theoretical Physics (SCTP) is highly appreciated
by all authors. We acknowledge  the support of King Fahd University of
Petroleum and Minerals under research group project R61306-1 and
R6130-2.


\begin{thebibliography}{99} %{wide}
\bibitem{Geim} A.K. Geim,  and  K.S.  Novoselov,  Nature Materials 6, 183 (2007).
\bibitem{1}H. Brody, Nature (London) 483, S29 (2012).
\bibitem{2}O. Klein, Zeitschrift f\"{u}r Physik 53, 157 (1929).
\bibitem{3}M.I. Katsnelson, K.S. Novoselov, and A.K. Geim, Nature Physics 2, 620 (2006).
\bibitem{4}S.Y. Zhou, D.A. Siegel, A.V. Fedorov, F. El Gabaly, A.K. Schmid, A.H. Castro Neto and A. Lanzara, Nature Materials 7, 259 (2007).
\bibitem{5}R. Costa Filho, G. Farias, and F. Peeters, Physical Review B 76, 193409 (2007).
\bibitem{6}Y. Zhang, T.-T. Tang, C. Girit, Z. Hao, M.C. Martin, A. Zettl, M.F. Crommie, Y.R. Shen, and F. Wang, Nature 459, 820 (2009).
\bibitem{7}E. McCann, Physical Review B 74, 1 (2006).
\bibitem{8}J.D. Bernal, Proceedings of the Royal Society A: Mathematical, Physical and Engineering Sciences 106, 749 (1924).
\bibitem{9}C. Bena, and G. Montambaux, New Journal of Physics 11, 095003 (2009).
\bibitem{10}S.B. Trickey, F. Müller-Plathe, and G.H.F. Diercksen, Physical Review B 45, 4460 (1992).
\bibitem{11}E. McCann, and V. Fal'ko, Physical Review Letters 96, 1 (2006).
\bibitem{12}E. McCann, D.S.L. Abergel, and V.I. Fal'ko, Solid State Communications 143, 110 (2007).
\bibitem{13}E. McCann, D.S.L. Abergel, and V.I. Fal'ko, The European Physical Journal Special Topics 148, 91 (2007).
\bibitem{14}Z. Li, E. Henriksen, Z. Jiang, Z. Hao, M. Martin, P. Kim, H. Stormer, and D. Basov, Physical Review Letters 102, 16 (2009).
\bibitem{15}M. Koshino, New Journal of Physics 11, 095010 (2009).
\bibitem{16}A.H. Castro Neto, F. Guinea, N.M.R. Peres, K.S. Novoselov, and A.K. Geim, Reviews of Modern Physics 81, 109 (2009).
\bibitem{17}M. Killi, S. Wu, and A. Paramekanti, Physical Review Letters 107, 2 (2011).
\bibitem{18}S. Das Sarma, S. Adam, E. Hwang, and E. Rossi, Reviews of Modern Physics 83, 407 (2011).
\bibitem{19}A. Lherbier, S.M.-M. Dubois, X. Declerck, Y.-M. Niquet, S. Roche, and J.-C. Charlier, Physical Review B 86, 075402 (2012).
\bibitem{20}E. McCann and M. Koshino, Reports on Progress in Physics. Physical Society (Great Britain) 76, 056503 (2013).
%\bibitem{21}M.I. Katsnelson, K.S. Novoselov, and a. K. Geim, Nature Physics 2, 620 (2006).
\bibitem{22}H. Bahlouli, E.B. Choubabi, and A. Jellal, Europhysics Letters 95, 17009 (2011).
\bibitem{23}H. Bahlouli, E.B. Choubabi, A. Jellal, and M. Mekkaoui, Journal of Low Temperature Physics 169, 51 (2012).
\bibitem{24}A. Jellal, E.B. Choubabi, H. Bahlouli, and A. Aljaafari, Journal of Low Temperature Physics 168, 40 (2012).
\bibitem{25}A. El Mouhafid and A. Jellal, Journal of Low Temperature Physics 173, 264 (2013).
\bibitem{26}I. Snyman and C. Beenakker, Physical Review B 75, 045322 (2007).
\bibitem{27}J.M. Pereira, P. Vasilopoulos, and F.M. Peeters, Physical Review B 79, 155402 (2009).
\bibitem{28}M. Barbier, P. Vasilopoulos, and F.M. Peeters, Physical Review B 82, 235408 (2010).
\bibitem{29}T. Tudorovskiy, K.J. A. Reijnders, and M.I. Katsnelson, Physica Scripta T146, 014010 (2012).
\bibitem{30}B. Van Duppen and F.M. Peeters, Physical Review B 87, 205427 (2013).
\bibitem{31}Ya. M. Blanter, and M. B\"{u}ttiker, Physics Reports 336, (2000).
\bibitem{33}B. Van Duppen, S.H.R. Sena, and F.M. Peeters, Physical Review B 87, 195439 (2013).
\bibitem{133}C. Bai, and X. Zhang, Physical Review B 76, 075430 (2007).
\bibitem{34}N. Gu, M. Rudner, and L. Levitov, Physical Review Letters 107, 156603 (2011).

\end{thebibliography}
\end{document}